\documentclass[11pt,Chicago]{uuthesis2e-2011Dec}
\usepackage{uuthesis-2011}
\usepackage{graphicx}
\usepackage{amsmath}
\usepackage{amsfonts}
\usepackage{flafter}
\usepackage{afterpage}
\usepackage{natbib209-thesis}
\usepackage{verbatim}

\usepackage{tikz}
\usetikzlibrary{shapes,arrows}
\usetikzlibrary{shapes,snakes}
\usepackage{url}

\author{Paul D. Nu\~nez}

\title{Towards optical intensity\\ interferometry for high\\ angular resolution\\ stellar astrophysics}
\author{Paul D. Nu\~nez}

\thesistype{dissertation}
\graduatedean{Charles A. Wight}
\department{Department of Physics and Astronomy}
\degree{Doctor of Philosophy \\\vspace{0.3cm} in \\\vspace{0.3cm} Physics}
\departmentchair{David Kieda}
\committeechair{Stephan LeBohec}
\firstreader{Carleton DeTar}
\secondreader{Kyle Dawson}
\thirdreader{David Dobson}
\fourthreader{David Kieda}
\chairtitle{Professor}
\submitdate{August 2012} 
\approvaldate{June $\mathrm{4^{th}}$ 2012} % Date your committee approved this dissertation (defense date)
\copyrightyear{2012}

\fourlevels

\begin{document}

\frontmatterformat
\titlepage
\copyrightpage
\phdapproval
\preface{abstract2}{Abstract} % uses abstract.tex
%\dedicationpage
{
\setlength{\footskip}{20pt}
\tableofcontents
\listoffigures
}

%\maketitle

%\tableofcontents

\maintext  

%QUESTION for thesis lady: FOOTNOTES?
%Weinstein no permission
%page order

\chapter{High angular resolution\\ astronomy and coherent\\ light}
\vspace{-0.9cm}
\section{Introduction}

Most stars are still detected as unresolved point sources at optical wavelengths, and our understanding at high angular resolution results from physical modeling of their light spectrum and variability. Stellar surface brightness distributions can be predicted to some degree, but stellar atmospheres (may) have convection zones, mechanically driven matter flows, radiation driven winds, accretion, and other complex phenomena that are difficult to investigate without the recourse to high angular resolution imaging. Imaging surface structures at near-optical wavelengths may provide direct evidence for many of these effects and is another mean to test our current understanding of stellar atmospheres and stellar evolution. However, it is only for a few stars that actual images have been obtained using different techniques. 

Recent results of optical high angular resolution astronomy, particularly Michelson interferometry, have started to reveal stars as extended objects and have increased our understanding of effects such as those mentioned above. Reconstructed images of stars with non-uniform radiance distributions have, in some cases, confirmed our understanding of stars, but have also surprised us in others, reminding us that we have still much to learn. However, most stars are still beyond the angular resolution of current methods, which is why we propose to revive an extremely successful, yet abandoned technique, namely Stellar Intensity Interferometry (SII). Being insensitive to atmospheric turbulence and imperfections in optics, a previous SII experiment pioneered by Robert Hanbury Brown \citep{hanbury_brown0} produced more scientific results in less than a decade than several amplitude (Michelson) experiments combined several decades later.   

Over the course of this work, we have realized that the techniques used to study stars are just as interesting as stars themselves, which is why some historical and technical background is given in Chapters 1-3. SII will be compared to other existing techniques such as Michelson interferometry and speckle interferometry, which allows one to obtain data that is similar to intensity interferometry (Fourier magnitude data). We will discuss how interferometric data allows one to obtain high angular resolution images, and how these techniques typically allow to measure the Fourier transform of the radiance distribution of the star. The main challenge for recovering images of stars with SII is the lack of phase information, which is discussed in Chapter 4. Some new results in phase retrieval techniques were obtained with a method relying on the theory of analytic functions. Provided sufficient Fourier coverage and signal-to-noise, this method will allow to obtain high angular resolution images of stars. 

On the other side of the photon energy scale, ground-based gamma-ray astronomy, discussed in Chapter 5, is a flourishing field which has allowed to detect very high energy (TeV) radiation emitted from galactic and extragalactic objects. In Chapter 5, a short detour from high angular resolution astronomy is taken to analyze a particular high energy emitting binary system ($LSI+61^{\circ}\,303$) consisting of a fast rotating main sequence star experiencing mass loss and a compact object whose nature is still subject of debate. This object displays a TeV light-curve that shows a modulation with the same periodicity as the binary, and almost begs for the development of a toy model in terms that describes the $\gamma$-ray attenuation. When fitted to the data, this model allows us to constrain some fundamental parameters of the system \citep{apj_paper}. However, many of the constraints that can be placed with TeV observations are still debatable, and optical high angular resolution data may  answer many questions on the nature of this object. In fact, the optical requirements for optical SII are adequately met with the large light collectors used in $\gamma$-ray astronomy, and this has prompted a recent revival of optical SII using IACTs \citep{holder2}. 

The recent success of $\gamma$-ray astronomy has motivated the construction of large IACTs. In the case of the Cherenkov Telescope Array project (CTA), it is anticipated that there will be nearly 100 telescopes that will be separated by up to $1\,\mathrm{km}$ \citep{cta}.  When used as an intensity interferometers, CTA will complement the science done with existing amplitude interferometers by observing at shorter wavelengths ($\sim 400\,\mathrm{nm}$) and increasing the angular resolution by nearly an order of magnitude. In Chapters 6-7, the capabilities of IACTs used for SII will be quantified via simulations, and the most important result of this work was to prove that high angular resolution images can be obtained from optical SII data obtained from these arrays by applying the phase retrieval techniques discussed in Chapter 4 \citep{mnras_paper}. 

There are also ongoing efforts for performing an intensity interferometry measurement. The final chapter (8) starts with a discussion of our current experimental efforts to measure intensity correlations from thermal sources in the laboratory and from stars with a pair of small Cherenkov telescopes. The detection of intensity correlations from a thermal source has proven to be elusive due to the extremely short coherence times and shot-noise dominated data. In order to gain some experimental understanding, intensity interferometry data were obtained by using a light source with a much longer coherence time by breaking the coherence of laser light (pseudo-thermal light). Here, angular diameters of small pinholes emitting pseudo-thermal light were measured.

\section{A brief history of high angular resolution\\ astronomy and intensity interferometry}

The problem of resolving stars dates back at least 400 years. Even before considering the problem of imaging, a very old problem is that of measuring the angular diameter of stars. Galileo placed an upper limit on the angular size of \emph{Vega} by measuring the distance he needed to be from a string in order for the star to be obscured \citep{galileo}. Galileo's reported angular size of $5$ arcseconds is most likely a measure of atmospheric scintillation. Newton also estimated the angular size of stars by assuming they were like distant suns and found the angular diameter of a first magnitude star like Vega to be 2 milliarcseconds (mas), which is actually very close to the accepted value of 3 mas. 
%which is hard to believe since the maximum angular resolution of the eye is of the order of 1 arcminute.

By the late 1800's it was already known that the maximum angular resolution was not ultimately determined by the size of the aperture, but by the atmospheric turbulence or ``seeing'' of $\sim 1$ arcsecond. This limits the practical size of a telescope to about $10\,\mathrm{cm}$ for resolving small objects. Interferometry was what really revolutionized  high angular resolution astronomy when H. Fizeau (1868) proposed to mount a mask on a telescope to observe interference fringes. This was implemented by M. Steph\'an in the Marseilles observatory with an $80\,\mathrm{cm}$ telescope, the largest instrument at the time (1874). Since they clearly saw interference fringes on every star they pointed at, they provided an upper limit of 0.16 arcseconds for the angular diameters of stars\footnote{The contrast of fringes is smaller for sources that subtend a larger angle in the sky as will be shown in section \ref{classical_coherence}.} and improved the angular resolution by an order of magnitude. The angular resolution would still have to be improved by several orders of magnitude in order to detect stars as extended objects. The first successful measurement of angular diameters was performed by Michelson and Pease (1920) at Mt. Wilson, and this is considered the birth of high angular resolution astronomy. In his preliminary experiments with a $36''$ telescope, Michelson measured the angular diameter of Jupiter's Galilean moons \citep{michelson_1891}. With an interferometer with a maximum baseline\footnote{We will use the term ``baseline'' to refer to the separation between points where the light signal is received.} of $6\,\mathrm{m}$, they first measured the angular diameter of the red giant \emph{Betelgeuse} to be $47\,\mathrm{mas}$ \citep{michelson_betelgeuse}, and in total measured the angular diameter of 6 giant stars, all within tens of milliarcseconds.  

Michelson was aware of one of the main difficulties in optical stellar interferometry, namely that of atmospheric turbulence effects. In his preliminary experiments, he noticed that fringes would drift (jitter) in time as a result of turbulence induced path differences between the two interfering beams. Fringes can drift on timescales of the order of milliseconds, and Michelson's work was truly remarkable considering that he measured the fringe contrast (visibility) with the unaided eye. Pease (1930) then built a larger $15\,\mathrm{m}$ version of their previous interferometer, but this project failed to give reliable results and was later abandoned (1937). This ``failed'' experiment clearly shows another main difficulty in Michelson optical interferometry, namely the need for very high (subwavelength) precision optics. It took over three decades for the next breakthrough to occur in optical interferometry. 

The understanding of interference phenomena in terms of coherence theory was a next crucial step for the advancement of high angular resolution astronomy. Michelson never mentioned the word ``coherence'', but he knew how the  fringe visibilities should behave as a function of the baseline for uniform disks, limb-darkened stars and binaries. The first investigations of coherence phenomena are due to Verdet (1865) \citep{mandel}, when he used the angular diameter of the sun to estimate the (transverse) coherence length of the sun to be $0.1\,\mathrm{mm}$ (see Section \ref{classical_coherence}). Between the 1920s and the 1950s several notorious scientists, including Weiner, van Cittert, Zernike, Hopkins and Wolf, participated in further investigations of coherence theory. The outcome of these studies was a method to quantify  the correlations between fluctuating fields at two space-time points and the dynamical laws which correlations obey in free space \citep{mandel}.  A particularly important result was to relate the degree of coherence, partially obtained from the visibility of interference fringes, to the Fourier transform of the radiance distribution of the star. This last statement is known as the van Cittert-Zernicke theorem (section \ref{correlation_interpretation}), and led to the development of aperture synthesis, which was first applied in radio astronomy by M. Ryle (1952). Radio interferometry became a flourishing field and is still responsible for most of the highest angular resolution images available today. The techniques learned during this period along with improved understanding of coherence theory quickly led to the development of intensity interferometry. 

Intensity interferometry was born in 1949 as a result of attempts to design a radio interferometer that could measure the angular scale of two of the most prominent radio sources, Cygnus A and Cassiopeia A. R. Hanbury Brown conjectured that the electronic noise recorded at two different stations was correlated, that is, that low frequency intensity fluctuations were correlated. At the time, Hanbury Brown thought that the main advantage of this technique over conventional radio interferometry was that it did not require synchronized oscillators at two distant receivers so that observations could be done with at longer baselines. With the help of R. Twiss, a formal theory of intensity interferometry was developed \citep{hanbury_brown1} (Chapter 3), and before long, they built the first radio intensity interferometer and eventually succeeded to measure the expected correlations and the angular sizes of Cygnus A and Cassiopeia A (1952). They realized however that they had ``used a sledge-hammer to crack a nut'' since the baseline was only of a few kilometers, and this could have been accomplished with conventional radio interferometry. Nevertheless, this experiment made them realize one of the most important advantages of intensity interferometry: they noticed that even when the sources were scintillating violently due to poor atmospheric conditions, the correlation was not affected. Correlations of intensity are essentially unaffected by atmospheric turbulence (see Section \ref{revival} for more details). 

In order to test the theory at optical wavelengths, R. Hanbury Brown and R. Twiss performed a series of laboratory experiments to measure the expected correlations \citep{hanbury_brown_lab}. In these experiments, two photomultiplier tubes were placed a couple meters from a pinhole that was illuminated by a narrow-band thermal source, and the correlation was then measured as a function of detector separation. These experiments were surrounded by controversy, and were followed by several experiments which seemed to contradict Hanbury Brown's and Twiss's results (e.g., \citet{adam}, \citet{brannen}). Most of these experiments used broadband light sources, which correspond to low spectral densities, and would have needed extremely long integration times in order to detect significant correlations \citep{hanbury_brown0}. These experiments were confirmed several times (e.g., \citet{morgan-mandel}), and the subsequent quantum understanding of Hanbury Brown's and Twiss's work led to the development of quantum optics. 

\begin{figure}[h]
  \begin{center}
    \includegraphics[scale=0.4]{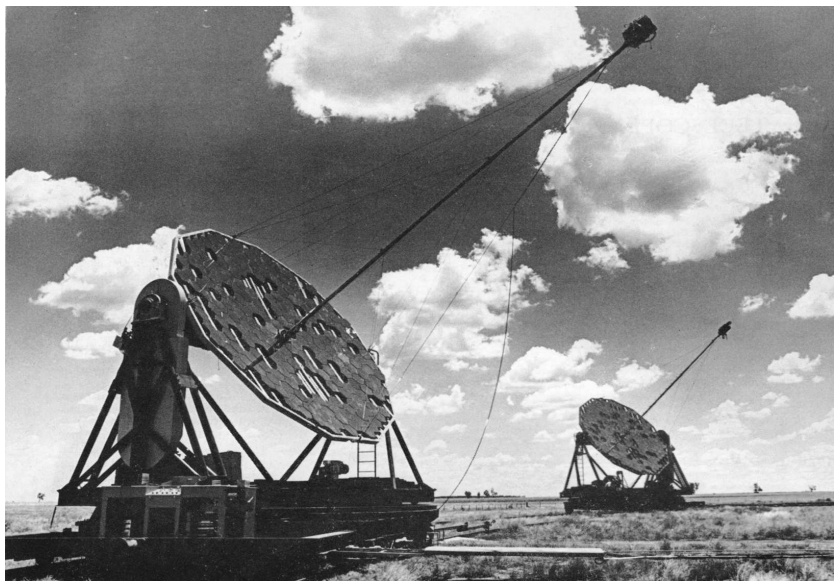}
    \caption[Light collectors used in the Narrabri stellar intensity interferometer]{\label{narrabri} Light collectors used in the Narrabri stellar intensity interferometer. Image from Sky \& Telescope, vol. 28, pp. 2-7, 1964.}
  \end{center}
\end{figure}

The following step was to construct an optical stellar intensity interferometer. Since large light collectors were needed, Hanbury Brown was concerned about the cost of two large optical telescopes. He soon realized that, for the same reason that measurements were insensitive to atmosphere induced path delays, there was no need for high precision optics, and ``light bucket'' type detectors could be used. In the small town of Narrabri (Australia),  two movable reflectors were placed on circular tracks (see Figure \ref{narrabri})  with a control station at the center were signals were correlated. The circular track allowed the baseline to be maintained perpendicular to the distant star. Between 1965 and 1972, the Narrabri stellar intensity interferometer measured the angular diameter of 32 bright stars, a considerable extension to the long standing catalog of 6 stars obtained previously by Albert Michelson.

The main work at Narrabri was completed when the diameters of the brightest stars in the southern hemisphere were measured, and the sensitivity limit of the instrument had been reached. Plans for a larger instrument were abandoned when it was shown that the same sensitivity could be reached with amplitude interferometry in $1/40$ of the time \citep{lipson}. Even so, it took over 30 years for amplitude interferometrists to overcome all the difficulties and to match the sensitivity of intensity interferometry. Even though stellar intensity interferometry was not further pursued, the same technique has been applied in other fields such as high energy particle physics to probe nuclei at very small scales\footnote{In this case by measuring correlations of particles with integer spin such as pions.} \citep{baym}. 

An interesting technique developed during this time by A. Labeyrie (1970) was that of \emph{speckle interferometry} \citep{speckle_interferometry}, where short exposures of a single telescope reveal atmospherically induced speckle patterns. Speckle patterns can be thought of as being generated by a random mask that changes at short (ms) time-scales, where each subaperture has a typical size ($\sim10\,\mathrm{cm}$) given by the atmospheric seeing. The interpretation of these speckle patterns allowed to obtain diffraction limited information as opposed to being limited by atmospheric seeing. This technique was used to obtain results on hundreds of fainter sources than those previously observed by Michelson, many of which were found to be binary stars.\footnote{Many of these observations were carried out by CHARA (Center for High Angular Resolution Astronomy) between 1977 and 1998, and the success of this program laid the groundwork for CHARA's entry into long baseline optical interferometry \citep{hartkopf}}

Labeyrie's work started a revival in optical stellar (amplitude) interferometry which resulted in the first successful beam combination of two telescopes separated by $12\,\mathrm{m}$  \citep{labeyrie_1975}. In this case, Labeyrie saw fringes on the bright star \emph{Vega}, and soon after, several projects developed the field that is now known as  \emph{optical long baseline interferometry}. Amplitude interferometry is currently producing very interesting results since technological advancements have made it possible to have baselines of the order of $\sim 100\,\mathrm{m}$ and to combine beams from more than two telescopes. This has resulted in several very high angular resolution images of rapidly rotating stars such as \emph{Vega} \citep{vega_image} and  \emph{Altair} \citep{altair_image}, and also the interacting binary $\beta$-\emph{lyrae} \citep{beta_lyrae}, and the eclipsing binary $\epsilon$-\emph{aurigae} \citep{epsilon_aurigae}. %Comment on phase retrieval?

\section{Theory of partially coherent light}

Interference phenomena are at the heart of high angular resolution astronomy observation, and therefore we start with a discussion on interference and its interpretation in terms of correlation theory. A plane wave can be formed from a point source located at infinity, and the study of interference phenomena allows us to measure deviations from a source being point-like. Light which has the same frequency and phase in all space is said to exhibit \emph{spatial coherence}, and light whose frequency and phase are well defined\footnote{A well defined frequency corresponds to a field expressed as a single plane wave as opposed to a superposition of plane waves of different frequencies.} over time is said to exhibit \emph{temporal coherence} \citep{mandel,lipson}. Interference can be clearly seen with plane monochromatic waves, which maximally exhibit spatial and temporal coherence, and the extent to which these phenomena can be observed is determined by the departure from these ideal conditions. 

\subsection{A classical discussion on coherence \label{classical_coherence}}

The amplitude of the electric field $E(t)$ of light at a fixed position can be expressed as the superposition of many monochromatic plane waves as

\begin{equation}
E(t)=\int E(\omega)e^{i\omega t}d\omega .
\end{equation}

If the electric field is non-zero during a time interval $\Delta t$, then $E(\omega)$ (which may be complex) has to have variations over a bandwidth $\Delta\omega$ comparable to variations of $e^{i\omega \Delta t}$ \citep{landau}, which occur when $\Delta\omega\sim 1/\Delta t$. That is 

\begin{equation}
  \Delta\omega\,\Delta t\sim 1.
\end{equation}

The time scale $\Delta t$ is known as the \emph{coherence time}, and is the time during which light can be considered approximately monochromatic with a defined phase. The length associated to this time is known as the coherence length given by

\begin{equation}
  \delta l\sim \frac{c}{\Delta \nu}=\left(\frac{\lambda}{\Delta\lambda}\right)\lambda.
\end{equation}

A similar argument can be made by expressing the electric field $E(x)$ at a particular time as a superposition of plane waves emitted from different angular positions $\theta$ of a far away source, i.e.,

\begin{equation}
  E(x)=\int E(\theta) e^{ikx\theta}d\theta.
\end{equation}

If the electric field is different from zero over a distance $\Delta x$, then $E(\theta)$ has to have variations over angles $\Delta\theta$ comparable to variations of $e^{ik\Delta x \theta}$, which occur when $\Delta\theta\sim 1/k\Delta x$. That is 

\begin{equation}
  k\Delta x\,\Delta\theta\sim 1.
\end{equation}

The length $\Delta x$ is known as the \emph{transverse coherence length} and is the length in which the wave is approximately planar. 

Another quantity of interest is the \emph{coherence volume}, given by 
\begin{equation}
  \Delta V\sim \Delta x^2 \Delta l\sim \frac{\lambda^4}{\Delta\theta^2\Delta\lambda}.
  \label{classical_volume}
\end{equation}

The physical significance of this volume will now be explained.

\subsection{Quantum interpretation of coherence}

It is interesting to note the connection with quantum mechanics by finding the volume of phase space $\Delta\mathcal{V}$ given by the uncertainty principle, that is

\begin{equation}
  \Delta \mathcal{V} \sim\frac{h^3}{\Delta p_x\Delta p_y\Delta p_z}.
\end{equation}

Within this volume, photons are indistinguishable, i.e. they belong to the same mode. If a source has a (small) angular size $\Delta \theta$, then $\Delta p_x=\Delta p_y\sim p \theta\sim \frac{h}{\lambda} \theta$, and 
$\Delta p_z\sim \frac{h}{\lambda^2}\Delta \lambda$. Therefore,

\begin{equation}
  \Delta \mathcal{V}\sim \frac{\lambda^4}{\Delta\theta^2\Delta\lambda}.
\end{equation}

This is equal to the coherence volume (eq. \ref{classical_volume}) found with classical arguments. In the quantum interpretation, the position of a photon is not defined within the coherence volume before a measurement takes place, and therefore interference phenomena can be seen. The classical example is given by the double slit experiment, where the wave-function can be described by the superposition of the photon having one position, or the other position, and the squared modulus yields a probability distribution with interference terms. In the quantum interpretation, the coherence volume is the region in which photons are indistinguishable \citep{mandel}. Even though they are indistinguishable, photons within the coherence volume are correlated (and not independent), since only symmetric states can occur between them. Brown and Twiss (1957) summarized the above by realizing that the physical significance of the coherence volume is that since photons within this volume can interfere\footnote{It is important to emphasize that photons do not interfere with each other, but with themselves, due to the probabilistic nature of their quantum state.}, they are indistinguishable. 

% It is very interesting that both the classical approach and a quantum approach give the same result. In fact, most of the results that will be discussed in this document can be described in semi-classical terms, where light can be regarded as a classical wave, and the concept of the photon only enters at detection time.  %Would be nice to comment more.

\subsection{Orders of magnitude for transverse coherence\\lengths and coherence times} \label{orders_of_mag}

As a first example, we can consider the sun, which is obviously not a point source, so we should not expect to detect spatially coherent plane waves. The angular size of the sun is $\sim 0.5^\circ$, so the transverse coherence length of the sun at $\lambda=500\,\mathrm{nm}$ is $\sim 0.05\,\mathrm{mm}$. This means we could observe interference fringes from the sun if slits were separated by less than $\sim 0.05\,\mathrm{mm}$. The most giant stars have angular diameters of tens of milliarcseconds, so that their transverse coherence length is of the order of tens of meters at $\lambda=500\,\mathrm{nm}$. Most stars have angular diameters of the order of $\sim 0.1\,\mathrm{mas}$, so that their transverse coherence length is of the order of a few hundred meters, hence the need for long baseline optical interferometry.

%A narrow filter, or an extremely narrow-band thermal source can have a bandwidth such that the coherence time is of the order of $\sim 10^{-8}\,\mathrm{s}$. 
The time resolution of fast photo-multiplier tubes is of the order of $\sim 10^{-9}\mathrm{s}$, which is several (3-4) orders of magnitude longer than the typical coherence time of a thermal light source, and poses a challenge for measuring temporal coherence effects (see Section \ref{photon_statistics}). Very coherent light sources such as lasers on the other hand, have much longer coherence times of the order of $10^{-4}\,\mathrm{s}$. 

\subsection{Interpretation in terms of correlations} \label{correlation_interpretation}

More understanding of interference phenomena can be gained by studying the correlation between partial beams of light that are separated in space and time. If two partial beams are superposed with a time delay between them that is much less than the coherence time, then the two are highly correlated. When a time delay of the order of the coherence time is introduced, the two beams no longer have the same frequency and phase, and therefore there is no correlation \citep{mandel, lipson}. 

A similar reasoning can be made with spatial coherence. A distant extended object, such as a star, can be thought of as a collection of points emitting spherical plane waves. The light emitted from different points in the object is statistically independent and not correlated. However, if light from the object is being detected at two points in space, the superposition of the light emitted from all points of the source is more correlated when the two detection points are close together. That is, two detectors that are close together receive essentially the same light signal. These statements can be made more quantitative by introducing the \emph{complex degree of coherence} between the electric field of light $E(\vec{r},t)$ at space-time points $(\vec{r}_1,t_1)$ and $(\vec{r}_2, t_2)$ as

\begin{equation}
  \gamma(\vec{r}_1, t_1;\vec{r}_2, t_2)=\frac{\left\langle E^*(\vec{r}_1, t_1)E(\vec{r}_2, t_2) \right\rangle}{\sqrt{\left\langle |E(\vec{r}_1, t_1)|^2 \right\rangle \left\langle |E(\vec{r}_2, t_2)|^2 \right\rangle}}
\end{equation}

The brackets $\langle ... \rangle$ indicate time averaging. The numerator in this expression arises naturally as an interference term when calculating the intensity of the superposition of two beams of light. The denominator is for normalization purposes. The complex degree of coherence is of central importance in interferometry, since it can be related to measurable quantities. In the case of amplitude interferometry, the complex degree of coherence can be related to the fringe visibility which is defined as

\begin{equation}
  V=\frac{I_{max}-I_{min}}{I_{max}+I_{min}}.
\end{equation}

In fact, if two monochromatic waves of the same frequency and with amplitudes $A_1$ and $A_2$ are combined, it is straightforward to show that the fringe visibility is\footnote{Eq. can be derived by first finding the intensity of the superposition of electric fields,  finding the maximum and minimum intensities, and noting that cross terms in the electric field correspond to the degree of coherence.} \citep{lipson}

\begin{equation}
  V=\frac{2A_1 A_2}{A_1^2+A_2^2}|\gamma(r_1, t; r_2, t)|. \label{visibility_eq}
\end{equation}

The visibility is the main observable in amplitude interferometry and allows us to measure the light coherence. Moreover, a measurement of the complex degree of coherence allows one to obtain information about the angular radiance distribution of the source as we will now show.

\subsection{Connection between source structure\\ and light coherence}

We will consider a monochromatic light source and choose to set the origin of coordinates at the light source. Points on the source are labeled by positions $\vec{x}'$, and the position of a point on the source with respect to a far away observer is 

\begin{equation}
 \vec{r}=\vec{x}-\vec{x}'.
\end{equation}

The observed electric field at $\vec{x}$ is

\begin{equation}
  E(\vec{x},t)=\int \frac{A(\vec{x}')}{|\vec{r}|}e^{ikr}d^2x',
\end{equation}
where $k$ is the wave number $2\pi/\lambda$, and we have omitted the time variation $e^{i\omega t}$ and the random phases induced by turbulence among other factors. Now we make the approximation %(these will be considered in more detail in section \ref{janvidas_work})

\begin{equation}
  |\vec{r}|\approx |\vec{x}|-\vec{x}'\cdot \frac{\vec{x}}{|\vec{x}|},
\end{equation}
so that 

\begin{equation}
  E(\vec{x},t)=\frac{e^{ikx}}{|\vec{x}|}\int A(\vec{x}')\,exp\left\{ik\vec{x}'\cdot \frac{\vec{x}}{|\vec{x}|}\right\}d^2x'.
\end{equation}

Now we calculate the time averaged correlation $\left\langle E(\vec{x}_i,t_i) E^*(\vec{x}_j,t_j)\right\rangle$ for the particular case when $t_i=t_j$. We should first note that 

\begin{equation}
  \left\langle A(\vec{x}')A(\vec{x}'')\right\rangle=I(\vec{x}')\delta(\vec{x}'-\vec{x}''),
\end{equation}
where $I(\vec{x}')$ is the light intensity at point $\vec{x}'$. This is because separate points on the source are not correlated over large distances. Now the time averaged correlation is

\begin{equation}
  \left\langle E(\vec{x}_i) E^*(\vec{x}_j)\right\rangle=C \int I(\vec{x}')\,exp\left\{ik\left(\vec{x}'\cdot \frac{\vec{x}_i}{|\vec{x}_i|}-\vec{x}'\cdot \frac{\vec{x}_j}{|\vec{x}_j|}\right)\right\}\, d^2x',
\end{equation}
where $C$ is a constant. When $|\vec{x}'|\ll |\vec{x}_i|$ and $|\vec{x}'|\ll |\vec{x}_j|$, we can write the angle $\vec{\theta}$ as 

\begin{equation}
  \vec{\theta}\equiv \frac{\vec{x}'}{|\vec{x}_i|}\approx \frac{\vec{x}'}{|\vec{x}_j|}.
\end{equation}

We can now express the correlation as

\begin{eqnarray}
\left\langle E(\vec{x}_i) E^*(\vec{x}_j)\right\rangle&=& \int I(\vec{\theta})\,e^{-ik\vec{\theta}\cdot(\vec{x}_i-\vec{x}_j)}\, d^2\theta \\
&=& \tilde{I}(\vec{x}_i-\vec{x}_j),
\end{eqnarray}
where $\tilde{I}(\vec{z}_i-\vec{z}_j)$ is the Fourier transform of the radiance distribution of the star. The Fourier transform goes from angular space to detector separation space. 

We have just shown that the complex degree of coherence is the normalized Fourier transform of the angular intensity distribution of the source \citep{lipson}. This last statement is known as the \emph{Van Cittert-Zernike theorem}. Therefore, by measuring the complex degree of coherence, or a related quantity, we can learn about the source structure. In Section \ref{correlation_interpretation} , we saw that the magnitude of the Fourier transform can be directly measured in amplitude interferometry with visibility measurements, and we will see that it can also be measured with intensity correlation measurements (Section \ref{intensity_correlations}). The phase of the Fourier transform is more elusive, and techniques have been developed for its measurement with amplitude interferometry (Section \ref{phase_in_michelson})  and its recovery with intensity interferometry (Section \ref{common_approaches}).

\chapter{Stellar astrophysics at high angular resolution}

Most stars are still merely detected as point sources and not as the extended and complex objects they truly are. One can only speculate as to what will be learned once more is known of this high angular resolution world. An analogy can be made with extragalactic astronomy, where one could ask what would be the status of the field if galaxies were regarded as unresolved point sources.  Some fundamental stellar parameters as well as several interesting effects can be studied with high angular resolution astronomy. In this chapter, a few representative topics of high angular resolution astrophysics are discussed. 

%An analogy can be made with extragalactic astronomy, where higher angular resolution allowed to view galaxies as distant collections of stars similar to our own galaxy, and dramatically changed the way we view the universe today.
\section{Angular diameters\label{angular_diameters}}

Stars can be characterized by measuring basic parameters such as the effective temperature, luminosity, chemical composition, mass and radius. A broad sample of these parameters provides effective constraints on stellar evolutionary models. Some of these quantities can be measured using traditional astronomical techniques. For example, the chemical composition can be measured by studying spectral lines. The effective temperature can be obtained with knowledge of the integrated light flux and the physical radius, which can be obtained by measuring the angular radius and the parallax (distance). In order to constrain the position of stars in the Hertzprung-Russell diagram, radii measurements with uncertainties of a few percent are necessary \citep{Kervella}. 

\begin{figure}[h]
 \begin{center}
 \includegraphics[angle=-90, scale=0.6]{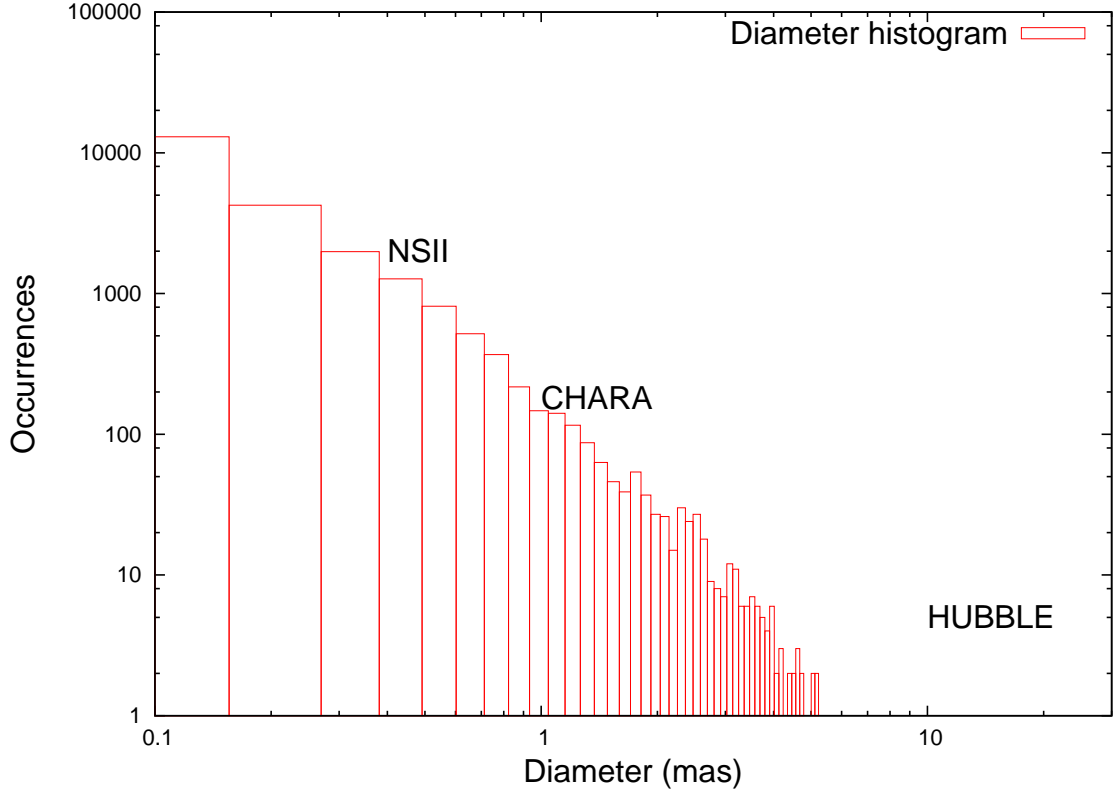}
 \end{center}
 \vspace{0.5cm}
 \caption[Number of stars for each estimated angular size from the JMMC stellar diameter catalog.]{\label{angular_size_histogram}Number of stars for each estimated angular size from the JMMC stellar diameter catalog. The approximate angular resolution at (near) optical wavelengths for NSII, CHARA, and the Hubble instruments is shown for reference.}
\end{figure}

The angular extent of stars is typically less than 1 milliarcsecond, and is only tens of milliarcseconds for even the most nearby giant stars. In Figure \ref{angular_size_histogram}, the  estimated angular diameter for 35000 stars in the JMMC \citep{catalog} catalog are shown, and we can see that the high angular resolution world really starts  below $\sim 10\,\mathrm{mas}$. Light received from most stars has transverse coherence lengths of several hundred meters (Section \ref{orders_of_mag}), so that model independent measurements of the physical size (at optical or near-optical wavelengths) can only be obtained through interferometric techniques \citep{lipson, chara_physics_today}, and there are still very few measurements of this basic fundamental parameter. At the

\noindent time of this writing, 242 stellar diameters have been directly measured, out of which only 24 correspond to main sequence stars \citep{chara_angular_diameters}. 

One of the main applications of angular diameter measurements is the understanding of stellar atmospheres. As was mentioned before, interferometric measurements can provide information of the effective temperature, and when combined with spectroscopic and spectrophotometric measurements, it can be used to create consistent models for stellar atmospheres \citep{lipson}. The challenge for stellar atmosphere modelers is to reproduce stellar line spectra from knowledge of these fundamental stellar parameters. This approach uses an angular diameter measurement, which may be extracted by fitting a uniform stellar disk model to the visibility measurements. However, stars are not uniform disks, and one such deviation from a uniform disk is limb darkening, which is the apparent decrease in surface brightness towards the edge of the star. An observer will see deeper into a partially absorbing atmosphere when viewing the star at its center than when observing the limb, and deeper layers are hotter (brighter) than outer layers. This effect can be modeled, and the correction can be applied to diameter measurements for a more precise determination of the effective temperature \citep{code}. A model-independent measurement of limb-darkening can be accomplished by analyzing visibility data beyond the first lobe if visibility data have a high enough signal-to-noise. As a result, the uncertainty in the effective temperature is no longer limited by interferometric measurements, but rather by photometric measurements. 

Other applications related to angular diameter measurements are the indirect calculation of stellar ages through isochrone fitting \citep{chara_angular_diameters}. Asteroseismology  and the study of stellar pulsation modes \citep{asteroseismology} will soon become possible with long baseline interferometry  due to the subpercent accuracy of angular diameter measurements.  %The next related application we will consider is the study of non-spherical fast-rotating stars. 

\section{Fast rotating stars \label{rotating_stars}}

Fast rotating stars are particularly interesting targets for SII, since they are normally hot. Rapidly rotating stars are typically young stars of spectral types O, B, and A; some are indeed rotating so fast that the effective gravity in their equatorial regions becomes very small (at critical rotation even approaching zero), and easily enables mass loss or the formation of circumstellar disks. Rapid rotation causes the star itself to become oblate, and induces gravity darkening. A theorem by \citet{von_zeipel} states that the radiative flux in a uniformly rotating star is proportional to the local effective gravity and implies that equatorial regions are dimmer than the poles. Spectral-line broadening reveals quite a number of early-type stars as rapid rotators and their surface distortion was already studied with the Narrabri interferometer, but not identified due to then insufficient signal-to-noise levels \citep{brown_1967}. Clearly, high angular resolution images will enable to see many of these interesting phenomena. 

\begin{figure}[h]
  \begin{center}
    \includegraphics[scale=0.3]{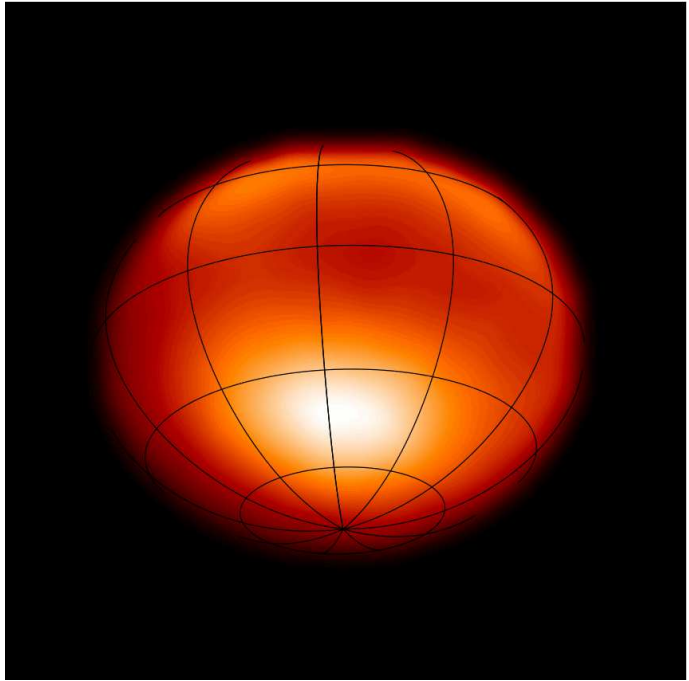}              
        %\caption{\label{altair} Surface of Altair obtained by the CHARA array.}                   
        \caption[Surface of $\alpha$-Cephei]{\label{altair} Surface of $\alpha$-Cephei. Stellar oblateness and pole brightenning can be seen. Courtesy of John Monnier.}                   
  \end{center}
\end{figure}

A number of these fast rotators have now been studied with amplitude interferometers. By measuring diameters at different position angles, the rotationally flattened shapes of the stellar disks are determined. For some stars, also their asymmetric brightness distribution across the surface is seen, confirming the expected gravitational darkening and yielding the inclination of the rotational axes. Aperture synthesis has permitted the reconstruction of images using baselines up to some $300\,\mathrm{m}$, corresponding to resolutions of $0.5\,\mathrm{mas}$ in the near-infrared H-band around $\lambda= 1.7\,\mu m$ \citep{zhao_2009}. In Figure \ref{altair}, an image reconstruction of $\alpha$-\emph{Cephei} obtained by the Center for High Angular Resolution Astronomy (CHARA) is shown \citep{acep}. In this image, we can clearly see the oblateness and pole brightening. Another interesting feature is that the bright spot does not appear to be exactly at the pole due to limb-darkening.  

With the relatively few stars that have been studied at high angular resolution, we have already been surprised at least once in the case of \emph{Vega}. This star has always been one of the standard calibration stars for the apparent visual magnitude scale. When \emph{Vega} was observed with the NPOI interferometer, it was learned that it is actually a fast rotating star viewed pole-on \citep{nature_vega}. 

Predicted classes of not yet observed stars are those that are rotating both rapidly and differentially, i.e., with different angular velocities at different depths or latitudes. Such stars could take on weird shapes, midway between a donut and a sphere \citep{macgregor}. There exist quite a number of hot rapid rotators with diameters of 1 mas or less. In fact, most hot ($T>10^{4\,\circ}\,\mathrm{K}$) stars in the JMMC stellar diameter catalog have diameters smaller than $1\,\mathrm{mas}$ (402 out of 418 hot stars). Clearly the angular resolution required to reveal such stellar shapes would be $0.1\,\mathrm{mas}$ or better, requiring kilometric-scale interferometry for observations around $\lambda= 400\, \mathrm{nm}$.  

A particularly interesting type of hot and fast rotating stars are \emph{Be} stars. These are rotating at near their critical velocities and have strong emission lines and infrared excess, which provide evidence for the presence of a circumstellar envelope due to mass loss. The detection of partially polarized ($\sim 1\%$) light also indicates the presence of a circumstellar disk in most of these stars \citep{stee}. The study of the kinematics of the circumstellar material permit to further understand the nature of the mass loss of these objects. For example, if circumstellar matter is radiatively driven, then angular momentum conservation predicts that the tangential velocity scales as $r^{-1}$. On the other hand, if the circumstellar disks are Keplerian, i.e., mechanically or viscosity driven, then tangential velocities scale as $r^{-1/2}$. Such kinematic studies\linebreak

\begin{figure}[h]
  \begin{center}
        \includegraphics[scale=0.6]{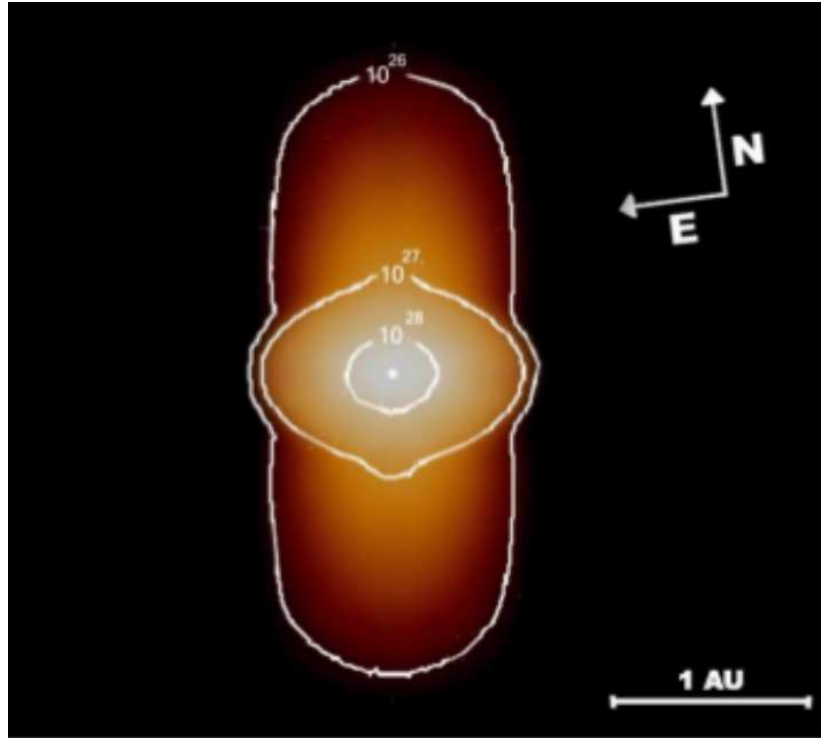}              
        \caption{\label{arae} $\alpha-Arae$ as imaged with the VLTI interferometer.}                   
  \end{center}
\end{figure}

\noindent can be done by noting that, for example, a fast rotating star viewed from its equator appears more red at one side, and more blue at the other. Therefore, images at different wavelengths reveal a shift in the image photo-center. Such studies are performed with the techniques of spectro-interferometry and spectro-astrometry \citep{oudmaijer}, and have revealed the presence of both an equatorial disk which is mechanically driven \citep{wheelwright}, and a polar stellar wind, which is radiatively driven \citep{stee}. Such is the case of $\alpha-Arae$, which was imaged with the VLTI array at $2.15\,\mu\mathrm{m}$ \citep{alpha_arae} as can be seen in Figure \ref{arae}.

\newpage

\section{Radiatively driven stellar mass loss}

Another use of imaging in connection with the astrophysics of hot stars is to quantify radiatively driven stellar 
mass loss. In radiatively driven stellar mass loss, matter is accelerated due to line scattering, which is the 
interaction of photons whose energy  matches a particular energy level spacing in an atom \citep{cak}.   One current method of measuring total
stellar mass loss is by analysis of P-Cygni\footnote{P-Cygni is a \emph{Be} star is the Cygnus constellation.} spectral line profiles, whose signature is
an asymmetry due to a blue shifted absorption. The blue shifted absorption in the spectral line is in turn due to the Doppler effect, i.e., as matter 
is accelerated by radiation, photons are red-shifted in the accelerated reference frame, and in an observers reference frame, only shorter wavelengths 
can meet the energy threshold for line scattering to occur. By analysis of P-Cygni spectral line profiles, only total 
mass loss rates have been measured so far\footnote{Radio (cm) wavelength observations can also yield mass loss rates.  The 
ionized winds of hot stars are free-free emitters and the flux scales as the density in the wind \citet{abbott_radio}.} \citep{puls}. With high angular resolution imaging, it will become possible to map out the distribution of mass loss 
across the stellar surface as we shall now describe. 

An interesting characteristic 
of mass loss in these types of objects is that radiative transfer provides a connection between the luminosity 
map and mass loss map in the star. Much of the theory
of radiatively driven stellar mass loss was developed by \citep{cak}, and the most important 
result that can be derived in connection to high angular resolution 
imaging is that the luminosity map $L(\theta_x, \theta_y)$ is related to the local mass loss rate  $\dot{M}(\theta_x, \theta_y)$ by a 
power law of the form 
$L(\theta_x, \theta_y)\propto \dot{M}(\theta_x, \theta_y)^{-\alpha}$. That is, an image of a star that is losing 
mass radiatively, provides a way to measure the mass loss rate at each point in  the star. The exponent $\alpha$ can be shown to be 
$2/3$ for hydrogen atmospheres, and has small deviations from this value when other elements are present \citep{puls}. Even though the work by 
\citep{cak} is still relevant today, and is appealing due to its simplicity, it gradually fails as the star's mass loss departs from a 
steady state and when winds are optically thick. For this reason, models that allow for departures from local thermodynamic equilibrium (LTE) 
have been developed \citep{model1, model2} and used to study total mass loss rates in bright stars. 

It would be very interesting to simulate the capabilities of IACT arrays for high angular resolution imaging of mass loss in hot stars, 
and in particular, to quantify the capabilities of imaging variations of mass loss rates across the stellar surface. From observations of variable 
features within P-cygni lines and spectropolarimetric studies, there is increasing indirect evidence for the existence of regions 
in the stellar surface which exhibit higher mass loss rates 
at scales comparable to the size of the star and with time scales of days \citep{clumps, clumps2}. The origin of these features is unknown, 
but possibly related to magnetic activity or stellar pulsation.  

Hot \emph{O} and \emph{B} type stars are known to have 
strong radiatively driven winds, and they happen to be ideal targets for SII due to their high spectral density. As an estimate of the 
number of sources that can be imaged, according to the JMMC stellar diameter catalog, $\sim 400$ hot ($T>10000^\circ\,\mathrm{K}$) stars can be imaged
in less than 10 hours of observation time with a future IACT array such as CTA. Even before investigating high angular resolution 
imaging capabilities, by studying the expected fidelity of the second lobe in the visibility function, we will already gain insight 
into the the capabilities of IACT arrays to measure mass loss. There are currently no published results which analyze second lobe data for hot
\emph{O} or \emph{B} type stars. 

In order 
to observe a localized region that is, for example, twice as 
bright as the rest of the star at blue wavelengths, it must experience a higher mass loss rate by a factor of 3 approximately. By extending 
existing non-LTE models such as the CoMoving Frame GENeral
 code (CMFGEN) \citet{model1}, we should be able to predict brightness contrast values more precisely. In view of 
some preliminary results obtained here, stellar mass loss maps can very likely be imaged. The extent to which stellar mass loss can be imaged needs to be 
further investigated if we wish to test stellar atmosphere models.

\section{Binary systems}

From the list of fundamental stellar parameters that was mentioned in Section \ref{angular_diameters}, a method for determining the stellar mass, arguably the
most important stellar parameter, was not discussed. The mass of a star essentially determines its fate. One way of finding stellar masses 
is through the determination of the orbital parameters of noninteracting binary systems. A considerable portion of the stars in our galaxy 
are in binary systems, so there is a large sample available to measure. In the case of noninteracting binary systems,  
the mass of the components does not change with time, so the determination of the mass allows us to test for stellar evolutionary models in general, 
i.e., not necessarily for stars belonging to multiple systems. 

When binary stars can be resolved, the orbital parameters can be found with a combination of spectroscopic and astrometric measurements. The spectroscopic
data permit the evaluation of the orbital velocities along the line of sight, whereas the astrometric (and parallax) data allow us to evaluate the 
orbital velocity in the plane perpendicular to the line of sight. However, in order to accurately measure the orbital parameters, a considerable portion of 
the orbit has to be observed, and the number of observable binaries is reduced to those having orbits observable within human time-scales. Binaries with short 
orbital periods have small angular sizes and are typically not resolved, so they have to be studied with long baseline interferometry techniques. Some of the 
orbital parameter measurements with long baseline techniques have been made with the Narrabri Stellar Intensity Interferometer (NSII), and more recently with
the Sydney University Stellar Interferometry (SUSI), and the Very Large Telescope Interferometer (VLTI). So far, interferometric studies have measured masses in
$\sim 15$ binaries with accuracies as small as a few percent \citep{quirrenbach}. 

As binary components get closer to each other, a whole new set of phenomena start to occur. Numerous 
stars in close binaries undergo interactions involving mass flow, mass transfer
and emission of highly energetic radiation: indeed many of the bright and variable X-ray
sources in the sky belong to that category. However, to be a realistic target for interferometry, and intensity
interferometry in particular, they must also be optically bright, which typically means B-star systems \citep{pasp}. 

There has been a recent interest in studying massive ($\sim 10M_\odot$) binaries across the whole electromagnetic spectrum. The classical \emph{Be} phenomenon, that was discussed in the previous section, has been associated to binary systems with hot massive stars. As mentioned above, \emph{Be} stars possess both a polar wind and an equatorial decretion disk, and the degree to which  each of these appear varies from one Be star to another. Binarity is thought to play a nonnegligible 
role, especially in the formation and/or truncation of the stellar disk \citep{millour}. 

One well-studied interacting and eclipsing binary is $\beta$ Lyrae ($m_v = 3.5$). The
system is observed close to edge-on and consists of a B7-type, Roche-lobe filling and mass-losing primary, 
and an early B-type mass-gaining secondary. This secondary appears to
be embedded in a thick accretion disk with a bipolar jet seen in emission lines, causing a
light-scattering halo above its poles. The donor star was initially more massive than the
secondary, but has now shrunk to about $3 M_\odot$ , while the accreting star has reached some $13
M_\odot$ . The continuing mass transfer causes the 13-day period to increase by about 20 seconds
each year \citep{harmanec}. 

\begin{figure}
\begin{center}
        \includegraphics[scale=0.5]{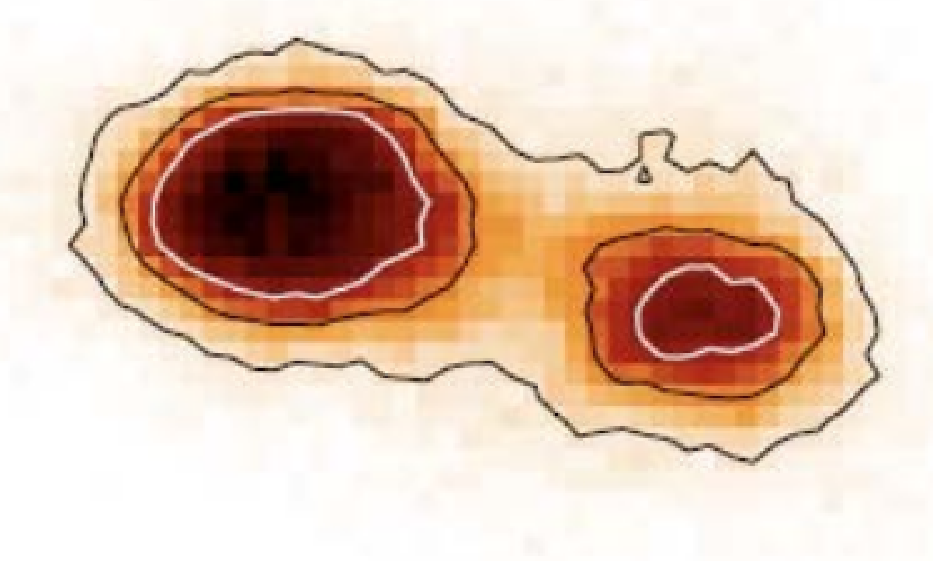}
\end{center}
\caption[First resolved near optical image of the binary system $\beta$-lyrae obtained by the CHARA array.]{\label{beta} First resolved near optical image of the binary system $\beta$-lyrae obtained by the CHARA array. The angular separation between 
components is $0.9\,\mathrm{mas}$}
\end{figure}

The first near-infrared optical image of the interacting binary system $\beta$-Lyrae was recently obtained by the CHARA group \citep{beta_lyrae}. With 
baselines up to $330\,\mathrm{m}$, the CHARA interferometer enabled the reconstruction of images at $2.2\,\mu\mathrm{m}$ and $1.6\,\mu\mathrm{m}$ which resolve both the donor star and the thick disk surrounding the mass gainer, located $0.9\,\mathrm{mas}$ away. A reconstruction obtained by the CHARA array is 
shown in Figure \ref{beta}. The donor star appears elongated, thus demonstrating the photospheric tidal distortion due to Roche-lobe filling. 

Another type of related objects are X-ray binaries. These systems
typically consist of a donor star and an accreting compact object, and can emit radiation as energetic as a few TeV. The high energy radiation
most likely originates from high accretion rates 
or shocks from to the interaction between the stellar and pulsar winds. In Section \ref{lsi_section}, we will discuss the case of the binary 
$LS\,I\,+61^\circ 303$, which consists of a hot \emph{Be} star and a compact object. This system may be too faint to be detected 
with interferometers, but it has been actively studied in the high energy (TeV) range by the VERITAS experiment \citep{lsi_veritas}. 

Here we have already started to quantify the capabilities of imaging binary systems with IACT arrays \citep{mnras}, and some detailed 
results are given in Section \ref{binary_sec}. Imaging the effects mentioned above will further our understanding of 
stellar evolution, the formation of compact objects, type Ia supernova, and 
even the formation of planetary systems around young stellar objects \citep{Karovska}.

\chapter{Intensity interferometry and light fluctuations}
\vspace{-0.9cm}

\section{Statistics of photo-electron detections}\label{photon_statistics}

Understanding the statistics of photo-electron detections is of central importance in intensity interferometry, since it ultimately determines the sensitivity of an intensity interferometry experiment. Here, we shall follow the discussion of chapter 9 in \citet{mandel}. 

The intensity of light is in general a fluctuating random variable. However, we shall first study the case in which the intensity $I(r, t)$ of the electromagnetic field is not a random variable, e.g. an ideal laser. The probability of detecting a photon in a small time $\delta t$ is proportional to the light intensity. Then the probability of detecting $n$ photons in a time interval $T$ is Poisson distributed, i.e.,

\begin{equation}
  \frac{dp(n, t, T)}{dT}=\frac{1}{n!}\left[ \eta\int_t^{t+T}I(r, t')dt'\right]^n\exp{\left[-\eta\int_t^{t+T}I(r, t')dt'\right]},
  \label{single_element}
\end{equation}
where the quantity in square brackets is the average number of detected photo-electrons, and $\eta$ is a constant that characterizes the detector. 

For a realistic light source (not an ideal laser), the intensity is actually a random variable whose distribution depends on the nature of the light source. Eq. \ref{single_element} is true for a single 
element of the ensemble of the intensity. Therefore $dp(n, t, T)/dT$ is not a Poisson distribution, but an average over equations of the form
of \ref{single_element}, that is

\begin{eqnarray}
  \frac{dp(n, t, T)}{dT}&=&\left<\frac{1}{n!}\left[ \eta\int_t^{t+T}I(r, t')dt'\right]^n\exp{\left[-\eta\int_t^{t+T}I(r, t')dt'\right]}\right>\\
  &=&\left< \frac{1}{n!}\mu^n e^{-\mu}\right>\\
  &=&\int_0^\infty \frac{1}{n!}\mu^n e^{-\mu}\frac{d\mathcal{P}(\mu)}{d\mu}d\mu \label{prob_detect}
\end{eqnarray}
where

\begin{equation}
  \mu\equiv \eta\int_t^{t+T}I(r, t')dt',
\end{equation}
and $d\mathcal{P}(\mu)/d\mu$ is the probability distribution for $\mu$. Therefore, in order to find the probability 
distribution $dp(n, t, T)/dT$, we need to find $d\mathcal{P}(\mu)/d\mu$ corresponding to the source under consideration. 

\subsection{The statistics of a thermal source with a slow detector}\label{slow_detector}

To find $d\mathcal{P}(\mu)/d\mu$, we can first find $d\mathcal{P}(I)/dI$, the probability distribution for the intensity (number of photons before 
going through the Poisson detector). In the case of a thermal source, consisting of many uncorrelated oscillators, the electric field ($E\propto\sqrt{I}$) is Gauss distributed because it is the sum of many independent random variates (central limit theorem). Since the electric field is a complex quantity $E=x+iy=Ae^{i\phi}$, the 
probability distribution of the real and imaginary parts is the product of both distributions, i.e.,

\begin{equation}
  \frac{d^2p(x,y)}{dxdy}=\frac{1}{2\pi\sigma^2}e^{-(x^2+y^2)/2\sigma^2}.
\end{equation}

This can also be expressed as a function of the magnitude and the phase by noting that $dxdy=AdAd\phi$

\begin{equation}
  \frac{d^2p(A,\phi)}{dAd\phi}=\frac{A}{2\pi\sigma^2}e^{-A^2/2\sigma^2}.
\end{equation}

Here the probability distribution for the phase is $dp(\phi)/d\phi=1/(2\pi)$, and the distribution for the amplitude is

\begin{equation}
  \frac{dp(A)}{dA}=\frac{A}{\sigma^2}e^{-A^2/2\sigma^2}.
\end{equation}

Now the distribution as a function of the intensity is found by noting that $dI/dA=2A$ and that $2\sigma^2=<I>$, so that 

\begin{equation}
  \frac{dP(I)}{dI}=\frac{1}{<I>}e^{-I/<I>}, \label{exponential}
\end{equation}
which is an exponential distribution.

We can calculate $d\mu/dI$ to find $dP(\mu)/d\mu$ to finally calculate the probability distribution \ref{prob_detect} for 
detecting $n$ photo-electrons. In general, 
this probability distribution is not Poissonian. 

We now consider the limiting case when $I(r,t)$ undergoes large variations 
in the time $T$, or equivalently, when the electronic resolution time is much larger than the coherence time $\tau_c$. Here $\mu$ has negligible fluctuations
and can be approximated by

\begin{equation}
  \mu=\eta\left\langle I(r,t)\right\rangle T.
\end{equation}

This in turn implies that $d\mathcal{P}(\mu)/d\mu\approx \delta(\mu-\left\langle I(r,t)\right\rangle T)$ (Dirac distribution), so that the 
distribution $dp(n, t, T)/dT$ is Poissonian and we have reproduced eq. \ref{single_element}, corresponding to the case of no fluctuations. As
soon as the distribution $d\mathcal{P}(\mu)/d\mu$ has a finite width, we start to see deviations from Poisson statistics. 

\subsection{The variance of the super-Poisson distribution\label{variance}}

We now calculate the variance of the probability distribution given by equation \ref{prob_detect}. Following \citep{mandel}, The average number of photons in a time interval $T$ is 

\begin{eqnarray}
  \left\langle n \right\rangle &=& \sum_{n=1}^\infty n\frac{dp(n,t,T)}{dT}\\
  &=& \int_0^\infty d\mu \sum_{n=1}^\infty n\frac{\mu^n}{n!}e^{-\mu} \frac{dp(\mu)}{d\mu}\\
  &=& \int_0^\infty \mu\frac{dp(\mu)}{d\mu}d\mu\\
  &=& \left\langle \mu \right\rangle.
\end{eqnarray}

Following similar arguments, 

\begin{eqnarray}
  \left\langle n^2 \right\rangle &=& \sum_{n=1}^\infty n^2\frac{dp(n,t,T)}{dT}\\
  &=& \left\langle \mu^2 + \mu \right\rangle,
\end{eqnarray}
so that the variance is 

\begin{eqnarray}
  \left\langle \Delta n^2 \right\rangle &=& \left\langle n^2+\left\langle n\right\rangle ^2-2n\left\langle n \right\rangle \right\rangle\\
  &=& \left\langle \mu^2 + \mu \right\rangle + \left\langle \mu \right\rangle^2-2\left\langle \mu \right\rangle^2\\
  &=& \left\langle \mu^2 \right\rangle + \left\langle \mu \right\rangle -\left\langle \mu \right\rangle^2\\
  &=& \left\langle n \right\rangle + \left\langle \Delta\mu^2 \right\rangle. \label{variance_general}
\end{eqnarray}

The fluctuations in the detected number of photons reflect the fluctuations in the light intensity integrated over the resolution time. In the case of the thermal source and a ``slow'' detector ($T\gg \tau_c$, where $\tau_c$ is the coherence time), the probability distribution for the integrated light intensity $\mu$ can be found by using the central limit theorem. The variance of the intensity can be calculated from eq. \ref{exponential} as

\begin{eqnarray}
  \left\langle \Delta I^2 \right\rangle &=&\left\langle I^2 \right\rangle - \left\langle I \right\rangle^2\\
                                &=& \int_0^\infty \frac{I^2}{\left\langle I \right\rangle^2} e^{-I/\left\langle I \right\rangle}dI  -
                                      \left( \int_0^\infty \frac{I}{\left\langle I \right\rangle^2} e^{-I/\left\langle I \right\rangle}dI \right)^2\\
                                &=& \left\langle I \right\rangle^2 .
\end{eqnarray}

Within the electronic time resolution $T$, there are $\tau_c/T$ possible intensity values, where $\tau_c$ is the coherence time. Therefore, from the central limit theorem, as the number of different intensity values ($T/\tau_c$) increases with electronic resolution time, the variance of the distribution of the integrated light intensity decreases as $\tau_c/T$. The variance of the integrated light intensity is then

\begin{equation}
   \left\langle \Delta\mu^2 \right\rangle= \left\langle \mu \right\rangle^2 (\tau_c/T).
\end{equation}

An important result that we have derived for the case of a slow detector is that 

\begin{equation}
  \boxed{\left\langle \Delta n^2 \right\rangle=  \left\langle n \right\rangle + \left\langle n \right\rangle^2\frac{\tau_c}{T}}. \label{mu_prob}
\end{equation}

Here it is clear that the statistics are no longer purely Poissonian because the variance is no longer equal to the mean. These statistics are commonly known as ``super-Poisson'' statistics, and we can see from eq. \ref{mu_prob}, that deviations from Poisson fluctuations can only be seen with detectors whose resolution time is not too far away from the coherence time. A photo-detector measures a Poisson part, whose fluctuations are described by the first term in eq. \ref{mu_prob}, usually called \emph{shot noise}, and a part related to intensity fluctuations, described by the second term in eq. \ref{mu_prob}, usually called \emph{wave noise}. 
%Comment on ease of radio observations ?

\subsection{A Monte-Carlo simulation of photon-electron detections}\label{monte_carlo}

\begin{figure}
  \begin{center}
    \rotatebox{-90}{\scalebox{0.45}{\includegraphics{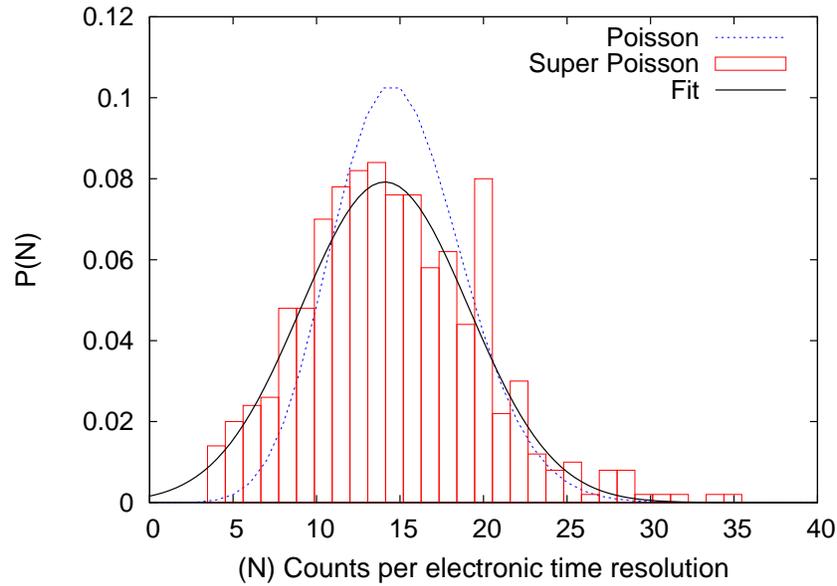}}}
    \vspace{1cm}
    \caption[Probability distribution of the number of photons per resolution time.]{\label{super_poisson}Probability distribution of the number of photons per resolution time. This corresponds to a source for which  an average of 15 photons per resolution time can be detected. Here the electronic resolution time is $T=10^{-8}\,\mathrm{s}$, and the coherence time is $\tau_c=5\times 10^{-10}\,\mathrm{s}$. The pure Poisson distribution is compared with a Gaussian fit, and deviations from Poisson statistics start to become visible.}
  \end{center}
\end{figure}

In Section \ref{slow_detector}, we studied the limiting case for a thermal source and a slow detector. We shall now consider the case in which deviations from Poisson statistics start to become visible. When the probability distribution for the integrated light intensity (eq. \ref{mu_prob}) is inserted in eq. \ref{prob_detect}, it is no longer straight-forward to derive an analytical expression for the probability of detecting $n$ photons in a time $T$. However, a Monte-Carlo simulation can be made computationally by generating random events with this probability distribution. We simulate a source that produces on average $1.5\times 10^9$ photo-electrons per second approximately. For a realistic photo-multiplier tube, the resolution time is taken to be $T=10^{-8}\,\mathrm{s}$. The coherence time for thermal light is typically much shorter than the resolution time, but for illustrative purposes, the probability distribution is shown for a coherence time  of $\tau_c=5\times 10^{-10}\,\mathrm{s}$ in Figure \ref{super_poisson}, so that deviations from Poisson statistics can be more easily seen. Here we took an observation time of $5\times 10^{-6}\,\mathrm{s}$, so that there are $500$ simulated measurements. The standard deviation of the distribution shown in Figure \ref{super_poisson} is approximately $5$ photons per resolution time, which is one more photon per resolution time than if the distribution were purely Poissonian. A more detailed simulation of an intensity interferometer is described in \citet{janvida}

\section{Intensity correlations\label{intensity_correlations}}

\begin{figure}
  \begin{center}
    \includegraphics[scale=0.45]{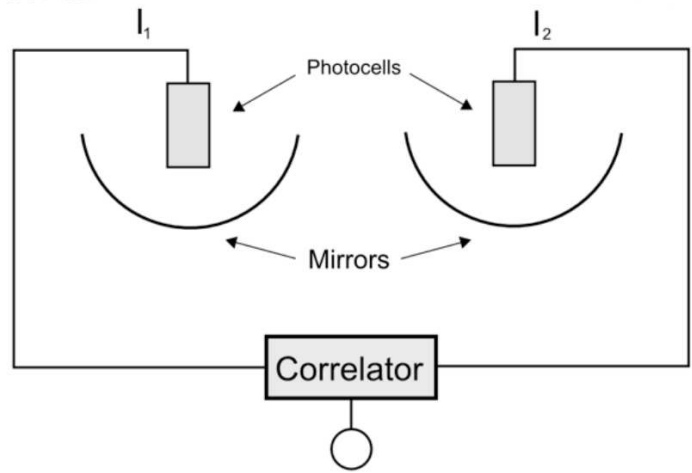}
    \caption[Schematic of the principle of an intensity interferometry experiment.]{\label{ii_schematic} Schematic of the principle of an intensity interferometry experiment. If both light collectors are within the coherence volume of light, then the current (intensity) fluctuations are correlated.}
  \end{center}  
\end{figure}

The wave noise can, in principle, be measured with a single detector. However, realistic sources have a smaller coherence time than the one considered in Section \ref{monte_carlo}, and can be as small as $10^{-14}\,\mathrm{s}$. Nevertheless, one can still see the effect by measuring the correlation between neighboring photo-detectors as illustrated in Figure \ref{ii_schematic}. The Poisson fluctuations are not correlated between detectors, but the fluctuations due to intensity variations are correlated when both detectors are located within the same coherence volume (see section \ref{classical_coherence}). This is because within the coherence volume, photons are indistinguishable and only symmetric states can occur between them, that is, they are correlated by the Bose-Einstein distribution. For the purpose of astronomical intensity interferometry, the correlation can be understood classically in the sense that both detectors are being driven by the same wave, and the wave has a definite frequency and phase within the coherence volume. Therefore, the wave noise is correlated between detectors in the same coherence volume.

There is additional information contained in intensity correlations as we shall now see. 
Intensity interferometry allows us to measure correlations of intensities between pairs of telescopes, averaged 
over the signal bandwidth (denoted as $\langle \ldots \rangle$). If we express the instantaneous intensity as $I=\bar{I}+\Delta I$, where 
$\Delta I=I(t+\delta t)-I(t)$, the measurable quantity, denoted as $\gamma^{(2)}$, is therefore

\begin{eqnarray}
  \gamma^{(2)} &=& \frac{\left\langle I_1 I_2 \right\rangle}{\left\langle I_1 \right\rangle \left\langle I_2 \right\rangle}  \label{corr0}\\
  &=& \frac{\left\langle (\bar{I}_1+\Delta I_1) (\bar{I}_2+\Delta I_2) \right\rangle}{\left\langle \bar{I}_1+\Delta I_1 \right\rangle \left\langle \bar{I}_2+\Delta I_2 \right\rangle}\\
  &=& \frac{\left\langle \bar{I}_1 \bar{I}_2 + \bar{I}_1 \Delta I_2 + \Delta I_1 \bar{I}_2 + \Delta I_1\Delta I_2\right\rangle}{\left\langle \bar{I}_1 \right\rangle \left\langle \bar{I}_2 \right\rangle}\\
  &=& 1+ \frac{\left\langle \Delta I_1\Delta I_2 \right\rangle}{\left\langle I_1 \right\rangle \left\langle I_2 \right\rangle}. \label{corr1}
\end{eqnarray}

In the previous expression we have used the fact that $\left\langle \Delta I \right\rangle=0$. Eq. \ref{corr1} 
expresses the fact that we measure correlations of intensity fluctuations. Now we make the connection between $\gamma^{(2)}$ and the degree
of coherence $\gamma^{(1)}\equiv\gamma$. Assuming that the electric fields are Gaussian random variates 
and making use of the \emph{Gaussian moment theorem}\footnote{Gaussian variates have the property 
that all higher order correlations can be expressed in terms of second order correlations. That is,\\ $\left\langle z^*_{i1} z^*_{i2} \ldots z_{iN} z_{j1} z_{j2} \ldots z_{jN} \right\rangle = \sum_{\mathrm{N!\,\, pairings}} \left\langle z^*_{i1} z_{j1}  \right\rangle \ldots \left\langle z^*_{iN} z_{jN}  \right\rangle$}, we can also rewrite eq. \ref{corr0} as

\begin{eqnarray}
   \gamma^{(2)}  &=& \frac{\left\langle E_1 E^*_1 \right\rangle \left\langle E_2 E^*_2 \right\rangle + \left\langle E_1 E^*_2 \right\rangle \left\langle E^*_1 E_2 \right\rangle}{\left\langle E_1 E^*_1 \right\rangle \left\langle E_2 E^*_2 \right\rangle}\\ 
  &=& 1 + \frac{ \left|\left\langle E_1 E^*_2 \right\rangle \right|^2 }{\left\langle| E_1 |^2 \right\rangle \left\langle |E_2|^2 \right\rangle}\\
  &=& 1 + |\gamma|^2.
\end{eqnarray}

A comparison with eq. \ref{corr1} yields the following important result.

\begin{equation}
  \boxed{\frac{\left\langle \Delta I_1\Delta I_2 \right\rangle}{\left\langle I_1 \right\rangle \left\langle I_2 \right\rangle}=|\gamma|^2}. \label{boxed_result}
\end{equation}

Since $\gamma$ is the (complex) Fourier transform of the radiance distribution of the object in the sky, measuring intensity correlations enables the squared modulus of this quantity to be measured \citep{hanbury_brown0}. Therefore, the phase of the Fourier transform is lost during the measurement process, and it must be recovered for model-independent imaging (chapter \ref{phase_recovery}).

\section{Signal-to-noise in intensity interferometry} \label{snr_section}

From the discussion of equation \ref{mu_prob} we learned that the variance of photon counts, or equivalently the intensity fluctuations, contain two contributions: Poissonian shot noise, and wave noise, and it is the latter contribution that is correlated between detectors. When the degree of coherence is maximum ($|\gamma|^2=1$), the signal-to-noise ratio per electronic resolution time  $T$ is then the ratio of the wave noise and the shot noise, i.e.,

\begin{equation}
  SNR=n\frac{\tau_c}{T}. \label{snr1}
\end{equation}

Here $n$ is the rate of detected photons within a certain optical bandwidth $\Delta \nu$, and $\tau_c$ is the coherence time. To further develop the previous expression, we can write the rate $n/T$ as 

\begin{equation}
  \frac{n}{T}=\int \frac{d^2n'}{d\nu dT} d\nu,
\end{equation}
so that for a small bandwidth $\Delta\nu=1/\tau_c$

\begin{equation}
  \frac{n}{T}=\frac{d^2n'}{d\nu dT}\frac{1}{\tau_c}.
\end{equation}

We should also note that when observing during a time $T_0$, the signal-to-noise increases as $\sqrt{T_0/T}$, so the signal-to-noise becomes

\begin{equation}
  SNR=\frac{dn'}{dT}\sqrt{T_0\Delta f/2},
\end{equation}
where $\Delta f$ is the electronic bandwidth and the factor of 2 is due to Nyquist's theorem. Now the rate of detected photons centered around a particular frequency can be written in terms of the rate of photons per area $N$ (centered around a particular wavelength), the detector area $A$, and the quantum efficiency $\alpha$ as 

\begin{equation}
  \frac{dn'}{dT}=NA\alpha,
\end{equation}
and the signal-to-noise for an arbitrary value of $|\gamma|^2$ is now \citep{hanbury_brown0}

\begin{equation}
  SNR=NA\alpha|\gamma|^2\sqrt{T_0\Delta f/2}. \label{snr2}
\end{equation}

It is important to emphasize that $N$ is a property of the light source, therefore the signal-to-noise depends on the brightness of the object at the observed wavelength. It is also important to note that the signal-to-noise is independent of the optical bandwidth. The sensitivity cannot increase indefinitely by increasing $A$,  since at some point the light collector will start to be large enough to resolve the light source and add uncorrelated intensities to the signal, therefore canceling the effect we want to measure. The integration time as well cannot be beneficially increased indefinitely since the finite point spread function of the optics results in integrating background light. Integrating background light places a limit on the minimum brightness the source can have, but does not pose a serious limitation for making precision measurements on a source that is much brighter than the night sky background. The electronic time resolution can in principle be as high as possible, but at some point the detected intensity fluctuations will be so fast that high precision optics are needed, therefore introducing additional technical difficulties associated with Michelson interferometry. A detailed discussion on the sensitivity of a modern stellar intensity interferometer is presented in Section \ref{sensitivity}.

\chapter{Phase recovery} \label{phase_recovery}
\vspace{-0.9cm}
\section{Alternatives for imaging}

The imaging problem in intensity in interferometry is then reduced to finding the phase of the complex degree of coherence. There are several alternatives: The first is to live with the fact that the phase is not directly measured, and recover images using parametric models. In many cases, even when the phase is partially known, data are fit to a parametric model and relevant physical quantities are extracted from the model.  

The second option is to measure third order correlations between intensity fluctuations, similar to what is done in amplitude interferometry with the phase closure technique, and also similar to what is done in speckle interferometry. The phase problem in amplitude interferometry is discussed in section \ref{phase_in_michelson}. 

 The last option consists of tackling the phase retrieval problem from the Fourier magnitude data, and is discussed in sections \ref{phase_retrieval_physics} 
and \ref{common_approaches}. At first glance, the problem seems quite hopeless since any phase one postulates is consistent with the (measured) Fourier magnitude. However, we show in section \ref{uniqueness}, that since the Fourier transform of a function with finite support\footnote{For example, a star a star $\mathcal{O}(\theta)$  of angular size $\Theta$ has finite support, i.e., $\mathcal{O}(\theta)=0\,\,\,\forall\,\,\,\theta\,\,\,s.t.\,\,\,|\theta|>\Theta$. } is analytic in the $(u,v)$ plane,  the phase can in principle be found by analytic continuation. We then present several approaches to the phase retrieval problem that make use of the theory of analytic functions in section \ref{common_approaches}.

\section{Phase retrieval problems in physics }\label{phase_retrieval_physics}

The phase retrieval problem and several related inverse problems arise in many fields of physics. Most of the phase retrieval problems arise when
a wave is scattered off an object, then the information of the object is contained in both the magnitude and the phase of the propagating wave. When only
the magnitude of the wave is measured, the phase is also needed to describe the object as accurately as possible. One of the earliest applications was in 
X-ray crystallography \citep{ladd}, where a periodic crystal creates a diffraction pattern corresponding to the squared modulus of the so called structure factor. The
structure factor is the Fourier transform of the electron density function, and since only the squared modulus is measured, the phase needs to be recovered
to determine the crystal structure. Another example unrelated to astronomy and optics is found in quantum mechanical scattering \citep{felcher}, where the observable is the probability amplitude or squared-modulus of the wave function. The phase problem arises when one seeks the functional form of the interaction potential 
from knowledge of the squared-modulus of the wave function. This can be understood within the Born approximation, where the first order correction to the 
wave function of the outgoing particle is proportional to the Fourier transform of the potential. Only the magnitude of this Fourier transform can
be measured, and the phase needs to be recovered to solve for the potential. Other examples where the phase problem is encountered include 
electron microscopy \citep{huiser} and  wave-front sensing \citep{gonsalves}. Here we concentrate on the application of the phase retrieval problem to astronomical intensity interferometry.

\section{Phase retrieval in amplitude interferometry} \label{phase_in_michelson}

To illustrate the fact that interference fringes contain phase information, consider the case of the following double slit experiment: Two 
narrow beams with a phase difference $\Delta$ between them interfere to form a diffraction pattern. Formally, the observed diffraction pattern is the 
intensity of the Fourier transform of the following radiance distribution

\begin{equation}
  \mathcal{B}(\theta)=e^{i\Delta/2}\delta(\theta+a/2)+e^{-i\Delta/2}\delta(\theta-a/2),
\end{equation}
where the slit separation is $a$. The observed diffraction pattern is therefore

\begin{eqnarray}
  |\gamma(x)|^2 &=& A|e^{i(kxa+\Delta)/2}+e^{-i(kxa+\Delta)/2}|^2\\
  &=& 2A|\cos{(kxa+\Delta)}|^2,
\end{eqnarray}
where $A$ is a constant specified by the detector characteristics. Therefore, the sinusoidal 
diffraction pattern is displaced by and amount $\Delta$. However, the problem is that besides the true phase difference
$\Delta$, there  are also phase differences induced by atmospheric fluctuations in time-scales of the order of a few milliseconds \citep{lipson}. Therefore,
atmospheric fluctuations have the effect of drifting fringes in time. 

The approach used in amplitude interferometry is to apply a technique known as \emph{phase closure} \citep{jennison}. To 
briefly illustrate this approach, consider an array of detectors that can be divided into closed loops of three detectors (triangles) $ijk$. The 
signal at each detector contains an atmospheric phase shift ($\Delta_{0,i}$, $\Delta_{0,j}$, $\Delta_{0,k}$). The phase at each detector is then
$\phi_i=\Delta_i-\Delta_{0,i}$, and the atmospherically modified coherence 
function between telescopes $i$ and $j$ is \citep{lipson}

\begin{equation}
  \gamma^a_{ij}=\gamma_{ij}e^{i(\Delta_{0,i}-\Delta_{0,j})},
\end{equation}
where $\gamma_{ij}=|\gamma_{ij}|e^{i(\Delta_i-\Delta_j)}$. Therefore, the product of the coherence functions along the closed loop is 

\begin{equation}
  \gamma^a_{ij}\gamma^a_{jk}\gamma^a_{ki}=|\gamma_{ij}\gamma_{jk}\gamma_{ki}|e^{\Delta_i+\Delta_j+\Delta_k}.
\end{equation}

The most important thing to note is that this quantity is independent of atmospheric turbulence as long as fringes are scanned in
timescales smaller than atmospheric fluctuations (milli-seconds). The sum of phases in the exponential is a measurable quantity
and is known as the \emph{closure phase}. For an array of $N$ receivers, there are $N(N-1)/2$ baselines and $(N-1)(N-2)/2$ independent triangles. 
Therefore, in a nonredundant array, there are $N-1$ more unknowns for the phase than there are closure phase equations. The procedure to 
find the phase consists in measuring the closure phase in each closed loop of the array,\footnote{Two noncollinear phase differences can 
be set to zero.} and the phase can be completely specified if there are enough redundant baselines.

%\section{Phase retrieval in speckle interferomertry (not done)}
%MISSING SPECKLE INTERFEROMETRY

\section{Uniqueness \label{uniqueness}}%Well defined problem

%Existence because every polynomial can be expressed as the square of another polynomial ?

As was stated before, the phase of the Fourier transform has to be recovered from magnitude information only in intensity interferometry. To gain some intuition on the phase recovery problem, first 
consider the one-dimensional case of an object $\mathcal{B}(\theta)$ of finite extent. The Fourier transform of the one-dimensional object  is an analytic function since 
it can be expressed as a z-transform, i.e.,

\begin{equation}
\gamma(z)=\sum_{n=0}^N\mathcal{B}(n\Delta\theta)z^n\Delta\theta, \label {polynomial}
\end{equation}
where $z\equiv \exp{(ik\,m\Delta x\,\Delta\theta)}$. $\gamma(z)$ is a polynomial in $z$ and therefore an analytic function. An analytic function (of order zero)
can in general be expressed  as the product of its zeros, so eq. \ref{polynomial} can be written as (Hadamard Factorization)

\begin{equation}
\gamma(z)=c\prod_j^N (z-a_j), \label{product}
\end{equation}
so that all of the information of $\gamma(z)$ is encoded in the roots $a_j$. In SII we have knowledge of $|\gamma(z)|^2=\gamma(z)\gamma(z^{-1})$,
where $\gamma(z)$ is a polynomial of degree $N$ in $z$, and $\gamma(z^{-1})$ is a polynomial of degree $N$ in $z^{-1}$. The phase recovery problem can
then be restated as finding $\gamma(z)$ from knowledge of $|\gamma(z)|^2$. The information contained in $\gamma(z)\gamma(z^{-1})$ is also contained in  

\begin{equation}
  Q(z)=z^{N}\gamma(z)\gamma(z^{-1}),
\end{equation}
which is a polynomial of degree $2N$ in $z$ (Hayes theorem \citep{Hurt}). The polynomial $Q(z)$ has roots $a_j$ and $a_j^{-1}$. The problem is 
then to find all the polynomials $\gamma(z)$, 
with non-negative coefficients where either $a_j$   or $a_j^{-1}$  is a root. If 
there are $N$ distinct roots, then there are no more than $2^{N}$ solutions. However, if there are $N'$ roots that satisfy $|a_j|=1$, then there are
$2^{N-N'}$ solutions \citep{Hurt}pg 30. For example, the Fourier transform of a step function (corresponding physically to a uniform disk-like star) has a corresponding 
z-transform of the form $\sum_n z^n$. The zeros of this function are all in the unit circle, so the solution is unique in this case. 

In general, given a polynomial $|\gamma(z)|^2$, the solution polynomial $\gamma(z)$ is not unique. However, all solutions are related to each other 
by a phase factor of the form\footnote{Provided there are no zeros in the origin.} $Az^B$ ($A,B\in \mathbb{C}$). In image space, this is equivalent to solutions differing by
 translations and scale factors (\citep{klibanov}). The set of solutions describing ``the same object'' are usually known as \emph{trivial associates}. 

%Include example?

The statement of analyticity of the Fourier transform of a one-dimensional object is actually much more general than the discrete case treated so far. The Paley-Weiner 
theorem \citep{Hurt} states that the Fourier transform of a one-dimensional function with bounded support is an analytic function. The proofs of uniqueness are ultimately 
based on the uniqueness of analytic continuation. That is, if we have knowledge of a function in a region of the complex plane, by analytic continuation we can
have knowledge of the function in the entire complex plane. \footnote{A known example of analytic continuation is found in classical electrodynamics, when we wish to
find the real part of the complex index of refraction, with knowledge of the imaginary part. These two quantities are related to each other by the Kramers-Kronig relations,
also known as the Hilbert transforms.} 

In the case of a two-dimensional function with compact support, its Fourier transform  $F(z_x,z_y)$ is fully analytic (see Plancherel-Polya theorem). In two 
dimensions, an analytic function can in principle be factorized (Osgood product) in a similar form as eq. \ref{product}, but the form of each factor and the number of factors
is unknown in general \citep{Hurt}. Zeros in two dimensions are not isolated, but rather form ``lines'' that uniquely define the function. This can be 
contrasted with the one-dimensional case, where the number of factors is known (eq. \ref{product}), and each factor 
corresponds to a root of the polynomial. The most important idea concerning uniqueness in two dimensions is that if $F(z_x,z_y)$ is irreducible, or cannot be written as the
product of two analytic functions, then it is uniquely determined by $|F(z_x,z_y)|$ (Sanz-Huan theorem). Going back to the discrete (polynomial) case, 
it has been noted \citep{Hurt} that most two-dimensional polynomials are irreducible, and therefore uniquely determined up to trivial associates by their
modulus. Moreover, even in cases when a polynomial $F(z_x, z_y)$ is irreducible, there is always a sufficiently small region around, say $(z_{x,0}, z_{y,0})$ , 
where there exists an irreducible polynomial.

\section{Common approaches to phase retrieval \label{common_approaches}}

The preceding section suggests that finding the zeros of the Fourier transform is a way of finding the solution. However, this a very unstable way of finding the 
solution \citep{klibanov}. Nevertheless, we shall see that care should be taken when extracting information of the phase when close to zeros of the Fourier transform. We shall 
now briefly describe some approaches to phase retrieval.

\subsection{Dispersion relations}

Explicit formulae for the phase rely on the theory of analytic functions. The so called dispersion relations in particular are derived from the 
Cauchy integral formula by promoting the position variable $x$ to be complex \citep{klibanov}. Suppose that the degree of correlation as a 
function of position $x$ can be expressed as

\begin{equation}
  \gamma(x)=|\gamma(x)|e^{i\phi(x)}.
\end{equation}

For the moment, we assume that this function does not contain any zeros. Now we take the $log$ of $\gamma(x)$ and use the 
Cauchy integral formula\footnote{\begin{equation}f^{(n)}(a)=\frac{n!}{2\pi i} \oint \frac{f(z)}{(z-a)^{n+1}}dz \nonumber \end{equation}}. The Cauchy integral along a large semicircle in the upper half complex plane is

\begin{equation}
  \log{\gamma(x)}=\frac{1}{2\pi i}\oint\frac{\log{|\gamma(x)|}}{x'-x}dx' - \frac{1}{2\pi} \oint \frac{\phi(x')}{x'-x} dx'.
\end{equation}

These integrals can be further simplified by making plausible assumptions of the asymptotic behavior of the magnitude and phase
as the radius $R$ of the semicircular path tends to infinity. For example, the magnitude can be assumed to 
decrease as $1/x^n$ for some $n \in \mathbb{R}$, 
and the phase can be assumed to be linear at infinity. The first term in the previous equation then becomes an integral along the real axis, and 
the second term becomes an angular integral on the semicircle that will not depend on $x$ since $R$ tends to infinity. Taking the real part of the
previous equation yields

\begin{equation}
  \phi(x)=\frac{P}{\pi}\int_{-\infty}^{\infty} \frac{log|\gamma(x)|}{x'-x}dx'+\alpha x + \mathrm{constant},
\end{equation}
where $P$ denotes the Cauchy principal value. The previous equation is also known as a logarithmic Hilbert transform. When $\gamma(x)$ does contain
zeros, there are additional contributions to the phase known as the ``Blaske phase''

\begin{equation}
  \Lambda(x)=\prod_j \frac{x-a^*_j}{x-a_j},
\end{equation}
where $a_j$ refers to the zeros of $\gamma(x)$, and the problem is again reduced to finding the zeros of $\gamma(x)$. 

\subsection{Cauchy-Riemann phase recovery\label{cr}}

%Remember that people have proposed methods of measuring phase differences along purely real and imaginary axes.

Most of my phase retrieval research has concentrated in this method, which relies only on the theory of analytic functions, and which does not reduce to finding the zeroes of $\gamma$. We shall first study the one-dimensional case \citep{mnras, holmes} and then treat the two-dimensional case to be used in SII analysis \citep{mnras}.

\subsection{The one-dimensional case}

 If we denote $I(z)=R(z) e^{i\Phi(z)}$, where $z\equiv\xi+i\psi$, we obtain the following relations from the Cauchy-Riemann equations\footnote{The C-R equations can be applied because ``I'' 
is a polynomial in z.}:

\begin{eqnarray}
\frac{\partial\Phi}{\partial\psi}&=&\frac{\partial\ln{R}}{\partial\xi}\equiv \frac{\partial s}{\partial \xi}\\ 
\frac{\partial\Phi}{\partial\xi}&=&-\frac{\partial\ln{R}}{\partial\psi}\equiv -\frac{\partial s}{\partial \psi},  \label{cauchy}
\end{eqnarray}
where we have defined $s$ as the log-magnitude. Notice the relation between the magnitude and the phase. By using the Cauchy-Riemann equations we can write the log-magnitude differences along the real and imaginary axes as:

\begin{eqnarray}
\Delta s_{\xi} &=&\frac{\partial s}{\partial \xi}\Delta\xi=\frac{\partial\Phi}{\partial\psi}\Delta\xi\\ 
\Delta s_{\psi}&=&\frac{\partial s}{\partial \psi}\Delta\psi=-\frac{\partial\Phi}{\partial\xi}\Delta\psi\\
\end{eqnarray}

If the log-magnitude were available along purely the $\xi$ or the $\psi$ axes, we could solve the previous two equations for the phase. 

However, notice that because $|z|=1$, 
we can only measure the log-magnitude on the unit circle in the complex space ($\xi,\psi$).  

In general, we can write the log-magnitude differences along the unit circle as

\begin{eqnarray}
\Delta s_{||}&=&\frac{\partial s}{\partial \xi}\Delta\xi+\frac{\partial s}{\partial \psi}\Delta\psi \label{path0}\\
            &=&\frac{\partial\Phi}{\partial\psi}\Delta\xi-\frac{\partial\Phi}{\partial\xi}\Delta\psi \label{path}\\ 
            &=&\Delta \Phi_{\bot}. \nonumber
\end{eqnarray}

Here $\Delta \Phi_{\bot}$ corresponds to phase differences along a direction perpendicular to $\Delta s_{||}$, that is, perpendicular to the unit circle in the $\xi-\psi$ plane. We are however interested in obtaining $\Delta\Phi_{||}$, so that we can integrate along the unit circle.

The general form of $\Phi$ can be found by taking second derivatives in eq. (\ref{cauchy}) and thus noting that $\Phi$ is a solution of the Laplace equation in the complex plane.

\begin{equation}
\frac{\partial^2\Phi}{\partial\xi^2}+\frac{\partial^2\Phi}{\partial\psi^2}=0.
\end{equation}

The general solution of $\Phi(z)$ in polar coordinates ($\rho, \phi$) is \citep{Jackson}

\begin{equation}
\Phi(\rho,\phi)=a_0+b_0\phi+\sum_{j} \rho^j  \left( a_j \cos{j\phi}+b_j\sin{j\phi}\right) \label{solution},
\end{equation} 
where terms singular at the origin ($\rho^{-j}$) have been omitted. These singular terms lead to ambiguous reconstructions including flipped images and have not been found to be essential for most reconstructions.  

Now taking the difference of $\Phi$ along the radial direction we obtain

\begin{equation}
\Delta\Phi_{\bot}(\rho, \phi)=\sum_j \rho^j((1+\frac{\Delta\rho}{\rho})^j-1)\left( a_j \cos{j\phi}+b_j\sin{j\phi}\right).
\end{equation}

Note from eq. (\ref{path}) that the length in the complex plane associated with $\Delta s_{||}$ is $\Delta\phi=|\Delta\xi+i\Delta\psi|$, and that the length associated with $\Delta\Phi_{\bot}$ is $\Delta\rho=|\Delta\xi+i\Delta\psi|$. Now setting $\rho=1$,  $\Delta\rho=\Delta\phi$, and for simplicity of presentation, expanding for small $\Delta\phi$, we obtain

\begin{equation}
\Delta\Phi_{\bot}(\phi)=\sum_j j\Delta\phi\left( a_j \cos{j\phi}+b_j\sin{j\phi}\right). \label{phi_bot}
\end{equation}

So now the coefficients $a_j$ and $b_j$ can be found using equations (\ref{path0}-\ref{path}) from the measured $\Delta s_{||}$, and thus $\Phi$ can be
found in the complex plane, with an uncertainty in $a_0$ and $b_0$. The
coefficients $a_j$ and $b_j$ can be calculated by performing the following integrals:

\begin{equation}
a_j=\frac{1}{2\pi\,j}\int_0^{2\pi} \frac{d\Phi_{\bot}}{d\phi} \cos{j\phi}\,d\phi 
\label{a}
\end{equation}

\begin{equation}
b_j=\frac{1}{2\pi\,j}\int_0^{2 \pi} \frac{d\Phi_{\bot}}{d\phi} \sin{j\phi}\,d\phi
\label{b}
\end{equation}

Note however that the previous expressions must exist, which is not the
general case. More explicitly, if the magnitude is zero, the log-magnitude
is singular. When imaging finite objects in image space, there will always be
zeros in the magnitude of the Fourier transform. In practice we are always
sample limited and nothing prevents us from calculating $a_j$ and $b_j$ approximately.

\subsection{One-dimensional examples}

\begin{figure}
  \begin{center}$
    \begin{array}{c}
    \vspace{0.5cm}
    \includegraphics[scale=0.36]{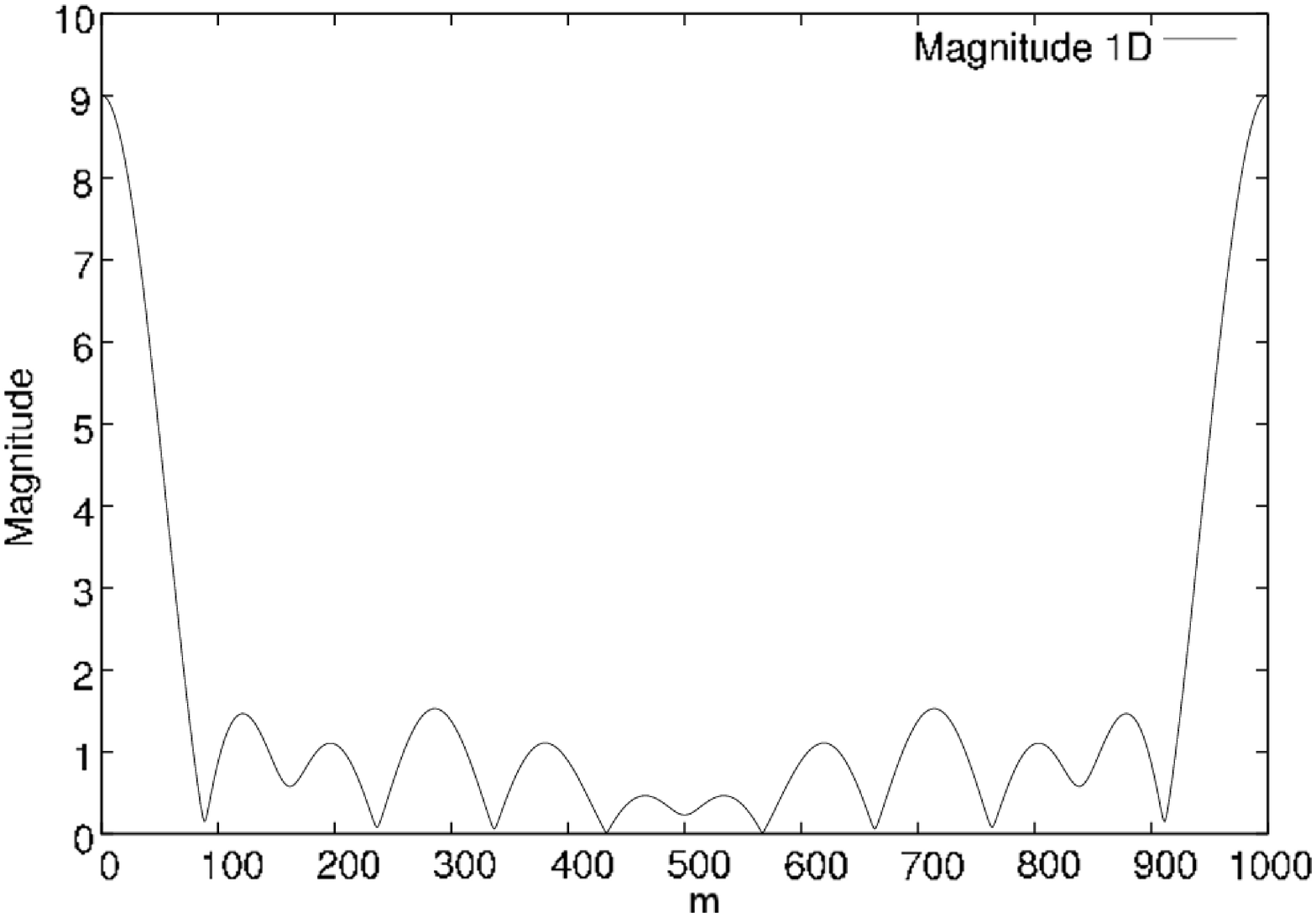}\\
    \vspace{0.5cm}
    \hspace{0.4cm}\includegraphics[scale=0.35]{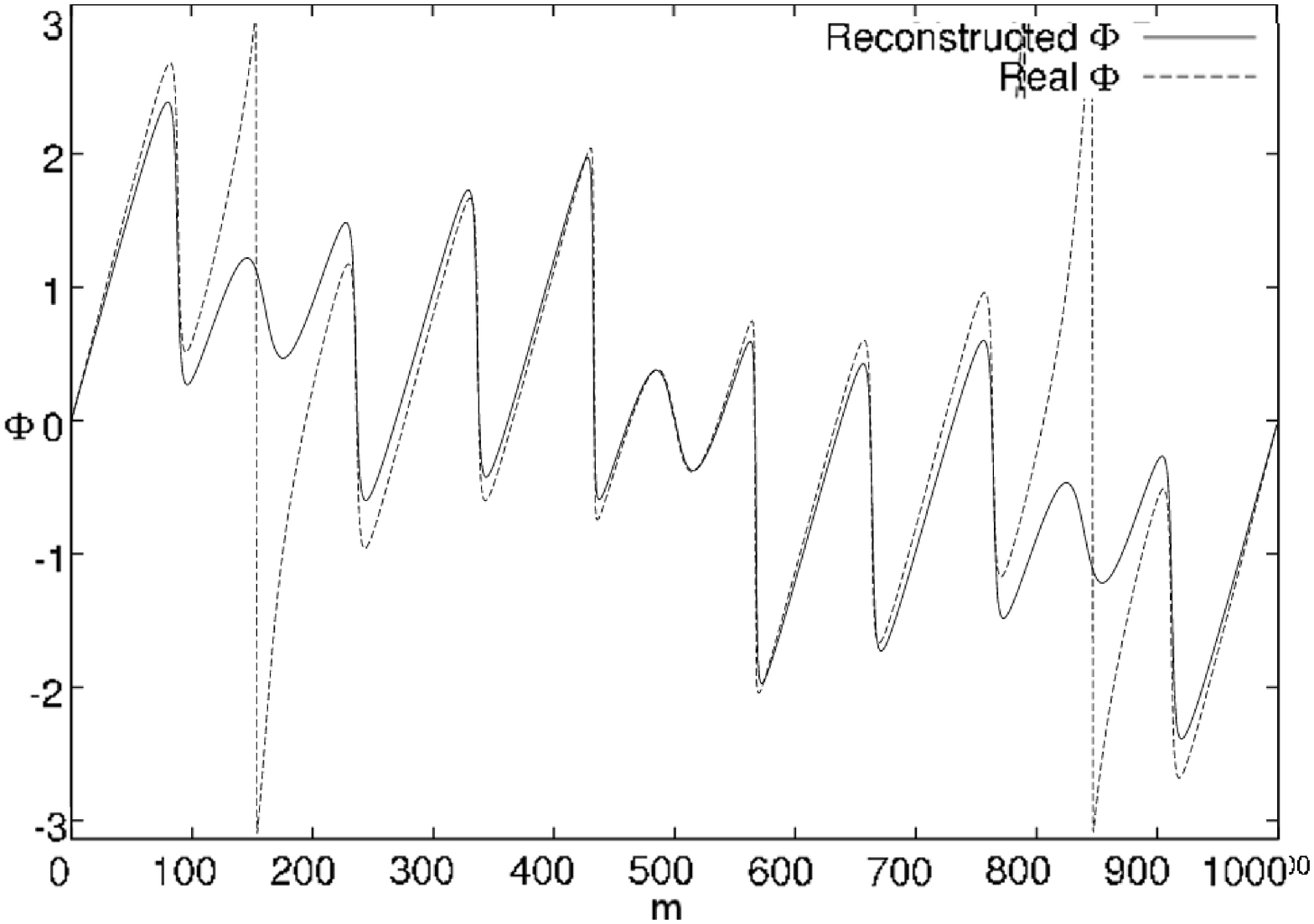}\\   
    \hspace{0.2cm}\includegraphics[scale=0.36]{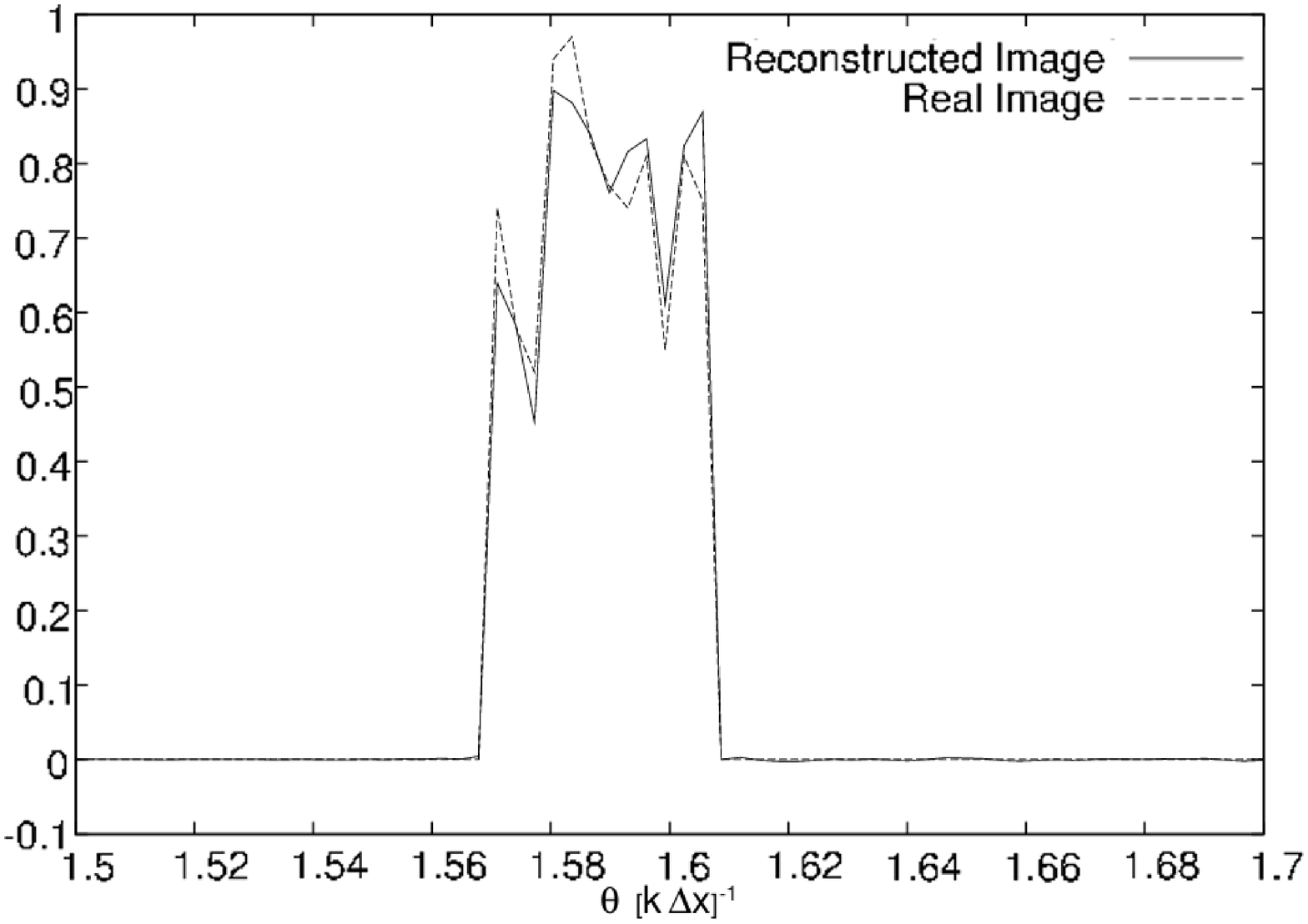}\\
    \end{array}$
  \end{center}
\caption[Example reconstruction of a random 1-dimensional image.]{\label{1d_example}Example reconstruction of a random one-dimensional image. The top figure is the magnitude of the Fourier transform of the original image. The middle figure is the phase of the original image compared with the reconstructed phase using the Cauchy-Riemann algorithm. The bottom figure (in arbitrary units of intensity) is the original image and the image using the estimated phase.}
\end{figure}

\begin{figure}
  \begin{center}$
    \begin{array}{c}
    \vspace{0.5cm}
    \includegraphics[scale=0.55]{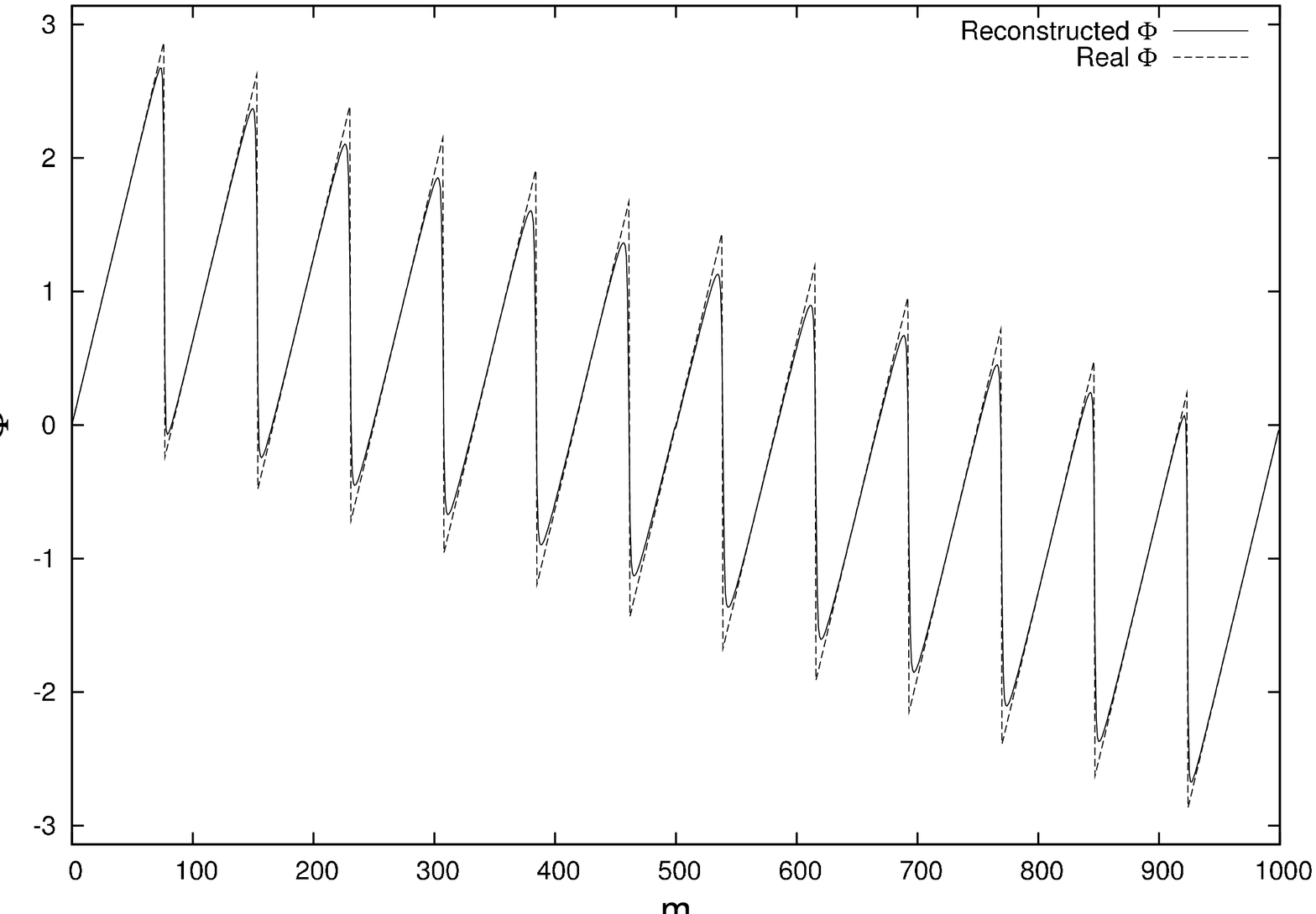}\\
    \vspace{0.5cm}
    \includegraphics[scale=0.55]{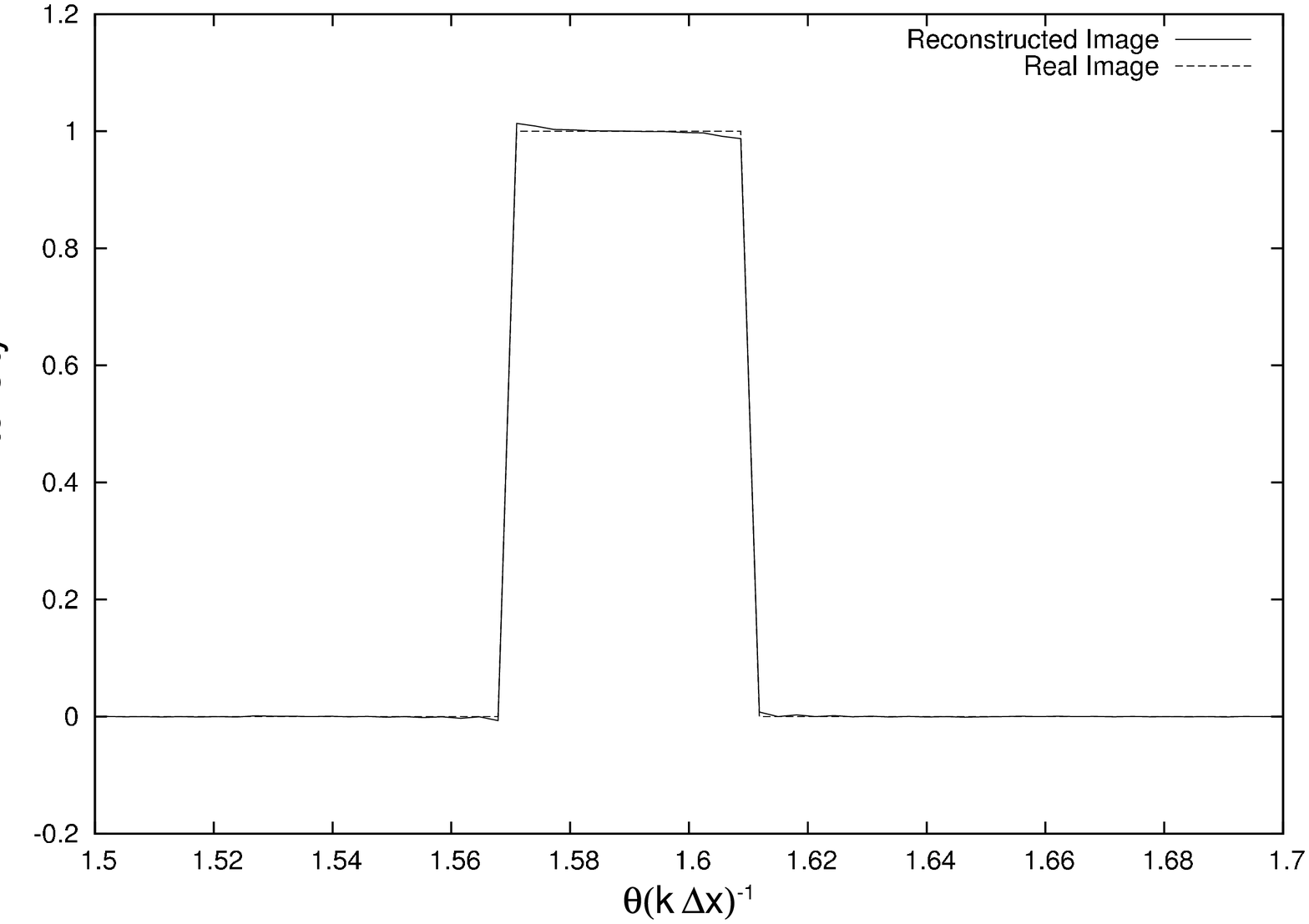}\\   
    \end{array}$
  \end{center}
\caption[Example reconstruction of a top-hat function.]{\label{top_hat}Example reconstruction of a top-hat function. The top figure displays the real and reconstructed phase using the Cauchy-Riemann phase reconstruction. The bottom figure displays the real and reconstructed image.}
\end{figure}

To illustrate the performance of the Cauchy-Riemann phase reconstruction, some \linebreak one-dimensional image reconstructions are shown below. These examples do not include noise or realistic sampling of data. In Figure \ref{1d_example}, the magnitude, phase, and reconstruction of a random image are shown. It should be emphasized that the only input in this example is the Fourier magnitude, and no prior information of the image for the reconstruction. A simpler example of a top-hat image reconstruction is shown in Figure \ref{top_hat}. The main limitation of the Cauchy-Riemann algorithm in 1-dimension is due to the fit of eq. \ref{phi_bot} by using \ref{a} and \ref{b}, which results in not accurately reproducing discontinuities in the phase. %Another approach (suggested by \citet{holmes}) finds the coefficients \ref{a} and \ref{b} by turning the problem into a linear system by minimizing a $\chi^2

\subsection{Two-dimensional case\label{2-dim}}

\begin{figure}
  \begin{center}
    \includegraphics[scale=0.4]{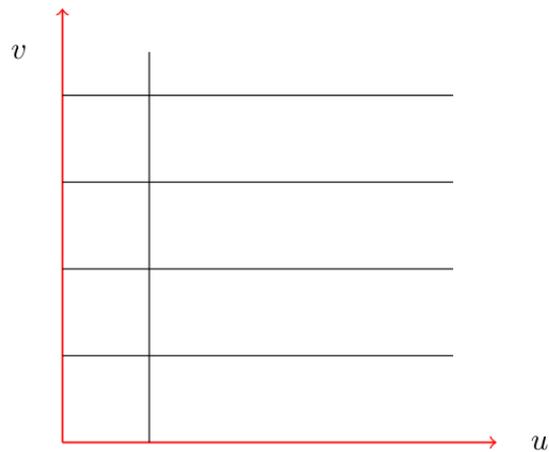}
    \caption[Schematic representation of two-dimensional phase reconstruction approach.]{\label{slices} Schematic representation of two-dimensional phase reconstruction approach. Several parallel slices are related to
      a single orthogonal slice.}
  \end{center}
\end{figure}

We can think of this one-dimensional reconstruction as a phase estimation along a single slice in the Fourier plane. A 
generalization to two dimensions can be made by following the same procedure for several slices as 
described in Figure \ref{slices}. In fact, the requirement that a two-dimensional complex function $(z_x, z_y)$ be analytic,
is equivalent to satisfying the Cauchy-Riemann equations in both $z_x$ and $z_y$ \citep{2-d_complex}. The direction of the slices 
is arbitrary, however for simplicity we reconstruct the phase along an
arbitrary set of perpendicular directions in the Fourier plane, 
and noting that one can relate all slices through a single orthogonal slice, i.e., once the phase at the origin is set to zero, 
the single orthogonal slice sets the initial values for the rest of the slices. 

 One can also require that the phase at a particular point 
in the complex plane be exactly equal when reconstructed along $z_x$ or $z_y$ since each reconstruction is arbitrary up to a constant (piston) and a
linear term (tip/tilt). However, imposing this requirement results in a severely over-determined linear system. More precisely, by imposing equality 
in $n^2$ points in the complex plane, and having $2n$ slices (each with an unknown constant and linear term), results in a linear system of $n^2$ equations
and $4n$ unknowns. Alternative methods of requiring slice consistency are a possible way of improving phase reconstruction, but are beyond 
the scope of this work. 

 The Cauchy-Riemann approach, with horizontal or vertical slices, and a single orthogonal slice, 
gives reasonably good results;  however, it is not the only
possible approach.  We have also investigated Gerchberg-Saxton phase
retrieval, Generalized Expectation Maximization, and other variants of the
Cauchy-Riemann approach. It is premature to conclude which of these approaches 
is best at this time, given the limited imagery and SNR levels that have been explored. However, the 
Cauchy-Riemann approach has shown to give better results in a number of cases \citep{Holmes.spie}.

\subsection{Two-dimensional examples}

\begin{figure}
  \begin{center}
    \includegraphics[scale=0.5]{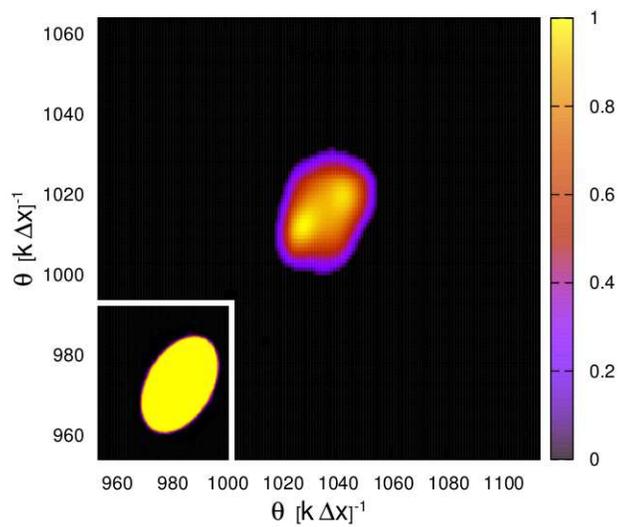}
  \end{center}
  \caption[Reconstruction of an oblate object, e.g., an oblate rotating star.]{\label{rotator_noiseless}Reconstruction of an oblate object, e.g., an oblate rotating star. Pristine image is shown in the bottom left corner}
\end{figure}

\begin{figure}
  \begin{center}
    \includegraphics[scale=0.5]{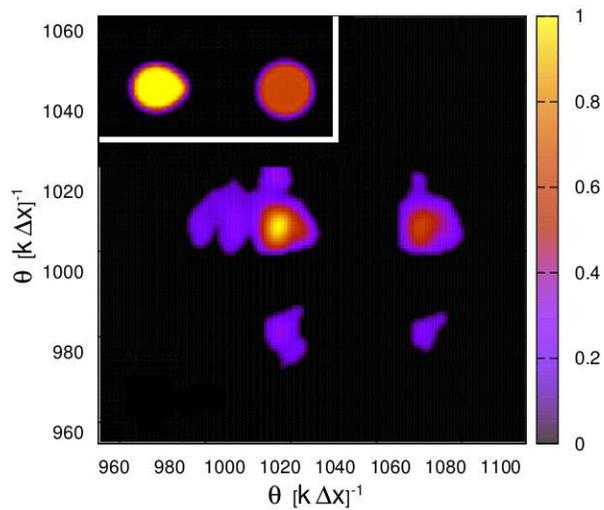}
  \end{center}
  \caption[Reconstruction of a binary object.]{\label{beta_noiseless}Reconstruction of a binary object. The pristine image (top left) is actually a simulated image of the well known binary system $\beta$-\emph{lyrae}.}
\end{figure}

\begin{figure}
  \begin{center}
    \includegraphics[scale=0.5]{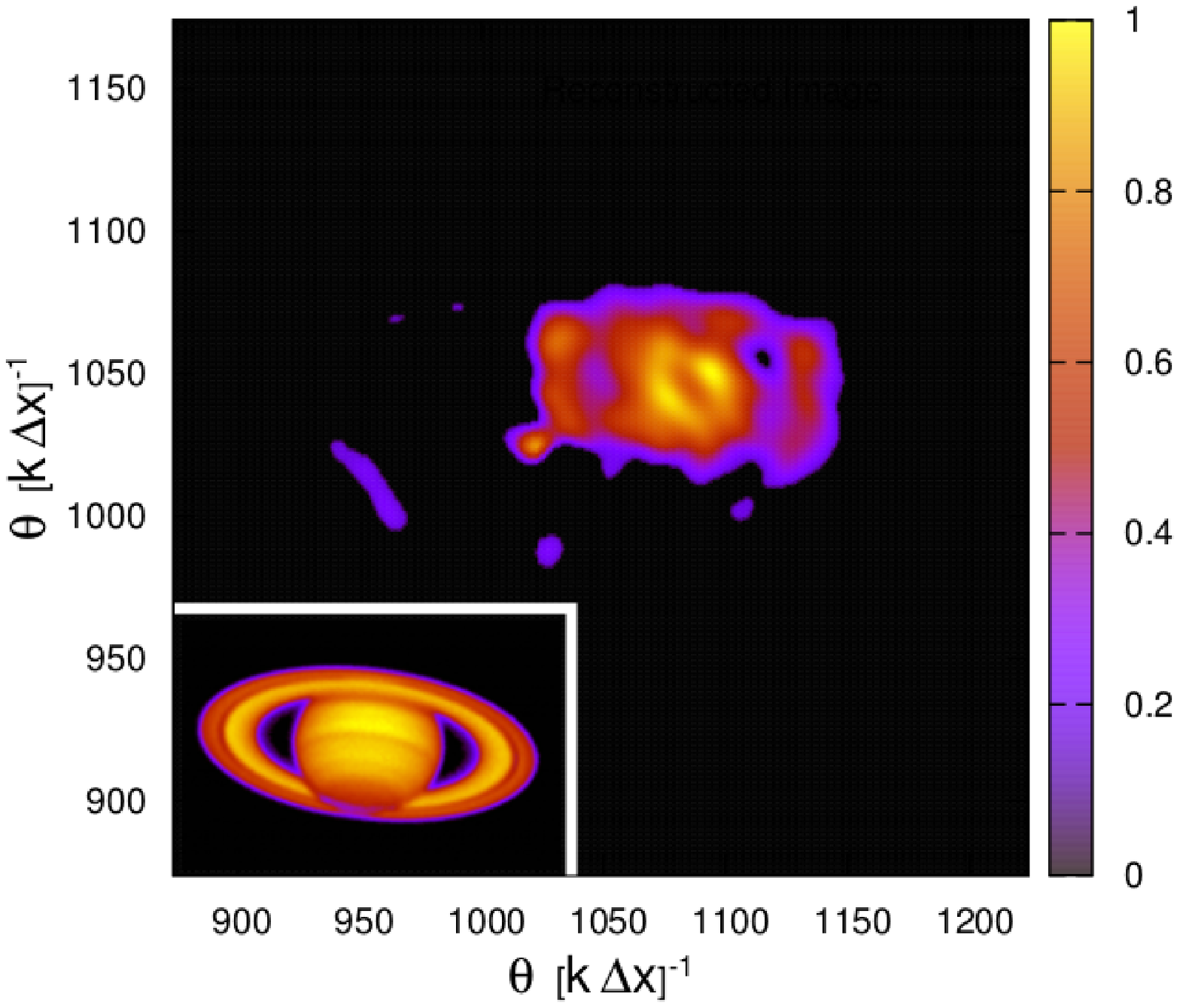}
  \end{center}
  \caption{\label{saturn_noiseless}Reconstruction of Saturn.}
\end{figure}

A few examples of two-dimensional image reconstructions are shown. Each of these examples takes the magnitude of the Fourier transform as the only input. In Figure \ref{rotator_noiseless}, the reconstruction of an oblate object, e.g. a fast rotating star is, shown. In Figure \ref{beta_noiseless}, the reconstruction of a simulated image of the binary $\beta$-\emph{lyrae} is shown. As a final example, an image of Saturn is reconstructed in Figure \ref{saturn_noiseless}. From the examples it can be seen that several main features are reconstructed approximately, and the quality of the reconstruction degrades with image complexity. More realistic examples are given in Chapter \ref{sii_with_iact}, as well as a more quantitative analysis of the reconstruction capabilities of this algorithm in the presence of noise, etc. 

\subsection{Error-reduction algorithm\label{gerchberg-saxton}}

The Gerchberg-Saxton algorithm, also known as the error-reduction algorithm, is an iterative procedure. Starting from a reasonable guess of the image
whose phase is unknown, the algorithm consists in going back and forth between image and Fourier space, and each time imposing general restrictions. Since 
the data consist of Fourier magnitude measurements, the restriction in Fourier space is that the magnitude corresponds to the data. The
restriction in image space can be as simple as requiring the image to be contained within some finite region. 

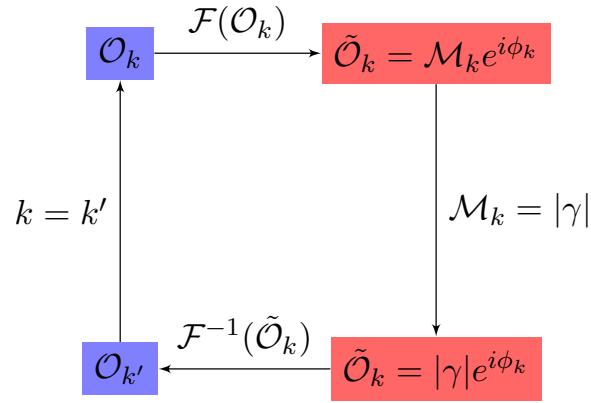
\begin{figure}[h]
\tikzstyle{arrow} = [draw, -latex']

\begin{center}
\scalebox{1.2}{
\begin{tikzpicture}
  \node(original)[rectangle, fill=blue!50]{$\mathcal{O}_k$};
  \node(Foriginal)[right of=original, node distance=3.5cm,rectangle, fill=red!60]{$\tilde{\mathcal{O}}_k=\mathcal{M}_ke^{i\phi_k}$};
  \node(Foriginal1)[below of=Foriginal,  node distance=3.5cm, fill=red!60]{$\tilde{\mathcal{O}}_k=|\gamma|e^{i\phi_k}$};
  \node(original1)[below of=original, node distance=3.5cm, fill=blue!50]{$\mathcal{O}_{k'}$};
  \draw[arrow](original)--node[above]{$\mathcal{F}(\mathcal{O}_k)$} (Foriginal);
  \draw[arrow](Foriginal)--node[right]{$\mathcal{M}_k=|\gamma|$} (Foriginal1);
  \draw[arrow](Foriginal1)--node[above]{$\mathcal{F}^{-1}(\tilde{\mathcal{O}}_k)$} (original1);  
  \draw[arrow](original1)--node[left]{$k=k'$}(original);
\end{tikzpicture}
}
\end{center}

\caption{\label{gs_routine}Schematic of the Gerchberg-Saxton error reduction algorithm.}
\end{figure}

Figure \ref{gs_routine} describes the Gerchberg-Saxton algorithm. Starting from an image $\mathcal{O}_k$, the first step consists in 
taking the Fourier transform to obtain something of the form $\mathcal{M}_ke^{i\phi_k}$. Now Fourier constraints can be applied, i.e., the magnitude is 
replaced by that given by the data, and the phase of the Fourier transform is kept. Now the inverse Fourier transform is applied and constraints can 
be imposed in image space. The constraints in image space can be very general, e.g. image positivity. However, if some apriori knowledge of the image
is available, stronger constraints can be applied, and the algorithm converges faster. For example, if the image is known to have a finite size, a 
\emph{mask} can be used, so that only pixels within the mask are allowed to have nonzero values. The performance of this algorithm 
depends strongly on the starting image, making it suitable for postprocessing purposes. Images using this algorithm are 

\newpage

\noindent presented in Chapter 7.

\section{Final remarks on phase recovery}

Phase retrieval is a field of research on its own right, and fully solving this mathematical problem has proven to be challenging indeed. The limitations of these algorithms, and  reasons why some algorithms work better than others are still not fully understood \citep{Hurt}.  However, it is clear that there is phase information contained in the Fourier magnitude, and perhaps one day, we will have full understanding of this mathematical problem. At this point, one is presented the following options: Either set on a quest to solve this problem, or use what is known so far to do science, e.g., astrophysics. I choose the later. The methods presented in the previous sections will be used Chapter \ref{sii_with_iact} to quantify the imaging capabilities of future Air Cherenkov Telescope Arrays (IACT).

\chapter{Air Cherenkov telescope arrays\\ and gamma-ray astronomy}

Imaging Air Cherenkov Telescope arrays are primarily used for $\gamma$-ray astronomy, which investigates some of the most violent phenomena in the universe. In this chapter, the subject of  $\gamma$-ray astronomy is briefly discussed. Even though the motivations for this field are entirely different from high angular resolution astronomy, they do share common interests for a few objects. One such object is the X-ray binary $LSI\,+61^{\circ}303$, which consists of a hot Be star, and a compact object, and was observed with the Very Energetic Radiation Imaging Telescope Array System (VERITAS). An analysis of $\gamma$-ray data allows us to constrain some fundamental parameters of the system \citep{apj_paper}, and many remaining questions can potentially be answered with long baseline optical interferometry.

\section{Highest energy gamma-ray sources}

The earth's atmosphere is constantly being bombarded by very energetic charged particles known as cosmic rays, whose energy spectrum essentially follows a power law which spans 12 orders of magnitude ($10^9-10^{21}\,\mathrm{eV}$). Their origin is unknown since their angular distribution is isotropic, and questions such as acceleration mechanisms and energy dependent composition (e.g., single protons or heavy nuclei ) are still subject of debate. The field of $\gamma$-ray astronomy was initially proposed for finding the origin of cosmic rays. Photons are not deflected by the complex magnetic fields that isotropize cosmic ray detection, and studying the spectral energy distributions of photons helps determine the nature of the particle acceleration mechanisms. 

\begin{figure}
        \begin{center}
        \includegraphics[scale=0.5]{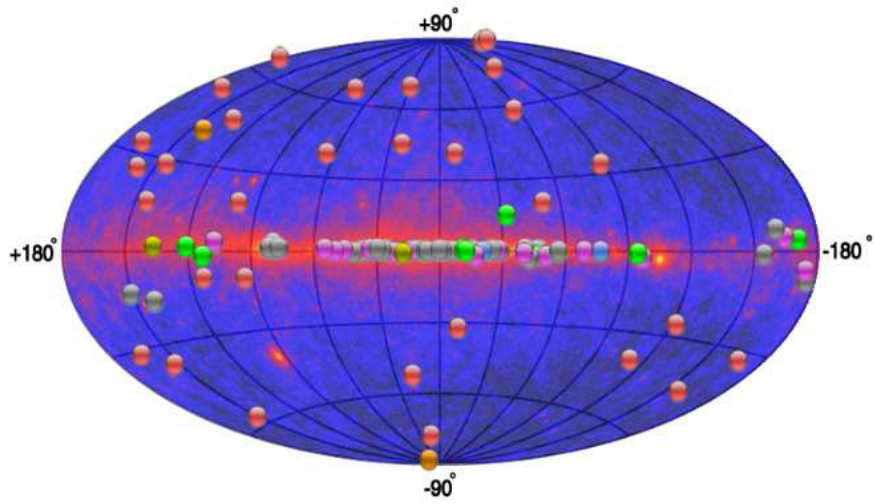}
        \end{center}
        \caption{\label{tevcat} Gamma-ray sky map taken from \emph{tevcat.uchicago.edu}. }
\end{figure}

It has been 100 years since the discovery of cosmic rays, and their origin is still unknown, or at least highly debated. However, $\gamma$-ray astronomy is a flourishing field, and after the detection of the Crab Nebula as the first TeV $\gamma$-ray source in 1989, more than $\sim 100$ high energy (TeV) sources have been discovered. Figure \ref{tevcat} shows the sky map of $\gamma$-ray sources, which are divided in two main categories: galactic sources, which can be seen to lie along the galactic plane, and extra-galactic sources. Extra-galactic sources include active galactic nuclei (e.g., M87 \citet{m87}), and more recently star-burst galaxies (e.g., M82 \citet{m82}). Galactic sources include supernova remnants, pulsar wind nebulae, X-ray binaries, and unidentified objects. Rather than studying the possible $\gamma-$ray emission mechanisms, which include high accretion rates, or inverse Compton scattering from electrons accelerated by shock-waves generated by supernova explosions, we will analyze data from a particular high energy emitting binary system, and constrain some of its fundamental properties in section \ref{lsi_section}. 

\section{Needs for gamma-ray astronomy}
\vspace{-0.0cm}
The extremely small wavelengths associated to $\gamma$-rays ($\sim 10^{-12}\,\mathrm{m}$) do not allow for them to be detected with traditional optics such as mirrors since interactions are at the nuclear level. At these very high energies, large amounts of stopping material are needed, and this acts essentially as a calorimeter. In the case of GeV $\gamma$-rays, whose flux is of the order of the order of $10^{-8}\,\mathrm{cm\,s^{-1}}$, enough material ($\sim 1\,\mathrm{m^2}$) can fit in a satellite for them to be detected from space. Such is the case of the recent Fermi satellite, which has been extremely successful at detecting nearly $1000$ sources.
% \vspace{-0.0cm}

As energies reach $1\,\mathrm{TeV}$, the particle flux is of the order of $10^{-13}\,\mathrm{cm\,s^{-1}}$, so very large areas ($\sim 100,000\,\mathrm{m^2}$) are needed as well as vast amounts of stopping matter ($1000\,\mathrm{g/cm^2}$), equivalent to $1\,\mathrm{m}$ of bricks! Detection from space becomes impractical, and in order to detect TeV $\gamma$-rays, the optically thick atmosphere is used to stop $\gamma$-rays, and large light ($\sim 100\,\mathrm{m}^2$) collectors detect the faint Cherenkov light produced as the electromagnetic particle showers propagate through the atmosphere. 
\vspace{-0.0cm}
\section{Imaging atmospheric Cherenkov technique}
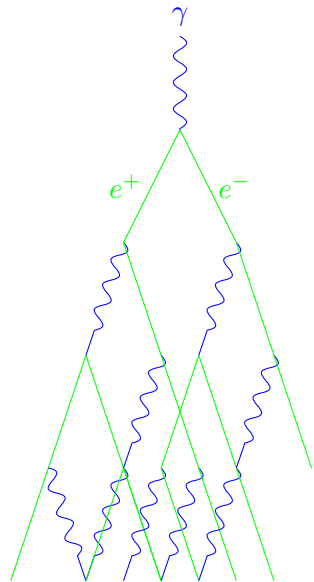
\begin{figure}
\begin{center}
\scalebox{1.0}{
        \begin{tikzpicture}
          \tikzstyle{level 2}=[sibling distance=15mm]
          \tikzstyle{level 3}=[sibling distance=10mm]
          \node(primary){\textcolor{blue}{$\gamma$}}
          child[blue]{edge from parent[snake=coil, line after snake=0pt, segment aspect=0]
            child[green]{
              child[blue]{edge from parent[snake=coil, line after snake=0pt, segment aspect=0]
                child[green]{
                  child[green]
                  child[blue]{
                    edge from parent[snake=coil, line after snake=0pt, segment aspect=0]
                  }
                }               
                child[green]{
                  child[blue]{edge from parent[snake=coil, line after snake=0pt, segment aspect=0]}
                  child[green]
                }
              }
              child[green]{
                child[blue]{
                  child[green]
                  child[green]
                  edge from parent[snake=coil, line after snake=0pt, segment aspect=0]
                }
                child[green]{
                  child[blue]{edge from parent[snake=coil, line after snake=0pt, segment aspect=0]}
                  child[green]
                }
              }
              edge from parent
              node[left]{\textcolor{green}{${e^+}$}}
            }
            child[green]{
              child[blue]{edge from parent[snake=coil, line after snake=0pt, segment aspect=0]
                child[green]{
                  child[blue]{edge from parent[snake=coil, line after snake=0pt, segment aspect=0]}
                  child[green]
                }
                child[green]{
                  child[blue]{edge from parent[snake=coil, line after snake=0pt, segment aspect=0]}
                  child[green]
                }
              }
              child[green]{
                child[blue]{edge from parent[snake=coil, line after snake=0pt, segment aspect=0]}
                child[green]
              }
              edge from parent
              node[right]{\textcolor{green}{${e^-}$}}             
            }
          };
        \end{tikzpicture}
}
\end{center}
\caption{\label{cascade} Schematic of electromagnetic cascade.}
\end{figure}

\vspace{-0.0cm}
When a $\gamma$-ray interacts with a nucleus at the top of the atmosphere, it induces an electro-magnetic cascade as illustrated in Figure \ref{cascade}. The interaction with the initial nucleus permits the creation of an electron-position pair, which then in turn emit radiation through Bremstrahlung when they encounter other charges. This process continues to develop and the shower continues to grow until particles reach an energy of a few hundred MeV and ionization dominates as an energy loss mechanism. At this point, $e^+e^-$ pairs are produced at a smaller rate, and the size of the electromagnetic shower starts to diminish. 
\vspace{-0.03cm}

\begin{figure}        
\begin{center}
\includegraphics[scale=1.0]{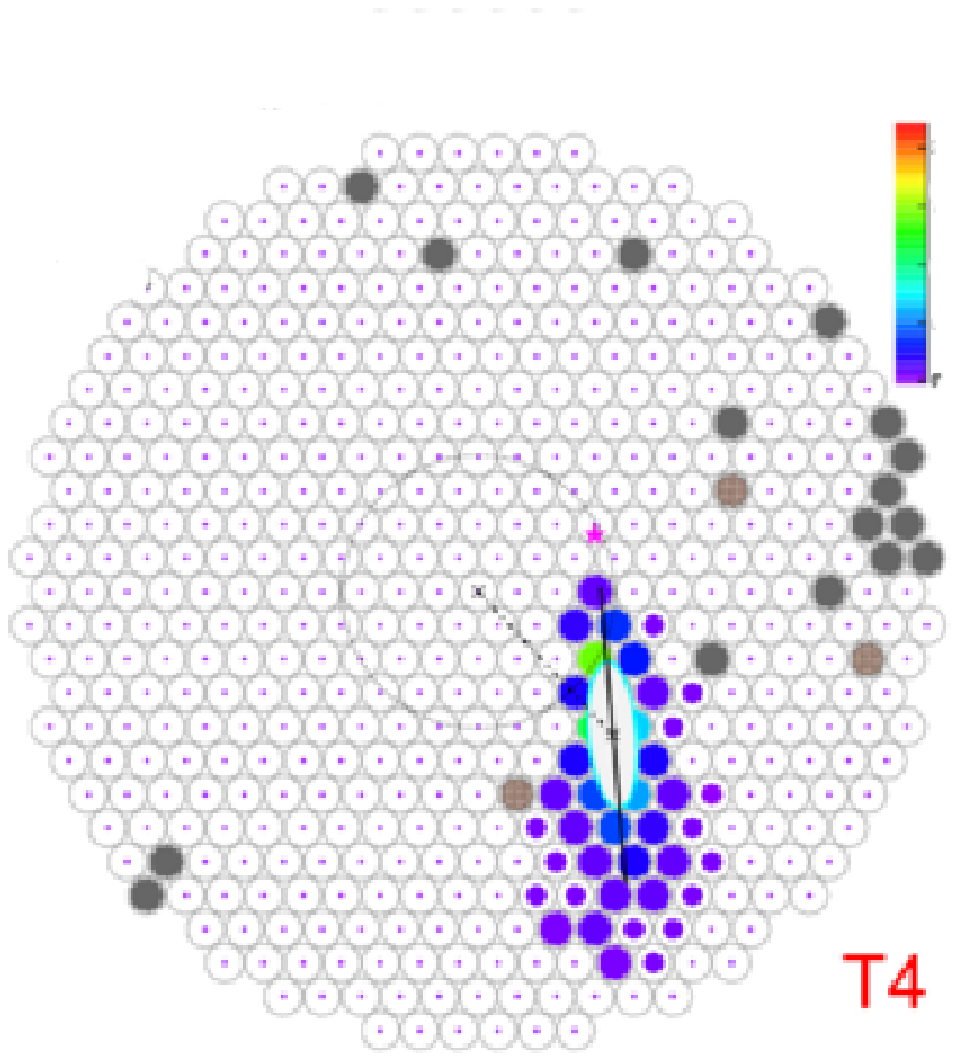}
\end{center}
\caption{\label{focal}Image of shower in the camera of one telescope of the VERITAS array which consists of 4 telescopes.}
\end{figure}

Since charged particles in the electromagnetic cascade travel faster than light in air, they emit Cherenkov radiation, analogous to the wake formed in water by a boat traveling faster than the speed of sound on the water surface. This light is seen as a ``streak'' of light in the focal plane of each telescope as shown in Figure \ref{focal}.

\section{Analysis}

Most of the recorded data ($99.9\%$) corresponds to cosmic ray induced showers, so much of the analysis has to do with discriminating $\gamma$-ray events from cosmic ray events. The main difference between the two is due to the fundamental nature of the interaction (QED vs. QCD), and is reflected in the shape of the shower: cosmic-ray showers are typically ``fatter'' since pions generated in a collision have large transverse momenta compared with $e^+e^-$ pairs. In order to determine the initial energy of the photon, the analysis depends strongly on accurate shower simulations. By using the shower images in several telescopes, the source of $\gamma$-rays can be found through a geometric reconstruction. That is, a line is traced through the major axis of the ``streak'' seen in each camera. The intersection of these lines points to the source. 

Once $\gamma$-ray-like events are selected, the background  needs to be subtracted. One way of accomplishing this is to first point the telescopes at the source, and then point away from the source, so that an estimate of the background can be found. This method is usually not practical since much time is spent looking at background. The way it is done in VERITAS is by pointing the telescopes at\footnote{Telescopes are not actually pointed at the source, but observations are made in ``wobble-mode'' CITE. A detailed description, although interesting, is beyond the scope of this document.} the source, and then selecting ``off-regions'' to estimate the background. % There are several types of off-regions, which vary in shape, size, and location, and the optimal type of off-region used may depend on the type of source being observed (e.g. point-like, extended). Clever background subtraction methods are constantly being developed.  

%\section{Stellar astrophysics with gamma-ray astronomy}
%\section{High energy emitting binary systems}

\section{X-ray binaries and $\gamma$-ray attenuation}
In the past few years, several high mass X-ray binaries have been detected as
gamma ray emitters \citep{Hess, Magic, Veritas}, causing an intensification of
observational and theoretical interest. High energy emitting binary 
systems consisting of a main sequence star and a compact object
are the only known variable galactic very high energy (VHE) sources, and their short
periods of days or weeks make them even more interesting observational
targets. These binary systems are starting to become astrophysical
laboratories in the sense that by increasing spectral coverage and statistics,
the nature of photon emission and absorption mechanisms is becoming more and
more constrained. Here we will mainly be concerned with high energy (TeV)
photons emitted from the vicinity of the compact object
and interacting with the background black body radiation and ejected material from the
companion star. Even though these systems can be incredibly complex, a simple model of the
absorption mechanisms and how they affect the system's 
light curve can still shed light on many aspects such as the masses
and the orbital parameters. 

One such example is the high energy emitting binary $LS\,I\,+61^\circ 303$ \citep{Massi}.  It was first detected
in the TeV range with MAGIC \citep{Magic} and further observed with VERITAS
 at flux levels ranging between 5\% and 20\% of the Crab Nebula
\citep{Veritas}. This source has been observed throughout most of the
electromagnetic spectrum starting with radio frequencies and extending to VHE
 gamma rays \citep{Leahy}. This broad spectral study indicates
that the system consists of a main sequence Be star of mass $M_1=12.5\pm2.5\,M_{\odot}$ \citep{Casares}, surrounded by a circumstellar disk \citep{Grundstrom, Paredes}, and a compact companion
separated by tens of solar radii at periastron. The compact companion can be either a
neutron star or a black hole \citep{Casares}, and its exact nature 
is still subject of investigation and debate \citep{Neronov}. The maximum VHE emission occurs close to apastron \citep{Veritas, Magic}, suggesting  that absorption plays an important role in the modulation. 

In the following sections, we consider the attenuation of gamma rays due to
interaction with background radiation and then consider the interaction with circumstellar material.
The $\gamma\gamma$ absorption mechanism in high energy emitting binaries was first pointed
out by \citet{Gould} and has been studied in the context of
observed  sources such as $LS\,I\,+61^\circ 303$ and $LS\,5039$ \citep{Dubus, Torres}. In the
attenuation due to pair production, the two variables that play a main role
are the concentration of background black body photons, and the energy threshold 
for pair production,  which in turn depends on the scattering
angle between the primary TeV photons and the low energy photons.

%The outline of the paper is as follows: First we describe some of the
%formalism needed to model the attenuation due to pair production in a high energy emitting binary
%system. Then we consider the particular case of LS I +61 303. We assume that the high energy
%radiation is emitted from the vicinity of the compact object and that its
%emission is isotropic with respect to the compact object and constant in time. The first analysis shows that the modulation due to pair production attenuation is not sufficient to account for VERITAS observations. For this reason, we include  additional interactions of VHE photons with circumstellar material. This model permits to constrain the orbital parameters, the mass of the compact object
%and the density of ejected material from the companion star.

\section{Interaction with background radiation}
\subsection{Radiative transfer equation}
In order to develop a gamma-ray attenuation model, we need to 
treat the general case of a binary system consisting of a 
 VHE emitting compact object and a main sequence star. The radiative transfer
 equation \citep{chandra} for the intensity $I(s,E)$, where $s$ is the distance
 traveled by a photon of energy $E$ from the emission point is

\begin{equation}
    \frac{dI(s,E)}{ds}=-(1+\cos\xi)\, n(s,\epsilon)\,\sigma(E,\epsilon, \xi)\,I(s,E) +    j(E,s)\,\,; 
    \label{radiative transfer}
\end{equation}
where $n(s,\epsilon)$ is the spectral density of background photons of energy $\epsilon$
emitted by the main sequence star, $\sigma(E,\epsilon,\xi)$ is the cross
section\footnote{Note that the term $(1+\cos\xi(s'))$ simply 
corresponds to the relative velocity between the incident and target photons} for the interaction between photons colliding at angle $\xi$, and
$j(E,s)$ is a source term which we will now describe.

\subsection{Neglecting the source term\label{secondaries}}

 Since the attenuation term is due to VHE photons creating $e^+e^-$ pairs, the
 source term is due to secondary gamma-rays in the
 electromagnetic cascade. The energy of these secondary gamma-rays is degraded
 by typically a factor of 4 with each interaction, and since VHE observations
 range between $\sim0.5$ TeV to $\sim 5$ TeV \citep{Veritas}, only those at
 the far end of the measured spectrum can contribute to the
 intensity at a fraction of their energy. However, as we shall see in section
 \ref{gamma-gamma}, only photons in the lower part of the spectrum are
 attenuated considerably, and feed the development of the electromagnetic cascade. Photons in the far end of the observed TeV spectrum
 are considerably more scarce since the spectrum is steep. Consequently, we neglect the source term.

Now we estimate the contribution of secondary gamma-rays with an
over-simplistic model which helps justify our neglection of the source term in
eq. \ref{radiative transfer}. We can estimate the effect of secondaries as an
increase in initial intensity $I(s_0+\Delta s, E)$ by $2I(s_0+\Delta s, 4E)$,
i.e., instead of having $I(s_0,E)$ in eq. \ref{formal solution} (defined below), we have

\begin{equation}
I(s_0+\Delta s,E)\rightarrow I(s_0+\Delta s,E)+2I(s_0+\Delta s,4E)P(s_0+\Delta s, 4E),
\label{increase}
\end{equation}
where a photon of energy $4E$ is assumed to produce an $e^+e^-$ pair, which in turn emitt a gamma-ray of energy $E$. $P(s_0+\Delta s, 4E)$ is the probability that the photon of energy $4E$ exists in the first place, and we have assumed an electromagnetic cascade toy model. Taking the intrinsic intensity to behave as a power law spectrum, $I(s_0+\Delta s,E)=I(E_0)\left(\frac{E}{E_0}\right)^{-\gamma}$, where $\gamma$ is the spectral index, eq. \ref{increase} simplifies as 

\begin{equation}
I(s_0+\Delta s,E)\rightarrow I(s_0+\Delta s,E)(1+2\times4^{-\gamma}P(s_0+\Delta s, 4E)).
\end{equation}

Since $\gamma\sim 2$ (see section 2.1) and $P(s_0+\Delta s, 4E)\leq 1$, $I(s_0+\Delta s, E)$ increases by a factor of $\sim9/8$ at most.

\subsection{Solution of the radiative transfer equation}

After neglecting the source term, the solution to the radiative transfer equation is  

\begin{equation}
I(s,E)=I(s_0,E)\,\exp\left\{-\int_{s_0, \epsilon}^{s,
    \infty}(1+\cos\xi(s'))\,n(s',\epsilon')\sigma(E, \epsilon',
  s')ds'd\epsilon'\right\}\,\,. 
\label{formal solution}
\end{equation}

Here $s_0$ is the emission point at the vicinity of the compact object (see
Figure \ref{orbit}), and $\epsilon$ corresponds to the threshold energy for
pair production,

\begin{equation}
\epsilon=\frac{m_e^2c^4}{E(1+\cos\xi(s))}.
\end{equation}

Note that the dependence of the
scattering angle $\xi$ in eq. \ref{formal solution} has been changed to a dependence on the path $s$. The problem then reduces to calculating the integral in the exponential of eq. \ref{formal solution}, also known as the optical depth $\tau(s,E)$ \citep{lightman}. In our calculation, we consider the main sequence star as point source, and in view of the results obtained by \citet{Dubus}, including the angular extension does not change our results significantly.  

The distribution of background black body photons can simply be taken as

\begin{equation}
n(r,E)=n_0(E)\frac{r_o^2}{r^2},
\label{background}
\end{equation}
where $r_0$ and $n_0$ are the radius of the Be star and the density of
background photons at this radius, i.e.,

\begin{equation}
n(r,E)=\frac{2\pi\,E^2dE}{c^3h^3}\left(\frac{r_0}{r}\right)^2\frac{1}{e^{E/kT}-1}.
\end{equation}

Here, the photon density has already been integrated over the half sphere (solid angle).

\section{The case of LS I+61 303 \label{lsi_section}}
\subsection{Attenuation}

There is debate as to what is the mechanism responsible for high energy emission. However, the aim of this paper is not to model the gamma ray emission but rather to investigate the effects of attenuation. This allows us to derive a few characteristics of the main sequence star environment and compact object orbit. 

\citet{Grundstrom} reported a temperature of 
$T\approx3\times10^{4}\,K$ and radius of $R\approx6.7R_{\odot}$ for the Be star. 
The black body distribution peaks at a few eV, and the threshold energy
for pair production with TeV incident photons is of the order of 1 eV, so that
most of the background photons may contribute to the attenuation,
provided the scattering angle is favorable. The
background photon density is found to be of the order of
$n_{\gamma}\sim10^{12}\,\rm{cm}^{-3}$ at a the radius of the star. The circumstellar
disk has been observed by \citet{waters} and by
\citet{Paredes}, who estimate the disc ion  density to be
$n_e\sim10^{13}\rm{cm}^{-3}$ at one stellar radius. The cross section for pair production is
of the order of $\sigma_{\gamma\gamma}\approx0.1\sigma_{T}$ at the threshold
energy. The cross section for interaction with hydrogen has a constant value of $\sigma_{\gamma
  H}\approx2\times10^{-2}\sigma_{T}$  above a few hundred MeV \citep{Heitler, Aharonian}. With these cross sections, a first estimate suggests that 
both interactions may result in comparable degrees of attenuation. However, there is a strong angular dependence in the $\gamma\gamma$ interaction, the extreme
case being when the both photons are emitted in the same direction, a configuration in which there is no VHE attenuation. Also the threshold energy is much higher when the incident and target photons are nearly parallel, so fewer
background photons contribute to attenuation. Consequently, $\gamma \gamma$ attenuation may not have strong modulation as a function of the orbital phase 
when compared with the modulation produced by interactions with the circumstellar material.

\subsection{Orbital parameters of $LS\,I\,+61^\circ 303$}
\begin{figure}[h]
  \begin{center}
    \scalebox{1.5}{
      \begin{tikzpicture}
        \draw [rotate=45] (0,0) ellipse (2cm and 1.5cm);
        \shade[ball color=green] (-0.7,-0.7) circle (0.08cm);
        \draw [dashed] (-2,-2) -- (2,2);
        \node at (-2, -2.1) {$_{\phi=0.3}$};
        \node at (2, 2){$_{\phi=0.8}$};
        \draw [dashed] (-0.7, -2.1) -- (-0.7, 2);
        \node at (-0.5, 2) {$_{\phi=0.035}$};
        \shade[ball color=red] (1.4, 1.4) circle (0.08cm);
        \node at (0, -2){$_{\phi=0.32}$};
        \draw [->, blue, semithick] (1.4, 1.35)--(1.4, -1.5);
      \end{tikzpicture}  
    }
  \end{center}
  \caption[Sketch of the orbital parameters of $LS\,I\,+61^\circ 303$.]{\label{orbit}Sketch of the orbital parameters of $LS\,I\,+61^\circ 303$. The
    arrow points to the observer.} 
\end{figure}
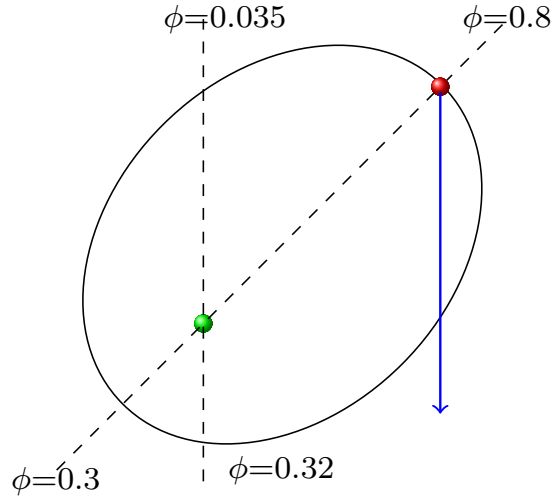

The orbital parameters for $LS\,I\,+61^\circ 303$ are still subject of research 
\citep{Aragona, Grundstrom, Casares} and are sketched in Figure
\ref{orbit}. The measurable 

\noindent quantities of interest 
are: The period $P$, the angle between the major axis of the
ellipse and the line of sight $\psi$, the projected
semimajor axis ($a_1\sin i$), corresponding to the ellipse of the Be star\footnote{The projected semi-major axis of the ellipse described by the compact object is typically labeled as $a_1\sin i$.}, the
eccentricity $\varepsilon$, the phase at periastron $\phi_0$, and the mass function $f(m_1,m_2)$,
which depends on the period and the radial velocity and relates the masses of
both objects and the inclination angle $i$. The most recent orbital solution
has been 
obtained by \citet{Aragona}, where $P=26.4960\,d$,
$\psi=40.5\pm5.7^{\circ}$, $a_1\sin i=8.64\pm0.52\,R_{\odot}$,
$\varepsilon=0.54\pm0.03$, $\phi_0=0.275$ and
$f(m_1,m_2)=0.0124\pm0.0022\,M_{\odot}$. It is important to remember that the
value of the angle $i$ depends on the mass of the compact object, and our
results may be used to constrain this quantity. Since
the mass of the compact object is a function of the inclination angle, we will take this mass to be
a free parameter of the model.

\section{The integrated flux of $LSI+61\,303^{\circ}$\label{flux section}}

Following observations of $LSI+61^\circ 303$ from 09/2006 to 02/2008, the VERITAS collaboration reported power law spectrum ($\frac{d\Phi}{dE}=\Phi_0\left(\frac{E}{TeV}\right)^{-\gamma}$) with a spectral index
of $\gamma=2.4\pm0.2_{stat}\pm0.2_{syst}$ at energies above $\sim0.5\,\rm{TeV}$,  and between phases $\phi=0.6$ and
$\phi=0.8$ \citep{Veritas}. Observations at lower energies made by Fermi between 08/2008 and 03/2009, indicate that the spectral index does not change significantly as a function of the orbital
phase \citep{Fermi}. Therefore, we assume a constant
intrinsic\footnote{By intrinsic we mean non attenuated by pair production.}
spectrum as a function of the phase at TeV energies. The integrated flux is then

\begin{equation}
F(\phi)=\int_{E_0}^{\infty}\frac{d^3N}{dEdtdA}
I(E,\phi)dE=F_0\int_{E_0}^{\infty}\left(\frac{E}{E_0}\right)^{-\gamma}
I(E,\phi)dE,
\label{integrated flux eq}
\end{equation}
where $E_0$ depends on the 
detection threshold energy of the detector, and $F_0$ is a normalization factor that is taken as 
a free parameter. 

\subsection{Light curve assuming only $\gamma\gamma$ interactions\label{gamma-gamma}}

\begin{figure}[h]
  \begin{center}
    \rotatebox{-90}{\includegraphics[scale=0.5]{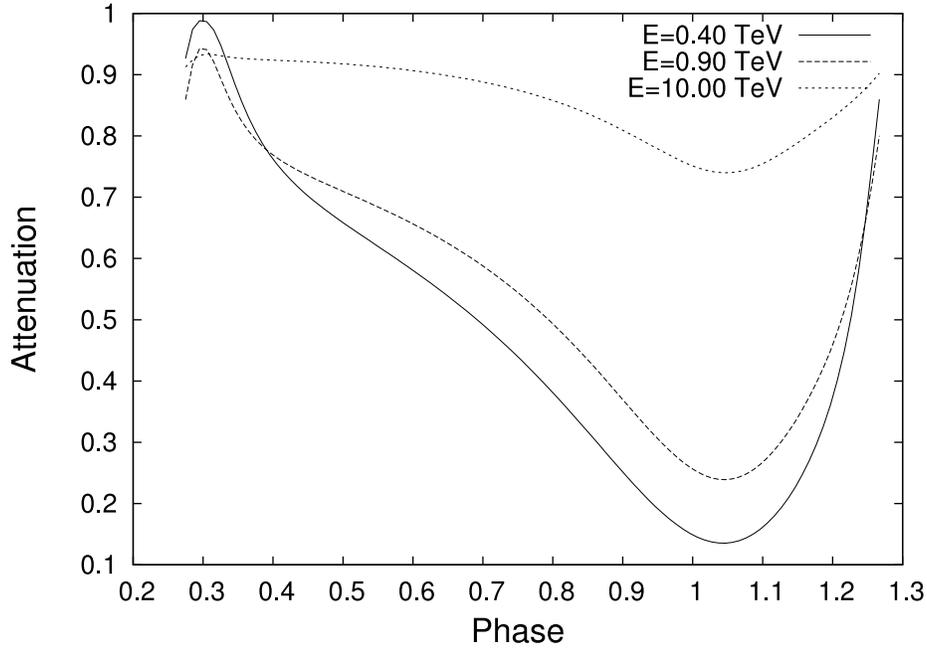}}
    \caption[Attenuation $e^{-\tau_{\gamma\gamma}}$ as a function of the orbital phase for 
      different incident photon energies ($\gamma\gamma$ interactions only)]{\label{att_gamma_gamma} Attenuation $e^{-\tau_{\gamma\gamma}}$ as a function of the orbital phase for 
      different incident photon energies ($\gamma\gamma$ interactions only). 
      A mass of $1.5M_{\odot}$ was assumed for the compact object.}
  \end{center}
\end{figure}

Figure \ref{att_gamma_gamma} shows the attenuation as a function of the orbital phase
for several different energies for the case of the compact object having the
canonical neutron star mass ($i\approx 64^{\circ}$ or $M\approx 1.5M_\odot$). In Figure \ref{att_gamma_gamma} we
essentially reproduce one of the results obtained by Dubus, except that the
orbital parameters used are the newer set obtained by \citet{Aragona}.
When only interactions with the background black body photons are taken into account, and the orbital plane is closer to being seen edge-on,
the optical depth  approaches a minimum when the compact
object is close to the main sequence star. This is especially the case for very 
high inclination angles, corresponding to the mass of the compact object being small, and
close to the Chandrasekhar mass. This behavior can be understood from
the angular dependence of the threshold energy in addition to the  
relative velocity of the incident and target photons approaching a minimum.  Also, at
high energies, the cross section for pair creation decreases as the inverse square of
the center of mass energy, decreasing the optical depth even more. That is,
even though the total density of background photons  
increases (as $1/r^2$) when the compact object approaches the Be star, a combination
of the previously mentioned factors dominates as can seen in Figure
\ref{att_gamma_gamma}. 
 
\begin{figure}[h]
  \begin{center}
    \rotatebox{-90}{\includegraphics[scale=0.5]{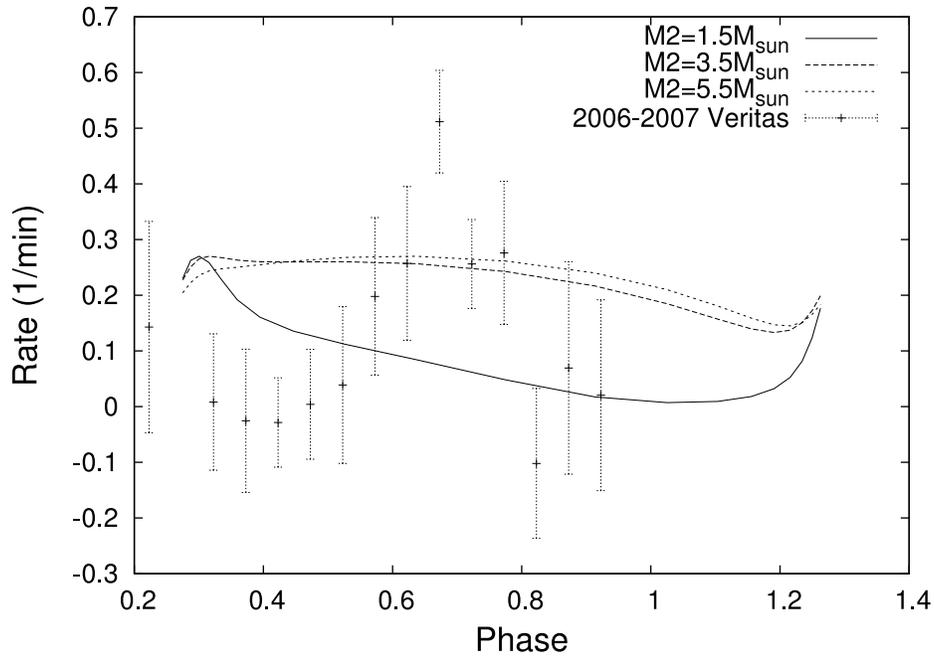}}
    \caption[Normalized light-curve for $\gamma\gamma$ interactions only]{\label{light_curve_gamma_gamma}Normalized light-curve for $\gamma\gamma$ interactions only. 
      Each curve corresponds to a different mass of the compact object.}
  \end{center}
\end{figure}

Figure \ref{light_curve_gamma_gamma} shows the normalized integrated flux
assuming different inclination angles and corresponding compact object masses. The VERITAS data shown in Figure
\ref{light_curve_gamma_gamma} \citep{Veritas2} were binned to show a single light-curve 
as opposed to monthly data. If we assume that the emission comes from the vicinity  
of the compact object, and is isotropic, and constant as a function of 
the orbital phase, then these  results lead us to conclude that there must be an additional  
attenuation mechanism at play.

\subsection{Light curve including $\gamma\gamma$ and $\gamma H$ interactions\label{gamma-h}}

The detailed structure of the circumstellar material surrounding a Be star in
the presence of the compact companion has been studied in detail by \citet{waters}, \citet{marti}\newline

\begin{figure}[h]
\begin{center}
  \rotatebox{-90}{\includegraphics[scale=0.5]{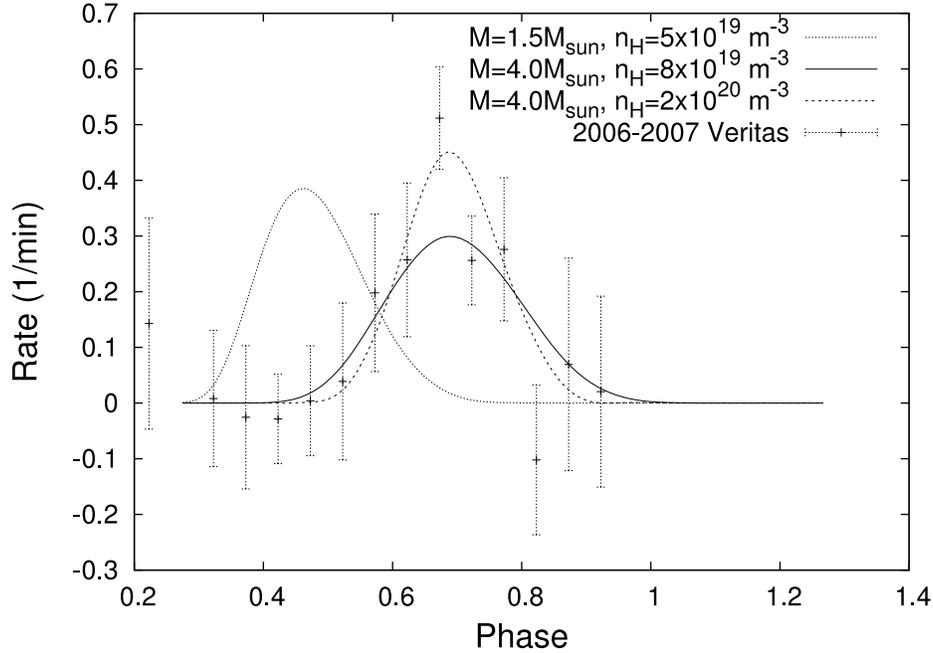}} 
  \caption[Light curve including an isotropic
    distribution of circumstellar material composed oh hydrogen.]{\label{light_curve_gamma_h} Light curve including an isotropic
    distribution of circumstellar material composed oh hydrogen. The mass of
    the compact object was set to $4.0M_{\odot}$ for the two curves whose peak is
    at phase $\sim 0.7$, and $1.5M_{\odot}$ for the curve whose peak is closer
    to phase $\sim 0.4$. The concentration hydrogen at $r\simeq100R_{\odot}$ for
    each curve is labeled in the top right-hand corner of the figure.}
  \end{center}
\end{figure}

\noindent and \citet{reig} among others. It is thought to have a main 
equatorial disk-like component and a polar wind. Typically, the parameters that describe the decretion disk include:  
The mass loss rate, the wind termination velocity, the half opening angle of the disk, and the 
radius of the disk. When comparing the quality of the data shown in Figure \ref{light_curve_gamma_gamma} and the complexity 
of the models that describe the circumstellar material, only an order of magnitude estimate of
the density of the material and its extension in the system can be achieved. With this in mind, 
we rather assume a simple isotropic distribution of material that decreases as
a power $q$ of the 
distance from the Be star ($n=n_0(r_0/r)^q$). We start by setting $q=2$ and then consider different radial dependences for comparison. Parameters found from existing models are 
 taken into consideration for our approximation. We assume that most of the 
material surrounding the Be star is composed of hydrogen, whose cross section with high energy photons 
is approximately \citep{Heitler} $\sigma_{\gamma  H}\simeq2\times10^{-2}\sigma_{T}$ and
roughly independent of the energy above a few hundred MeV.

 We can now add this contribution 
to the optical depth and obtain the light curves shown in Figure \ref{light_curve_gamma_h}, and thus constrain 
 the mass of the compact object and the density of the circumstellar material
 at a distance of $r\approx100R_{\odot}$  (characteristic
 order of magnitude of the system). For the case of a constant cross
 section and a $1/r^2$ distribution of hydrogen, the optical depth can
 actually be found analytically as can be seen in the supplementary Section \ref{appendix}, and the
 behavior at the phase near emission peak is roughly Gaussian. To find the best fit values we take
 the mass, characteristic density and normalization factor as free
 parameters. As mentioned in the previous section, higher masses for the
 compact object shift the emission peak to higher values of the phase as shown
 in Figure \ref{light_curve_gamma_h}. From this figure it is clear that the emission
 peak corresponding to a canonical $1.5M_{\odot}$ neutron star 
 is only marginally supported by observations.

We take the inclination angle, characteristic density, and normalization factor as free
 parameters. We find the inclination angle to be $i< 28^\circ$ ($M_2>3M_\odot$)
 at the $89\%$ confidence level (CL) in the
context of this model, and $i<34^{\circ}$ ($M_2>2.5M_{\odot}$) at the $99\%$
CL. These limits are not in good agreement with the
neutron star scenario generally favored for the broad-band spectrum it
implies\footnote{See \citet{Neronov} for more details}. However, our results are   
still consistent with other observational constraints ($10^{\circ}<i<60^{\circ}$) \citep{Casares} obtained from optical spectroscopy. As for
the circumstellar material, if we assume the characteristic extension to be
$r_0\approx100R_\odot$, consistent with more sophisticated models
\citep{Torres}, then the density of hydrogen in the disk is found to be $2.0\times 10^{13}\rm{cm}^{-3}\leq n_H\leq 1.9\times 10^{15} \rm{cm}^{-3}$ at the $99\%$ CL and $n_H=\left(2.7\pm ^{11.3}_{2.1}\right)\times 10^{14}\rm{cm}^{-3}$ at the $68\%$ CL. 

By integrating the volume density along the line of sight to the compact object at apastron, we find a column density of $1.9\times10^{26}\rm{cm}^{-2}\leq N_H \leq 1.8\times 10^{28}\rm{cm}^{-2}$ at the $99\%$ CL, which is much higher than results found elsewhere in the literature \citep{waters, marti, esposito}. In particular, when we use the column density found by X-ray observations $N_H=(5.7\pm 0.3)\times 10^{21}\rm{cm}^{-2}$ \cite{esposito}, we find a reduced $\tilde{\chi}^2$ of $3.06$ (11 degrees of freedom), corresponding to a $\chi^2$ probability $P(\tilde{\chi}^2\geq 3.06)=0.04\%$. A rough estimate suggests that by including $\sim 10\%$ of helium, the column density would be reduced by a factor of $\sim 2$, which is not sufficient to achieve compatibility with X-ray results.  

Density profiles in Be stars typically have radial dependences of $1/r^q$, where $2.3<q<3.3$ \citep{Lamers}, depending on the opening angle of the disk. Therefore we expect our constraint on the density to constitute a lower bound\footnote{This is assuming that the disk and orbit lie in the same plane.}. We perform our calculation with $q=3$ and note that our results do not change considerably.  %, that is, $n_0\approx(8\pm6)\times10^{19}m^{-3}$. 

The hydrogen density also corresponds to a mass loss rate of $\dot{M_1}\approx10^{-7}\Omega \frac{V_{\mathrm{wind}}}{\mathrm{1km\,s^{-1}}} M_\odot \rm{yr}^{-1}$, where $\Omega$ is the solid angle. Typically accepted values for the mass loss rate are in the range of $\sim 10^{-7}M_{\odot}\rm{yr}^{-1}$ to $10^{-8}M_{\odot}\rm{yr}^{-1}$, 
 as have been reported by \citet{snow} and \citet{waters}
 among others. A first glance at our result for the mass loss implies that it
 does not agree with the observations, i.e., setting $\Omega=4\pi$ and
 $V_{wind}\sim 100\rm{km\,s}^{-1}$ \citep{waters}. However, if we relax the assumption of
 an isotropic distribution of hydrogen, our result implies that small solid
 angles are favored as well as small velocities for the stellar wind. Small
 solid angles are consistent with the thin disk scenario that is most commonly
 accepted. Small velocities of the order of a few $\rm{km\,s^{-1}}$ are however not
 consistent with what is found elsewhere in the literature,
 e.g., \citet{waters}, and the wind indeed has higher velocities, this would imply that the system may have been observed while in a state of high mass loss rate.  

\section{Discussion on $LSI+61\,303^{\circ}$}

Since, in the TeV range, the interaction with matter is approximately independent of the energy, 
and since, as Figure \ref{light_curve_gamma_gamma} shows, $\gamma\gamma$ interactions are insufficient 
to account for the orbital modulation, then the intrinsic nonattenuated differential spectrum
 is essentially the same as the observed spectrum (a power law of
spectral index $-2.4$). However, the intrinsic TeV luminosity is several orders
of magnitude higher than the measured luminosity. Taking the distance to the
source to be approximately $1.8\,\rm{kpc}$ \citep{Steele}, we find the intrinsic luminosity to be $L\approx5\times10^{37}\rm{erg\,s^{-1}}$ when the hydrogen density is of the order of $\sim5\times 10^{13}\rm{cm}^{-3}$. This intrinsic luminosity is comparable to that suggested by  \citet{Bottcher} for \emph{LS 5039}, the only other known TeV binary thought to contain a black hole. 

It is interesting to compare this intrinsic luminosity to the Eddington
luminosity\footnote{At the energies considered here, the cross section for
  inverse Compton is $\sim 0.1\sigma_T$}
$L_{Edd}\approx1.3\times10^{39}(M_2/M_\odot)\rm{erg\,s^{-1}}$, which is comparable to $L$, and implies that 
radiation may be beamed in our direction. It is also
interesting to calculate the accretion rate that would be needed in order to
obtain the intrinsic luminosity: By taking $L\approx GM_2\dot{M_2}/R$, where $R$
is of the order of the Schwarzschild radius ($2GM_2/c^2$), we find
$\dot{M_2}\approx 2\times 10^{-8}M_\odot \rm{yr}^{-1}$. This rate is  comparable with the observed mass loss rate of $\sim 10^{-8}M_\odot \rm{yr}^{-1}$. The
fact that the accretion rate is comparable to the measured mass loss rate, suggests that
the flow of matter can be quite complicated, e.g., an increase in the accretion
rate would strip most of the circumstellar mass, leading to time variability. This 
may explain
the fact that no VHE detections have been reported since 2008.

Still assuming the intrinsic luminosity to be constant in time, we can estimate the amount of hydrogen needed  to attenuate the source to below the detectability threshold. We find that the density must increase from $\sim5\times10^{13}\rm{cm}^{-3}$ to $\sim5\times10^{14}\rm{cm}^{-3}$ at the characteristic distance of $100R_{\odot}$. This amount of hydrogen in turn leads to much higher mass loss rates than those observed, and it may also imply a stronger activity of the source. 

It is worth mentioning that the attenuation model is not the only possible way to account for  the modulation. For example, there is also the possibility of the emission being anisotropic, and the modulation resulting from a geometrical effect. This possibility is described in detail by \citet{Neronov}, were a shocked pulsar wind with a large Lorentz factor is thought to be the cause of emission.\parskip 0pt
\section{Final remarks on $\small{LSI+61\,303^{\circ}}$'s\\ TeV data analysis}

For the case of $LS\,I\,+61^\circ303$,
we find that attenuation due to $\gamma\gamma$ interactions with the
background radiation does not account for the observed high energy
flux modulation as a function of the orbital phase, namely a narrow peak
near apastron. This effect leads us to investigate some
properties of the ejected material from the Be star, and the inclination angle of the orbit. We find the
angle of the orbit to be $i<34^\circ$ ($M>2.5M_\odot$) at the $99\%$ confidence level, 
suggesting that the compact object is a black hole rather than a neutron star. We
also find the density of hydrogen in the disk to be 
$2\times 10^{13}\rm{cm}^{-3}\leq n_H\leq 2\times 10^{15} \rm{cm}^{-3}$ at the $99\%$ CL (at $100R_\odot$), which accounts for most of the
observed gamma ray absorption.  
If the compact object is indeed a black hole as our analysis
suggests, then the gamma ray emission is likely to be powered by accretion \citep{Neronov}.
Also, a black hole scenario might be even more complicated due to
the possibility of VHE emission originating from termination of jets, therefore we
cannot exclude the possibility of the modulation being due to geometrical effects.  Current VHE data do not allow us to constrain the system much more
than what we have already done, and the fact that VHE detections have not been
reported since the VERITAS \citep{Veritas} and MAGIC \citep{Magic} detections
where made, makes the problem even more puzzling. A possible explanation might
originate from a complex matter flow. This is suggested by the fact that the accretion rate needed to explain an intrinsic
nonattenuated luminosity, is comparable to the measured mass loss rate of the Be star.

An inconsistency arises when comparing our results with those derived from X-ray observations. We find the column density to be $1.9\times10^{26}\rm{cm}^{-2}\leq N_H \leq 1.8\times 10^{28}\rm{cm}^{-2}$ ($99\%$ CL), which is only compatible with X-ray results at the 0.04\% confidence level. Such an incompatibility may imply that pair production in the stellar wind is not the cause of the modulation. Consequently, our estimates on the mass and column density may not be valid. An alternative explanation by \citet{Neronov} suggests that the modulation is due to a geometrical effect. Here a shocked pulsar wind is thought to flow along a cone with a large Lorentz factor, producing beamed radiation which can be seen when the cone passes through the line of sight.
\newpage
\section{Prospects for $LSI+61\,303^{\circ}$\\ at high angular resolution}
Imaging at high angular resolution will allow us to further understand the nature of this object, and more objects of this class. The angular size of the Be star is approximately $0.3\,\mathrm{mas}$, and the angular separation between components may be larger by an order of magnitude depending on the inclination of the orbital plane and the orbital phase. Therefore, only interferometric techniques allow us to resolve this system. Radio observations with the Very Large Baseline Interferometer (VLBI) show structure at the milliarcsecond scale \citep{Massi} and show evidence of a precessing jet associated to the compact object. However, more information about the circumstellar environment of the Be star can be obtained by going to shorter near-infrared wavelengths since Be stars are known to have expanding dust shells, viscous disks, and/or strong radiatively driven winds. Current instruments such as CHARA, whose angular resolution can be as good as $0.3\,\mathrm{mas}$ at $550\,\mathrm{nm}$, could use their largest $330\,\mathrm{m}$ baseline to obtain spectro-intereferometric data, where a shift in the image photocenter as a function of the wavelength may allow us to constrain the kinematics of the circumstellar matter\footnote{If the Be star is observed edge-on, then one side should be blue shifted, and the other should be red-shifted since it is fast rotating. The measured phase of the complex visibility would be consistent with a nonsymmetric object.}. If observations are done in the K band ($\sim 2200\,\mathrm{nm}$), the angular resolution will decrease to $\sim 1.3\,\mathrm{mas}$, but it would be interesting to measure the interferometric visibility across the $H_\alpha$ emission line, which is associated to the cool circumstellar environment. If a decrease in the visibility is evident, this would imply that the $330\,\mathrm{mas}$ baseline resolves the circumstellar environment, if no decrease is evident, an upper limit to the physical extension can be found. 

To obtain a fully reconstructed optical image, much better baseline ($(u,v)$) coverage is necessary, and going to shorter wavelengths may be beneficial in terms of angular resolution.  At these short wavelengths, information of the stellar shape and temperature distribution can be obtained. In order to image features ranging between $0.3-3\,\mathrm{mas}$, an instrument would require baselines ranging between a few tens of meters to a few hundred meters. In terms of angular resolution, this is within the capabilities of future intensity interferometers, whose simulated results show that imaging stellar shapes and temperature distributions is indeed possible (see Chapter \ref{sii_with_iact}). However, this object is just barely within the brightness detectability limit with intensity interferometry, and more detailed simulations are needed in order to determine if this is a suitable target.

\newpage
\section{Supplement: Optical depth for constant\\ cross section and $1/r^2$ density  distribution}\label{appendix}

\begin{figure}
  \begin{center}
    \scalebox{3.0}{
      \begin{tikzpicture}
        \draw (0,0) circle (0.5cm);
        \shade[ball color=green] (0,0) circle (0.05cm);
        \shade[ball color=red] (0.35,0.35) circle (0.05cm);
        %\draw [dashed] (0.35,0.35) -- (2.5,0);
        \draw [->,dashed] (0.35,0.35) -- node[above]{\tiny{to obs.}}(1.5,0.35);
        %\shade[ball color=blue] (2.5,0) circle (0.05cm);
        \draw (0.55cm, 0cm) arc (0:45:0.55cm);
        \draw [dashed] (0,0) -- (1.5, 0);
        \draw (0.65cm, 0.15cm) node {\tiny{$\theta$}};
        \draw [->](-0.1,-0.1)--(1, -0.1);
        \draw [->](-0.1, -0.1)-- (-0.1,1);
        \node at (1,-0.25){\tiny{x}};
        \node at (-0.2,1){\tiny{y}};
      \end{tikzpicture}
    }\\\vspace{1.5cm}
    \scalebox{0.6}{
     \includegraphics{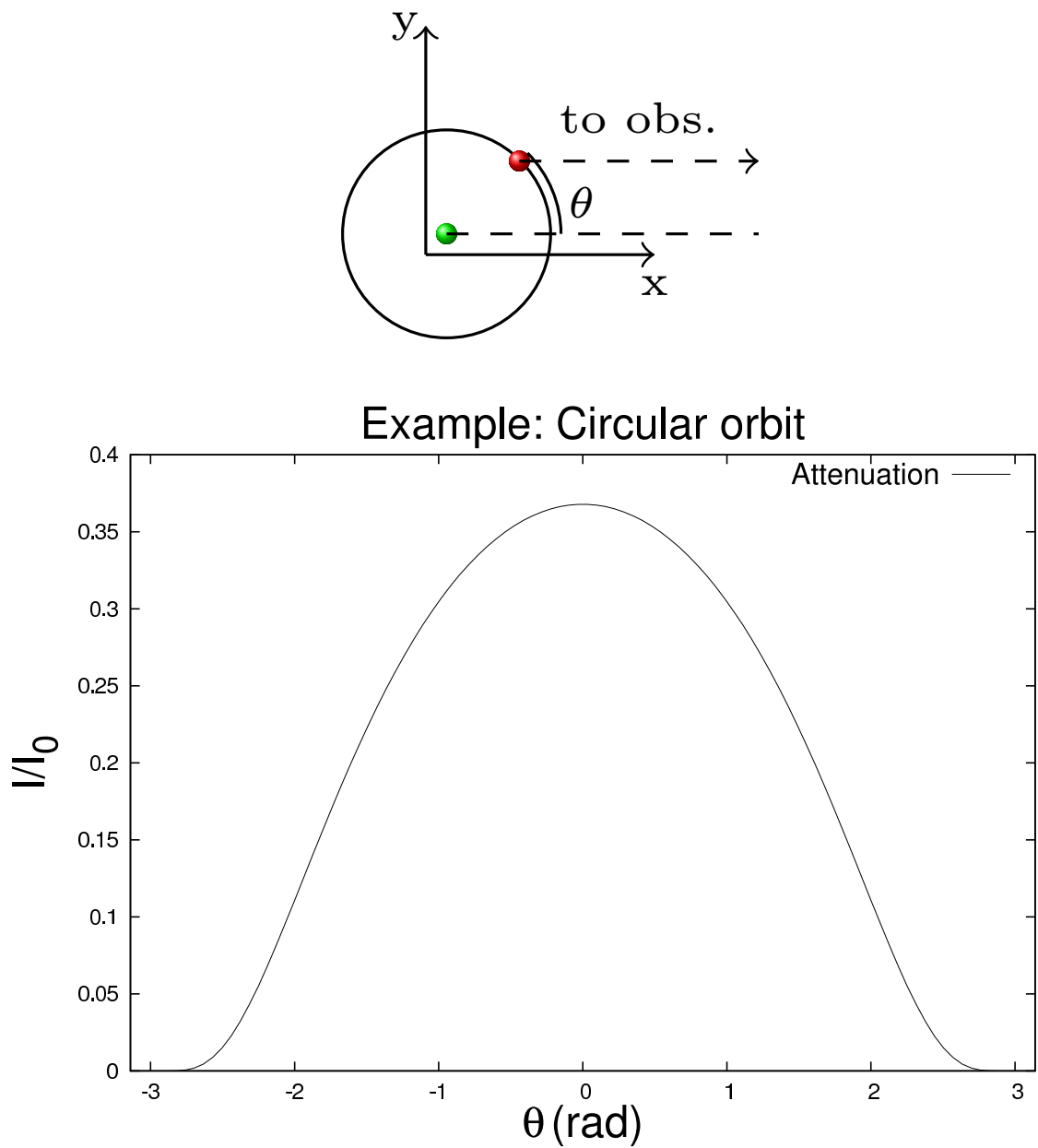}
    }
  \end{center}
  \caption{\label{circle}  Upper figure: Coordinate system used for calculating the optical depth  (eq. \ref{first approximation}). Lower figure: Light curve for a circular orbit.}
\end{figure}

Using a $1/r^2$ distribution of hydrogen, the cross section $\sigma_H$
accounting for interactions between VHE photons and hydrogen, and the system
of coordinates shown in Figure \ref{circle} (corresponding to an orbital plane
seen edge on) , we can calculate the integral for the optical depth to be

\begin{eqnarray}
  \int_{x_i}^{\infty}\frac{n_0r_0^2\sigma_H}{x^2+y_i^2+z_i^2}dx 
  &=&\left[\frac{n_0r_0^2\sigma_H}{\sqrt{y_i^2+z_i^2}}\tan^{-1}\left(\frac{x}{\sqrt{y_i^2+z_i^2}}\right)\right]^{\infty}_{x_i}\\
  &=& \frac{n_0r_0^2\sigma_H}{\sqrt{y_i^2+z_i^2}}\left(\frac{\pi}{2}-\tan^{-1}\left(\frac{x_i}{\sqrt{y_i^2+z_i^2}}\right) \right),
  \label{first approximation}
\end{eqnarray}
where $x_i$, $y_i$, and $z_i$ are functions of the orbital angle $\theta$, and $r_0$
is the characteristic radius of the hydrogen disk. For the case of a circular orbit as seen edge on
(Figure \ref{circle}), we can easily see the limiting behavior of the
intensity as a function of the orbital angle. That is, expanding around $\theta\sim0$ reveals that the attenuation around this region behaves like a Gaussian.

\begin{equation}
For\,\,\theta\sim0:\,\,I(r_i, \theta)= I_0(\theta, r_i)e^{-\frac{n_0r_0^2\sigma_H}{r_i}(1+\theta^2)}.
\end{equation}

Similarly, expanding around $\theta\sim\pi/2$ reveals that the attenuation behaves like a decreasing exponential

\begin{equation}
For\,\,\theta\sim\pi/2:\,\,I(r_i, \theta)=I_0(\theta, r_i)e^{-\frac{n_0r_0^2\sigma_H}{r_i}\theta}.
\end{equation}

For a more complicated geometry of $LS\,I\,+61^\circ 303$, it is now just a matter of inserting the appropriate expressions for $x_i(\theta)$, $y_i(\theta)$ and $z_i(\theta)$.

\begin{comment}
\begin{figure}
  \begin{center}
    \begin{equation}
    \begin{array}{cc}
          \scalebox{2}{
            \begin{tikzpicture}
              \draw (0,0) circle (0.5cm);
              \shade[ball color=green] (0,0) circle (0.05cm);
              \shade[ball color=red] (0.35,0.35) circle (0.05cm);
              %\draw [dashed] (0.35,0.35) -- (2.5,0);
              \draw [->,dashed] (0.35,0.35) -- node[above]{\tiny{to obs.}}(1.5,0.35);
              %\shade[ball color=blue] (2.5,0) circle (0.05cm);
              \draw (0.55cm, 0cm) arc (0:45:0.55cm);
              \draw [dashed] (0,0) -- (1.5, 0);
              \draw (0.65cm, 0.15cm) node {\tiny{$\theta$}};
              \draw [->](-0.1,-0.1)--(1, -0.1);
              \draw [->](-0.1, -0.1)-- (-0.1,1);
              \node at (1,-0.25){\tiny{x}};
              \node at (-0.2,1){\tiny{y}};
            \end{tikzpicture}}&\hspace{1cm}\rotatebox{0}{\scalebox{0.25}{\includegraphics{figures/circular.eps}}}\\          
      \end{array}
    \end{equation}
  \end{center}
  \caption{\label{circle}  Coordinate system used for calculating the optical
    depth  (eq. \ref{first approximation}).}
\end{figure}
\end{comment}

\chapter{Air Cherenkov telescope arrays as SII receivers} \label{sii_with_iact}
\vspace{-0.9cm}

\section{A revival of SII\label{revival}}

Even though Stellar Intensity Interferometry (SII) was abandoned in the 1970s,
there has been a recent interest in reviving this technique, mainly due to the
unprecedented $(u,v)$ plane coverage that future imaging air Cherenkov telescope (IACT) arrays will
provide \citep{cta}. The possibility of probing stars at the 
submilliarcsecond scale and visible wavelengths has motivated new developments in instrumentation and
simulations, the latter being the focus of this chapter. 

Recent results obtained with amplitude (Michelson) interferometry have started to
reveal stars as extended objects (e.g., \citealt{coast, chara_review}), and with nonuniform light 
intensity distributions in the milliarcsecond scale. 
Such interesting results can be further investigated with SII taking advantage of the longer (km)
baselines and relative ease of observing at shorter (blue) wavelengths.
For example, measuring stellar diameters at different wavelengths, will make it possible to further investigate the
wavelength dependence of limb darkening,
\citep{Mozurkewich} and thus constrain stellar atmosphere models. Radii measurements with uncertainties of a few percent, along 
with spectroscopic measurements are necessary  
to constrain the position of stars in the HR diagram  (e.g., \citealt{Kervella}). With the
methods described in this chapter, we show that diameters can in principle be measured
with accuracies better than $1\%$ when using realistic  array
configurations for future experiments such as CTA (Cherenkov Telescope
Array). As another example we can consider 
fast rotating B stars, which are ideal candidates for imaging oblateness,
pole brightening \citep{altair_image, von_zeipel}, radiatively driven mass loss \citep{Friend}, and perhaps even pulsation modes
\citep{pulsation}. The impact of rotation on stellar evolution is nontrivial, and 
several studies have been made in the subject (e.g., \citealt{martin, maeder}). Images of rotating stars have become available 
in the past few years (e.g., \citealt{altair_image, vega_image}),  and 
measurements of oblateness with accuracies of a few percent have been made. 
We will show that this is comparable to what 
can be achieved with SII using large arrays of Cherenkov telescopes. 
There is also the case of interacting binaries, for which we can measure angular separation,
diameters, and relative brightness. It may even be possible to measure mass transfer \citep{verhoelst}. Measurements of the angular separation
in binaries is crucial for determining the masses of stars. These masses must be found to within $\sim 2\%$ \citep{Andersen}
in order to test main sequence models. With the methods described in this chapter, we show that angular 
separations can be found to within a few percent from reconstructed images. % A more convenient and accurate analysis (beyond the scope of this study) does not require imaging, and should yield sub-percent uncertainties, so that testing main sequence models will become possible with SII. 

In preparation for a large-scale SII observatory deployment, several 
laboratory experiments are in progress \citep{stephan.spie}. Their main goal is to measure light intensity correlation between two receivers. 
It is also worth mentioning the \emph{StarBase} \citep{starbase} observatory
(located in Grantsville, Utah) which consists of two $3\mathrm{m}$ light receivers separated by $24\,\mathrm{m}$ and which will be used
 to test high time resolution digital correlators,
band to measure the second order degree of coherence for a few stars (see chapter \ref{experiment}). Various analog and digital correlator
technologies \citep{Dravins.timescale} are being implemented, and cross correlation of streams of photons with
nanosecond-scale resolution has already been achieved.

Intensity interferometry, unlike amplitude interferometry, relies on the
correlation between intensity fluctuations 
averaged over the spectral band at electronic (nanosecond) time resolution.
These averaged fluctuations are much slower than the (femtosecond) 
light wave period. This correlation is directly related to the 
complex degree of coherence $\gamma_{ij}$ as \citep{lipson}

\begin{equation}
|\gamma_{ij}|^2=\frac{<\Delta I_i \,\Delta I_j>}{<I_i> <I_j>}.
\end{equation}

Here, $<I_i>$ is the time average of the intensity received at a particular
telescope $i$, and $\Delta I_i$ is the intensity fluctuation. Measuring 
a second-order effect results in lower signal-to-noise ratio when compared to amplitude interferometry \citep{holder2}. This
sensitivity issue can be dealt with by using large light collection areas 
(such as those available with air Cherenkov telescopes), longer exposure times
and baseline redundancy.  

The low frequency fluctuation can be interpreted classically as the beat formed by
neighboring Fourier components. Since SII relies on low frequency
fluctuations, which are typically several orders of magnitude smaller than the
frequency of optical light,  it does not rely on the phase difference between light waves, 
but rather in the difference between the relative phases of the two components 
at the detectors \citep{hanbury_brown0}. The main advantage is the relative insensitivity to
atmospheric turbulence and the absence of a requirement for sub-wavelength precision
in the optics and delay lines \citep{hanbury_brown0}. 

The complex mutual degree of coherence $\gamma$ is proportional to the Fourier
transform of the radiance distribution of the object in the sky (Van Cittert-Zernike theorem).
However, since with SII, the
squared-modulus of $\gamma$ is the measurable quantity, the main
disadvantage is that the phase of the Fourier transform is lost in the
measurement process. The loss of phase information poses a severe difficulty,
and images have in the past been reconstructed from the bispectrum technique, 
using monolithic apertures  (e.g., \citealt{lawrence}).
The imaging limitations
can be overcome using a model-independent phase recovery technique. Even though
several phase reconstruction techniques exist \citep{fienup}, we concentrate on a two
dimensional version of the one dimensional analysis introduced by Holmes and Belen'kii (2004), which is based on the theory of analytic functions,
and in particular the Cauchy-Riemann equations.  

Following recent successes in Gamma ray astronomy, a next generation Cherenkov telescope array is in a preparatory stage.
This project is currently known as CTA (Cherenkov telescope array) \citep{cta}, and will contain between 50 and 100 telescopes with 
apertures ranging between $5\,\mathrm{m}$ and $25\,\mathrm{m}$. In this chapter we investigate the sensitivity 
and imaging capabilities of SII implemented on such an atmospheric Cherenkov 
telescope array. We start with a discussion of sensitivity (section \ref{sensitivity}), followed by a discussion of simulating 
noisy data as would be realistically obtained with such an array (section \ref{simulation}). Since data have
a finite sampling in the $(u,v)$ plane, we discuss our method of fitting an
analytic function to the data in order to estimate derivatives which are
needed for phase reconstruction (section \ref{fit}). We then proceed to quantify the reconstruction quality using several criteria. We
start with the simple case of uniform disks (section \ref{disks}) and progressively increase the degree of
image complexity by including oblateness (section \ref{oblate_sec}), binary
stars (section \ref{binary_sec}), and  obscuring
disks and spots (section \ref{complex}).

\section{Sensitivity} \label{sensitivity}

The signal to noise ratio (SNR) for an intensity correlation measurement depends on the
degree of correlation $\gamma$, the area $A$ 
of each of the light receivers, the spectral density $n$ (number of photons per unit area per 
unit time, per frequency), the quantum efficiency $\alpha$, the electronic bandwidth $\Delta f$, 
and the observation time $t$. The SNR can be expressed as \citep{hanbury_brown0}

\begin{equation}
SNR=n(\lambda, T, m_v)\,A\;\alpha\;\gamma^2\sqrt{\Delta f t/2}. \label{snr}
\end{equation}

This expression can be found from the results presented in section \ref{variance}, and eq. \ref{mu_prob} in particular. The SNR is 
essentially the ratio between the wave noise, which is correlated
between neighboring detectors, and the shot noise, which is uncorrelated between detectors. 

\begin{figure}[h]
  \begin{center}
    \rotatebox{-90}{\includegraphics[scale=0.5]{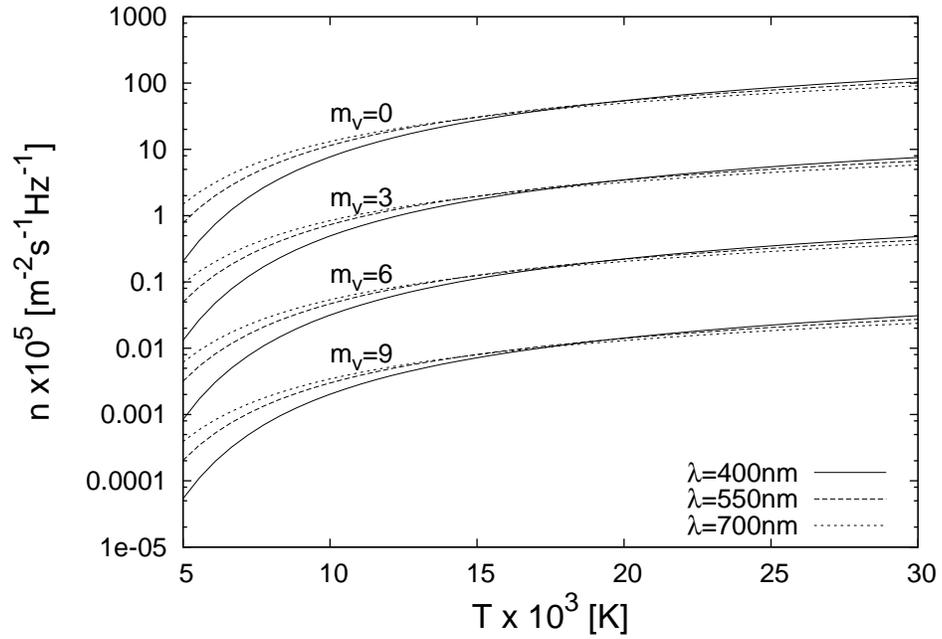}}
  \end{center}
  \vspace{0.5cm}
  \caption[Spectral density as a function of temperature for several different visual magnitudes and observed wavelengths.]{\label{n_vs_T}Spectral density as a function of temperature for several different visual magnitudes and observed wavelengths. Atmospheric absorption as a function of the wavelength is not taken into account.}
\end{figure}

The spectral density $n$ is related to the visual magnitude $m_v$ of the star as well as its temperature $T$ 
and observing wavelength $\lambda$. The spectral density $n(\lambda, T, m_v)$ is the number of black body photons 
per unit area, per unit frequency and per unit time. The dependence of the visual magnitude $m_v$ is found by 
recalling that the flux for a $0^{th}$  magnitude star with a temperature of $9550^{\circ}\mathrm{K}$ observed at $550\,\mathrm{nm}$  
is $3.64\times 10^{-23}\mathrm{W m^{-2}Hz^{-1}}$ \citep{flux}. This in turn corresponds to a spectral density of
$10^{-4}\,\mathrm{m^{-2} s^{-1} Hz^{-1}}$. The 
spectral density as a function of temperature (for different visual
 magnitudes and observing wavelengths) is shown in Figure \ref{n_vs_T}, and we see that at constant visual magnitude 
and observing wavelength, higher temperatures correspond to higher spectral densities. We find that the increase 
in temperature $\Delta T(\lambda, T, \Delta m_v)$ is approximately $\frac{\lambda k T^2}{h c}\Delta m_v$ for the 
range of temperatures and wavelengths considered in Figure \ref{n_vs_T}. For example, at $400\,\mathrm{nm}$, a decrease of 1 visual 
magnitude is equivalent to increasing the temperature of the star from $5000\,\mathrm{K}$ to $5700\,\mathrm{K}$. Therefore, bright and hot targets 
are the most easily detected with SII.

\begin{figure}[h]
  \begin{center}
    \rotatebox{-90}{\includegraphics[scale=0.55]{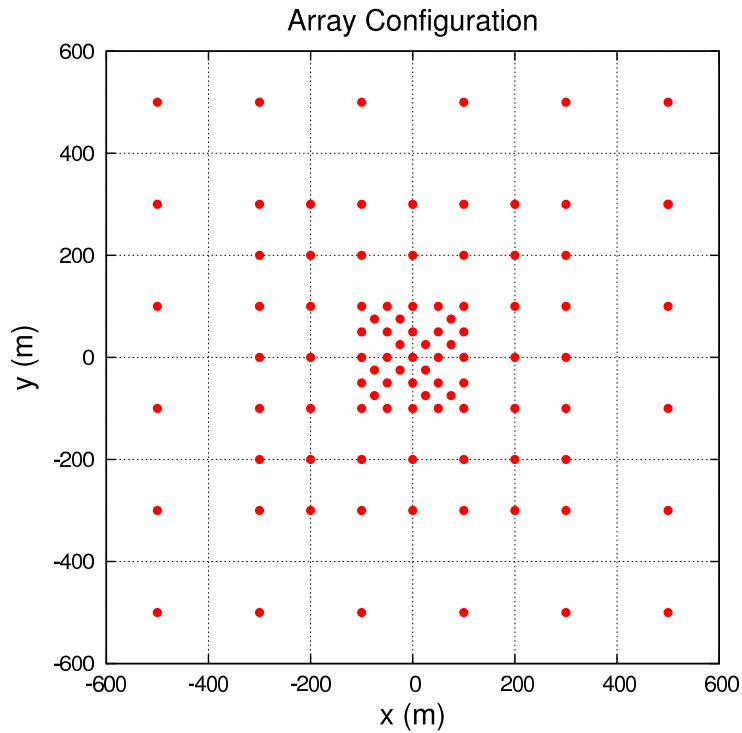}}
    \vspace{0.5cm}
  \end{center}
  \caption{\label{array} Array configuration used for our analysis.}
\end{figure}

\begin{figure}[h]
\begin{center}
  \rotatebox{-90}{\includegraphics[scale=0.55]{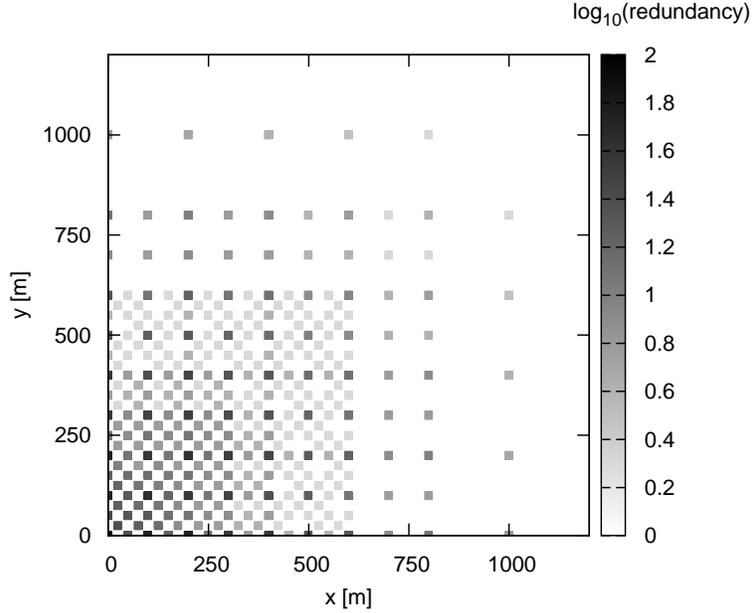}}
\end{center}
\caption[Histogram of CTA baselines.]{\label{baselines} A total of 4656 nonzero baselines are available in the array design used in 
this study. Gray scale measures the $log$ of the baseline redundancy. Since the array shown in 
Figure \ref{array} is almost symmetric with respect to $x$ and $y$, only a quadrant of the $(u,v)$ plane is displayed.}
\end{figure}

We use a preliminary design of the CTA project as an array configuration \citep{cta_simulation}, which is shown in
Figure \ref{array}. This array contains $N=97$
telescopes and $N(N-1)/2=4646$ baselines (many of which are redundant) which are shown in Figure \ref{baselines}. Each detector is
assumed to have a light collecting area of $100\,\mathrm{m}^2$  and a
light detection quantum efficiency of $\alpha=0.3$. Using a $\lambda/D$ criterion, we find that 
the largest baselines of $1.5\,\mathrm{km}$ resolve angular
scales of $\sim 0.05\,\mathrm{mas}$ at $400\,\mathrm{nm}$.
The smallest $48$  baselines of $35\,\mathrm{m}$ resolve angular scales 
of $\sim 2\,\mathrm{mas}$ . However, we show in section \ref{disks}, 
that the largest angular scales that can be realistically \emph{imaged} with our analysis, 
in a model independent way, are more determined by 
 baselines of $\sim 70\,\mathrm{m}$. This is because the estimation
of derivatives of the phase (needed for phase recovery) degrades as the number of 
baselines is reduced. Baselines of $70\,\mathrm{m}$ resolve 
angular scales of $\sim 1.2\,\mathrm{mas}$ at $400\,\mathrm{nm}$.

\begin{figure}[h]
  \begin{center}
    \rotatebox{-90}{\includegraphics[scale=0.4]{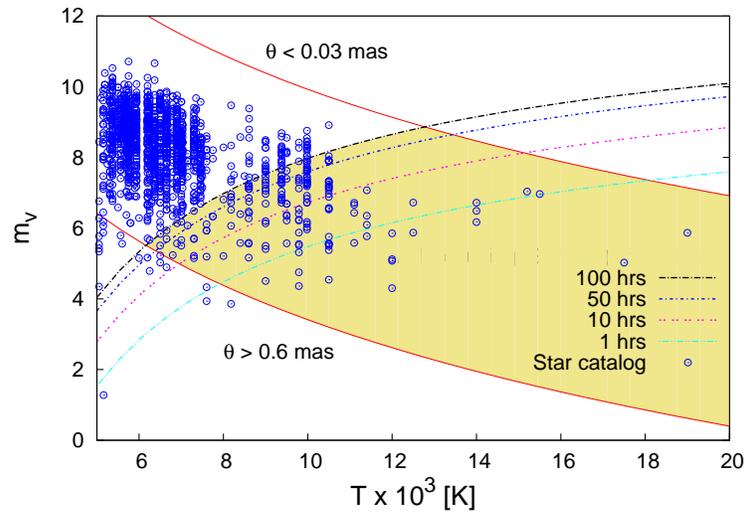}}
  \end{center}
  \vspace{0.5cm}
  \caption[Sensitivity curves for CTA.]{\label{MvsT} The four parallel curves indicate the maximum detectable visual magnitude 
as a function of the temperature for several exposure times. Each of these four curves corresponds to
5 standard deviation measurements and $|\gamma|=0.5$. Also shown is the (shaded) region corresponding
to angular radii between $0.03\,\mathrm{mas}$ and $0.6\,\mathrm{mas}$, and observable within 
$100\,\mathrm{hrs}$ of observation time. The positions in $(T,m_v)$ space of 2000 stars from the 
JMMC stellar diameters catalog are included. }
\end{figure}

These order-of-magnitude considerations are taken into account when performing
simulations and image reconstructions, i.e., the minimum and maximum size of
pristine images that can be reconstructed by data analysis, do not go far beyond these limits. More precise array resolution 
limits are presented
in section \ref{disks} (diameters ranging between $0.06\,\mathrm{mas}\,-\,1.2\,\mathrm{mas}$) . By combining 
these angular scales with the SNR (eq. \ref{snr}),  
we obtain Figure \ref{MvsT}. This figure shows the highest visual magnitude, for which photon correlations (with $|\gamma|=0.5$)
can be detected (5 standard deviations), as a function of the temperature, and for several different exposure times. Also 
shown in Figure \ref{MvsT}, is the shaded region 
corresponding to angular diameters between $0.03\,\mathrm{mas}$ and $0.6\,\mathrm{mas}$ \footnote{These curves of constant angular size can be found approximately by recalling that the visual magnitude $m_v$ is related to a calibrator star of visual magnitude $m_0$ by: $(m_v-m_0)=-2.5\log{F/F_0}$. Here, $F$ and $F_0$ refer to the flux in the visual band . To express $m_v-m_0$ as a function of the angular
size, note that flux is proportional to $\theta^2T^4$, where $\theta$ is the angular size and $T$ is the temperature of the star.}, and observable 
within $100\,\mathrm{hrs}$. From the Figure we 
can see how correlations of photons from faint stars can be more easily detected if they are hot. To quantify the number of stars
for which photon correlations can be detected with the IACT \linebreak

\newpage
\noindent array, we use the JMMC stellar diameters catalog \citep{catalog}. We 
find that $\sim 1000$ (out of $\sim33000$) stars from the JMMC catalog can be detected 
within $1\,\mathrm{hr}$, correlations from $\sim 2500$ stars can be detected within 
$10\,\mathrm{hrs}$, and $\sim 4300$ can be detected within $50\,\mathrm{hrs}$. In Figure \ref{array},
we show a random sample of 2000 stars (out of $\sim33000$) 
from the JMMC catalog. Interstellar reddening may play a role in reducing the number of measurable targets

\section{Simulation of realistic data} \label{simulation}

Pristine images of disk-like stars, oblate stars, binaries, or  featured stars, are first generated. The original ``pristine'' 
image consists of $2048\times2048$
pixels corresponding to $\sim10\,\mathrm{mas}\times10\,\mathrm{mas}$ of angular extension and a
wavelength of $\lambda= 400\,\mathrm{nm}$. The
Fourier transform of the image is then performed via an FFT algorithm and
normalized so that its value is one at zero baseline. This results in a
Fourier transform sampled every $\mathrm{\sim(8\,\mathrm{m})/\lambda}$, 
i.e., $2\times 10^7$ cycles per radian field-of-view at a wavelength $\lambda$ of $400\,\mathrm{nm}$. We then find 
the squared-modulus of the degree of coherence between the members of each pair of telescopes.
This is obtained from a linear interpolation of the squared Fourier magnitude in the finely
sampled FFT. Diurnal motion is not taken into account in the simulations. Diurnal motion plays 
a significant role in increasing the $(u,v)$ coverage when exposure times are long. 
As a consequence there is less $(u,v)$ coverage in the simulations since 
projected baselines do not drift with time. The smaller
$(u,v)$ coverage is however compensated by smaller statistical error in the correlation measurements. 

\begin{figure}[h]
  \begin{center}
    \includegraphics[scale=0.45, angle=-90]{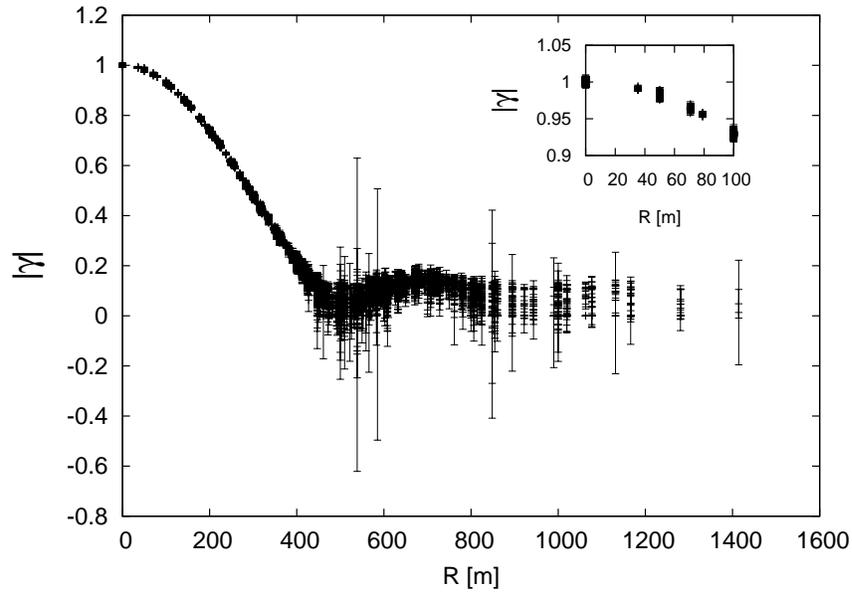}
  \end{center}
  \vspace{0.2cm}
  \caption[Example of simulated data for a $3^{rd}$ magnitude 
uniform disk star ($T=6000^{\circ}K$) of radius $0.1 \,\mathrm{mas}$ and $10\,\mathrm{hrs}$ of observation time.]{\label{example_simulation} Example of simulated data for a $3^{rd}$ magnitude 
uniform disk star ($T=6000^{\circ}K$) of radius $0.1 \,\mathrm{mas}$ and $10\,\mathrm{hrs}$ of observation time. Here we show
$|\gamma|$ (instead of the directly measured $|\gamma|^2$) as a function of telescope separation.}
\end{figure}

The final step in the simulation phase is the addition of noise to the
correlation at each baseline. This noise was found to be
Gaussian by performing the time integrated product of two random streams of
simulated photons as detected by a pair of photo-multiplier tubes. The standard
deviation of the noise added to each pair of telescopes is calculated
from eq. \ref{snr}. In this study we take the signal bandwidth to be $\Delta
f=200\,\mathrm{MHz}$. An example of simulated data as a function of telescope separation 
is shown in Figure \ref{example_simulation}. This corresponds to a $3^{rd}$ magnitude 
uniform disk star ($T=6000^{\circ}\mathrm{K}$) of radius $0.1 \,\mathrm{mas}$ 
and $10\,\mathrm{hrs}$ of observation time. The software used for the simulations, as well as the 
analysis\footnote{See sections \ref{fit} and \ref{phase_rec_sec} for details on the analysis. Some 
variants of the analysis software were developed in \textbf{MATLAB}. All software is available upon request.}, was developed in \textbf{C}.  

In section \ref{simple}, the capabilities for reconstructing simple stellar images, with mostly uniform radiance distributions, are discussed in detail. Then
 the degree of image complexity is increased by generating pristine images of stars with nonuniform radiance distributions, 
e.g., limb-darkening and star spots. These simulated images
correspond to black-bodies of a specified temperature containing an arbitrary number
of ``star spots'' of variable size, temperature, and location at the surface of the spherical star in this three-dimensional model. The simulated 
stellar surface is then projected onto a plane, so that
spots located near the edge of the visible half-sphere appear more elongated than those located near the center. Additionally, limb-darkening is 
included by assuming that the stellar atmosphere has a constant opacity (more details of the simulated images are presented in section \ref{results}). Then 
 the image reconstruction capabilities are quantified in section \ref{complex}.

\section{Data analysis}

%DIAGRAM??

\subsection{Fitting an analytic function to the data} \label{fit}

The estimation of derivatives of the Fourier log-magnitude is at the heart of the Cauchy-Riemann phase
recovery algorithm (section \ref{cr}), and is thus greatly simplified when data is known on
a square grid rather than in a `randomly' sampled way as is directly available
from observations. Once simulated data are available (or observations in the future), an analytic function is fitted to
the data. 

An analytic function can be expressed as a linear combination of basis functions. When data
$f(x_i)\equiv|\gamma(x_i)|^2$ are available at baselines $x_i$, with uncertainty
$\delta f(x_i)$, the coefficients of the basis functions can be found by minimizing the
following $\chi^2$:

\begin{equation}
\chi^2=\sum_i\left[\frac{\left(f(x_i)-\sum_ka_kg_k(\alpha R(x_i))\right)}{\delta f(x_i)}\right]^2. \label{chi2}
\end{equation}

Each $a_k$ is the coefficient of a basis function $g_k$.  The constant $\alpha$ is a scaling factor,
 and $R$ is a rotation operator. The scaling factor and rotation angle are found by first performing a two-dimensional 
Gaussian fit. Finding the appropriate scale and rotation angle has the advantage of reducing the number of
basis elements needed to fit the data. 

Basis functions that tend to zero
at very large baselines, where data are scarce (see Figure \ref{baselines}), are ideal. For this reason, we use 
Hermite functions. There are situations where data are more easily fit with a different set of basis functions, e.g., 
a binary with unresolved members, where the data is purely periodic. In such a situation, data do not rapidly tend to zero at large baselines, so 
the Hermite function fit may contain a large number of elements and result
in high frequency noise where data are scarce.\footnote{A basis set consisting of products of sines and cosines is more
suitable in this situation} The best choice of basis functions may therefore depend on the structure of the object. However, 
for consistency, we use the Hermite fit for all the objects that we analyze,
and find that it gives reasonably good results. 

The $\chi^2$ minimization problem can be turned into a linear system by setting the set of partial
derivatives $\frac{\partial \chi^2}{\partial a_k}$ to zero. We typically start
with a small number of basis elements, say eight, and only increase the number
of basis elements if the optimized reduced $\chi^2$ is greater than some predefined value. 

\subsection{Cauchy-Riemann phase reconstruction} \label{phase_rec_sec}

 Since the image is real, the Fourier magnitude 
is symmetric with respect to the origin in the $(u,v)$ plane. Lack of phase information results 
in the reconstructed image being arbitrary up to a translation and reflection. 

\section{Imaging capabilities for simple stellar objects }\label{simple}

In order to perform a model-independent image reconstruction, 
the phase of the Fourier transform needs to be recovered from
magnitude information only \citep{lipson}. The Cauchy-Riemann phase
reconstruction technique is discussed in section \ref{cr}, and we will use this to obtain the results presented below.

We investigated the imaging capabilities for simple objects,\footnote{For preliminary study see \citet{nunez.spie}.} namely uniform
disk-like stars, oblate rotating stars, binaries, and more complex
images. For most of the objects that we consider, image reconstruction is 
not necessary, i.e., from the Fourier magnitude alone, one can extract radii, 
oblateness, relative brightness in binaries, etc. Estimation of these 
parameters is probably more accurate when extracted directly from Fourier
 magnitude data only, especially if some apriori knowledge of the original 
image is available. However, measuring simple parameters from reconstructed 
images is the first step in quantifying reconstruction capabilities with IACT 
arrays. We assume no apriori knowledge of the images that are being 
reconstructed, and then do a statistical study of the uncertainties of the reconstructed parameters.

\subsection{Uniform disks \label{disks}}

\begin{figure}[h]
  \begin{center}
    \rotatebox{-90}{\scalebox{0.6}{\includegraphics{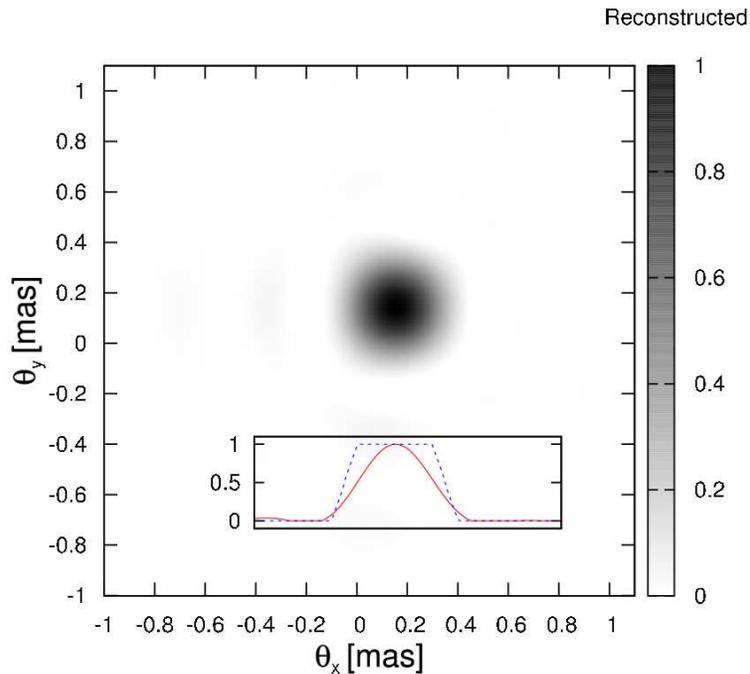}}}
  \end{center}
  \vspace{-0.5cm}
  \caption[Example of a reconstructed uniform disk of radius $0.1\,\mathrm{mas}$.]{\label{radius_a} Example of a reconstructed uniform disk of radius $0.1\,\mathrm{mas}$. 
    Also shown is a slice of the reconstructed  image (solid line) compared to 
    a slice of the pristine image convolved with the PSF of the array (dashed line).}
\end{figure}

\begin{figure}[h]
  \begin{center}
    \rotatebox{-90}{\scalebox{0.4}{\includegraphics{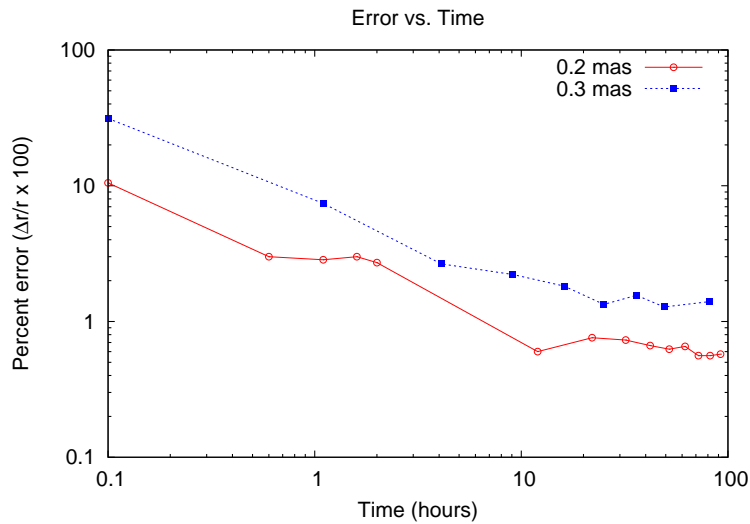}}} 
  \end{center}
  \vspace{0.5cm}
  \caption[Percent error as a function of time for two reconstructed radii ($m_v=6$).]{\label{uncertainty_b} Percent error as a function of time for two reconstructed radii ($m_v=6$). This 
    error was found by performing several reconstructions for each  radius and exposure time, 
    and then taking the standard deviation of the reconstructed radius.}

\end{figure}

In order to quantify the uncertainty in the reconstructed radius, we 
simulate data corresponding to $6^{th}$ magnitude stars ($T=6000^{\circ}\,\mathrm{K}$) with disk radii up to
$1\,\mathrm{mas}$ for 50 hours of exposure time\footnote{\label{long_exposures} This brightness and exposure time 
correspond to uncertainties in the 
simulated data of a few percent. Such long exposure times can be reduced to 
a few hours as is shown in Figure \ref{uncertainty_b}}. An example of such a reconstruction is shown in Figure
\ref{radius_a}, where the radiance is shown in arbitrary units between 0 and
1. For a uniform disk, the reconstructed phase is null in the first lobe, and we find that the rms deviations from the true
phase are approximately $0.19\,\mathrm{rad}$ in the null zone.  A first look at the reconstruction in Figure \ref{radius_a} reveals
that the edge of the reconstructed disk is not sharp, so a threshold in the
radiance was applied for measuring the radius. The radius is measured
by counting pixels above a threshold and noting that the area \linebreak

\noindent of the disk is
proportional to the number of pixels passing the threshold. After experimenting
with different radii, we chose the threshold for measuring the radius to be
0.2. We can now compare the simulated and
reconstructed radii as is shown in Figure \ref{radius_b}, where each point in the
Figure corresponds to an individual simulation (including noise) and
reconstruction. Further optimization in the threshold for measuring the radius 
should further improve the precision.

\begin{figure}[h]
  \begin{center}
    \rotatebox{-90}{\scalebox{0.5}{\includegraphics{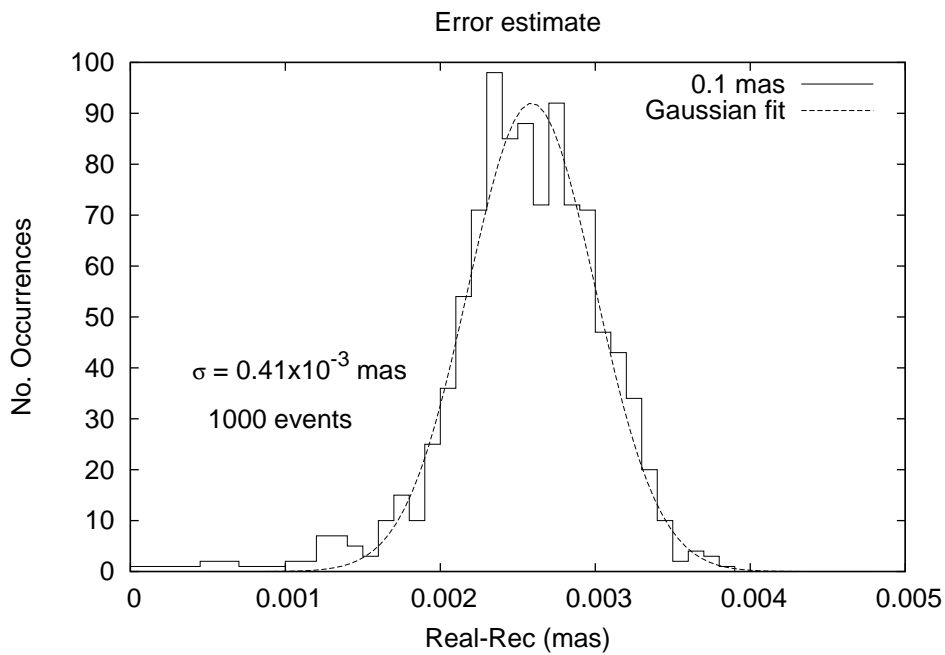}}}
  \end{center}
  \vspace{0.5cm}
  \caption{\label{uncertainty_a} Histogram of real radius minus reconstructed
  radius for 50 hours of exposure time on a $6^{th}$ magnitude star ($T=6000\,\mathrm{K}$) of $0.1\,\mathrm{mas}$ radius. }
\end{figure}

Figure \ref{radius_b} clearly shows that stellar radii ranging from $0.03\,\mathrm{mas}$
to $0.6\,\mathrm{mas}$ can be measured with uncertainties ranging between subpercent and 
a few percent (Figures \ref{uncertainty_a} and \ref{uncertainty_b}). It can 
be seen from Figure \ref{radius_b}, that the
uncertainty increases linearly as a function of the pristine (simulated) radius. This
is due to a decrease in the number of baselines that measure a high degree of correlation
 when the angular diameter increases. As the pristine radius $\theta$
decreases, the distance to the first zero in the correlation increases as
$\theta^{-1}$, so the number of telescopes contained within the airy
disk increases as $\theta^{-2}$. Consequently, decreasing the pristine radius is
equivalent to increasing the number of independent measurements by a factor of
$\theta^{-2}$. Since the uncertainty decreases as the square root of the number of independent
measurements, the error decreases linearly with the radius. For radii above
$0.6\,\mathrm{mas}$, there are simply not enough baselines to constrain the Fourier
plane information for image reconstruction. For radii greater than $0.6\,\mathrm{mas}$,
the distance to the first zero in the degree of correlation is of the order of $100\,\mathrm{m}$, 
but only baselines at $35\,\mathrm{m}$ and $50\,\mathrm{m}$ are 
capable of measuring the Fourier magnitude with more than 3 standard deviations (see eq. \ref{snr}). In 
Figure \ref{uncertainty_b} we show the relative percent error (RMS of a statistic) 
as a function of time for two radii, where it can be seen that a
 relative error of a few percent is achieved after only a few hours.

\subsection{Oblate stars\label{oblate_sec}}

\begin{figure}[h]
  \begin{center}
    \rotatebox{-90}{\includegraphics[scale=0.6]{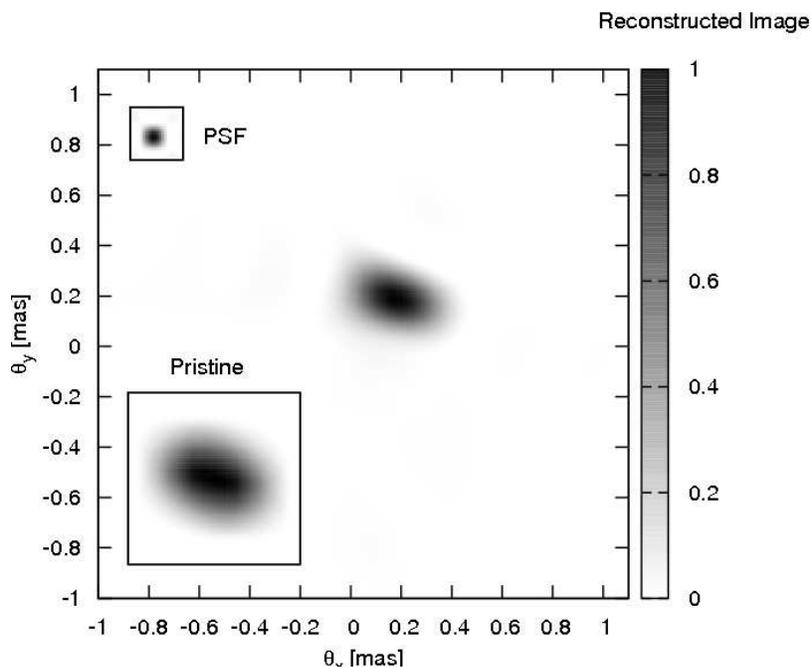}}
  \end{center}
  \caption{\label{oblate_example_a} Simulated and reconstructed oblate rotator of magnitude 6 and 50 hours
  of observation time.}
\end{figure}

For oblate stars we use the same magnitude and exposure
parameters that are used for disk-like stars. Uniform oblate stars can be
described by three parameters: the semimajor axis $a$, the
semiminor axis $b$, and the orientation angle $\theta$. Judging from the
limitations obtained from reconstructing disks, we produce 
pristine images whose values for $a$ and $b$ are random numbers less than $1\,\mathrm{mas}$, and choose $1\leq a/b \leq2$ . The value
of the orientation angle $\theta$ also varies randomly between $0^\circ$ and
$90^\circ$. A typical image reconstruction can be seen in Figure
\ref{oblate_example_a}.

After applying a threshold on pixel values as was done for disk shaped stars,
the reconstructed parameters are found by calculating the inertia
tensor of the reconstructed image. The eigenvalues and
eigenvectors of the inertia tensor provide information for the reconstructed values of $a$, $b$ and
$\theta$. To do this, we make use of the relation between the matrix eigenvalue and the semimajor/minor axes
$I_{xx}=\frac{1}{4}a^2M$, where $M$ is the integrated brightness. A similar relation for $I_{yy}$ holds
for the semiminor axis $b$.

\begin{figure}[h]
  \begin{center}
    \rotatebox{-90}{\scalebox{0.5}{\includegraphics{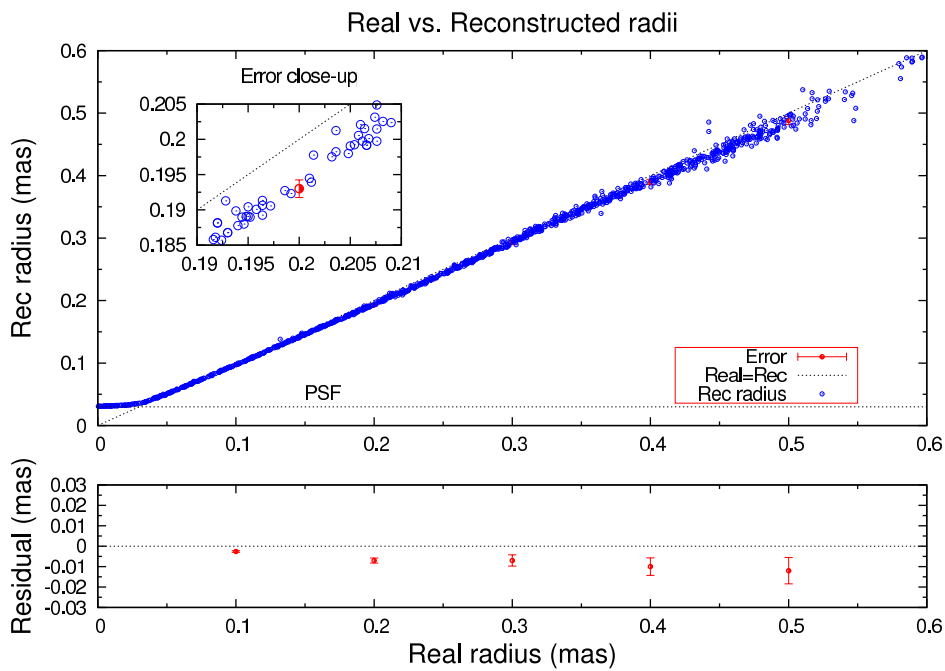}}}
  \end{center}
  \vspace{0.5cm}
  \caption[Simulated vs. Reconstructed radii for magnitude 6 stars 
with 50 hours of observation time.]{\label{radius_b}  Simulated vs. Reconstructed radii for magnitude 6 stars 
with 50 hours of observation time (see footnote \ref{long_exposures}). The 
top sub-figure shows the uncertainty for a $0.2\,\mathrm{mas}$ measurement. The 
bottom sub-figure shows the residual (Reconstructed-Real) along with the uncertainty in the radius.}
\end{figure}

\pagebreak

\newpage

\begin{figure}[h]
  \begin{center}
    \rotatebox{-90}{\includegraphics[scale=0.4]{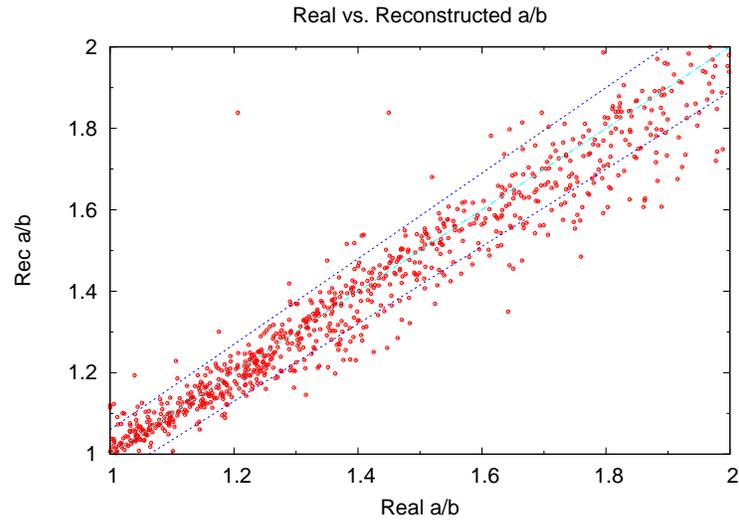}}
  \end{center}
  \vspace{0.5cm}
  \caption[Real vs. reconstructed $a/b$ for oblate
  stars.]{\label{oblate_example_b} Real vs. reconstructed $a/b$ for oblate
  stars. The distance between the two linear envelopes is 2
  standard deviations.  All pristine images
  that have either $a>0.5\,\mathrm{mas}$ or $b>0.4\,\mathrm{mas}$ are not included}
\end{figure}

The resulting reconstructed semimajor/minor axes as a function of their real values have a similar
 structure as the scatter plot for reconstructed radii shown in Figure
\ref{radius_b}, and are well reconstructed up to $0.5\,\mathrm{mas}$ within a few percent. 
In Figure \ref{oblate_example_b}, it can be seen how the uncertainty in the reconstructed oblateness $a/b$
increases with increasing oblateness. As with disk shaped stars (section \ref{disks}), 
the uncertainty in the reconstructed semimajor/minor axes decreases as the square root 
of the number of baselines measuring a high degree of correlation. Therefore, the 
uncertainty in the reconstructed semimajor/minor axes is proportional to $\sim \sqrt{ab}$, and the error in the reconstructed oblateness 
is proportional to $\sim \sqrt{a/b+a^3/b^3}$. 

\begin{figure}[h]
  \begin{center}
    \rotatebox{-90}{\includegraphics[scale=0.4]{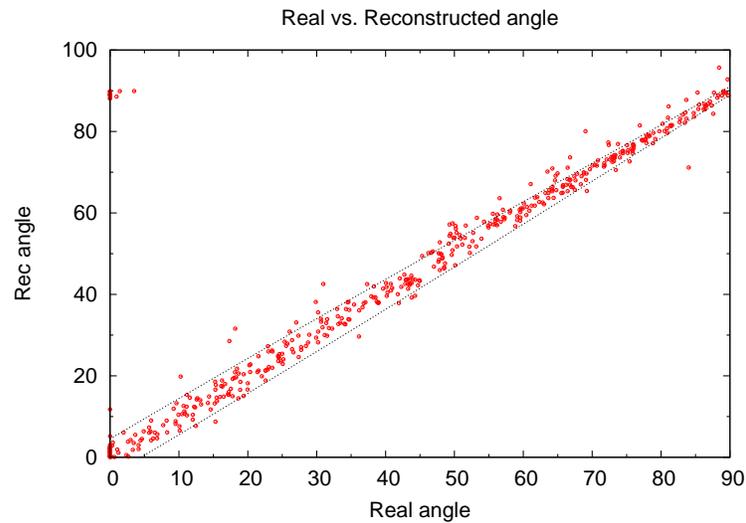}}
  \end{center}
  \vspace{0.5cm}
  \caption[Real vs. reconstructed orientation angle for oblate stars.]{\label{theta_oblate} Real vs. reconstructed orientation angle for oblate stars. All pristine images
    that have either $a>0.5\,\mathrm{mas}$, $b>0.4\,\mathrm{mas}$ or $a/b<1.1$ are not included.  
    We also note that reconstructed angles are always less than $45^\circ$
    due to the fact that the Fourier magnitude data is the same for pristine
    images flipped about the x, y or x and y axes. Therefore, for all 
    pristine angles larger than $45^\circ$, we replace the reconstructed angle by
    $\theta_{rec}'=90^\circ-\theta_{rec}$. }
\end{figure}

The reconstructed orientation angle as a function of the pristine angle is shown in Figure \ref{theta_oblate}, and
several factors play a role in the uncertainty of the reconstructed value. For a fixed value of $a$ and $b$, 
the orientation of the telescope array with respect to the main lobe of the Fourier magnitude 
determines the number of baselines that measure a high degree of correlation. The number of 
baselines that measure a high degree of correlation is greater when the main lobe of the Fourier 
magnitude is aligned with the $x$ or $y$ direction of the array (see Figure \ref{array}), and is smaller by a factor of $\sim \sqrt{2}$ (assuming a uniform grid of telescopes)
when its main axis is at $45^{\circ}$ with respect to the array. However, the uncertainty (proportional to spread of points) 
in Figure \ref{theta_oblate} does not appear to be symmetric at $0^{\circ}$ and $90^{\circ}$, and is smaller at $90^{\circ}$. This due to the
way the phase is reconstructed, i.e., due to the slicing of the Fourier plane 
along the $u$ or $v$ directions (see section \ref{2-dim}). In the case of Figure \ref{theta_oblate}, the $(u,v)$ plane is
sliced along the $u$ direction, with a single orthogonal reference slice along the $v$ direction. The 
main lobe of the Fourier magnitude  
of an oblate star has more slices passing through it when it is elongated along the $v$ 
direction (corresponding to an orientation angle of $90^{\circ}$ in image space), yielding a better reconstruction. This is in contrast to
the orthogonal case of $0^{\circ}$, where the main lobe of the Fourier magnitude has a smaller number of slices passing through it.

\subsection{Binary stars\label{binary_sec}}

\begin{figure}[h]
  \begin{center}
    \rotatebox{-90}{\includegraphics[scale=0.55]{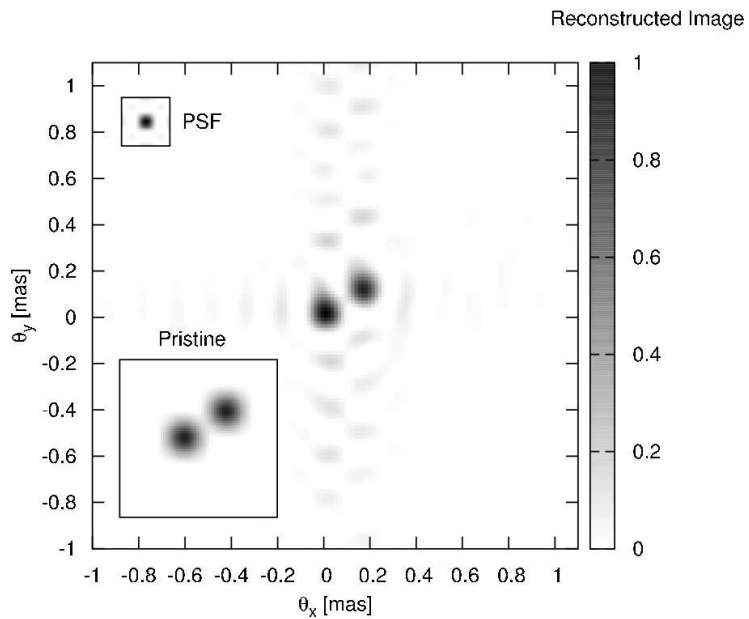}} 
  \end{center}
  \caption{\label{binary_example_a} Simulated and reconstructed binary of magnitude 6
    and 50 hours of observation time.} 
\end{figure}

Simulated data are generated for $5^{th}$ magnitude binary stars ($T=6000^{\circ}\,\mathrm{K}$), and an exposure of 
50 hours after noting that the uncertainty in the degree of correlation is of the order of a few percent (eq. \ref{snr}).
Binaries stars are parametrized by the radii $r_1$ and $r_2$ of each star, 
their separation $d$, position angle $\theta$, and relative brightness in arbitrary units between 0 and 1. 
We generate pristine images with random parameters within
the following ranges: radii are less than $0.3\,\mathrm{mas}$, angular separations are
less than $1.5\,\mathrm{mas}$, the relative brightness per unit area is less than or equal to 1, and
the orientation angle is less than $90^\circ$. A typical reconstruction can
be seen in Figure \ref{binary_example_a}.

To measure the reconstructed parameters we identify the two brightest spots (regions) whose \linebreak

\newpage
\noindent pixel values exceed 
a threshold of 0.2. We then find the radius for each bright spot and its centroid position.
 Identifying spots is a non trivial task
in noisy reconstructed images and our analysis sometimes fails to identify
the ``correct'' reconstructed spots. For example, a common issue is that
close reconstructed spots that are faintly connected by artifacts, are sometimes 
interpreted as a single spot. It should be again stressed that image reconstruction may not be
the best way to measure reconstructed parameters. For example, the data can just as well be
fit by the general form of the Fourier magnitude of a resolved binary system (containing only a few parameters). 

\begin{figure}[h]
  \begin{center}
    \rotatebox{-90}{\includegraphics[scale=0.4]{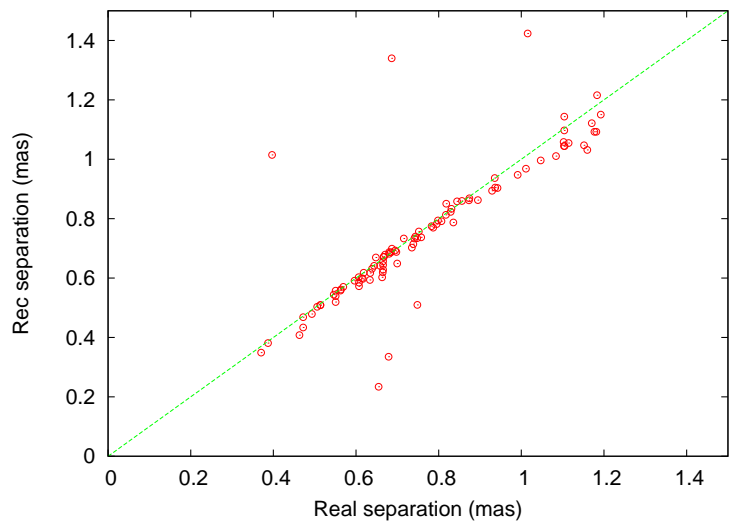}}
  \end{center}
  \vspace{0.5cm}
  \caption[Real vs. reconstructed angular separation in
    binary stars.]{\label{binary_example_b} Real vs. reconstructed angular separation in
    binary stars. Binary stars whose relative brightness is less than 0.3 are not included in this plot. }
\end{figure}

In Figure \ref{binary_example_b}, we show reconstructed angular separations
as a function of their real values. The reconstructed values of the angular separation are found
to within $\sim 5\%$ of their real values and  cannot be much 
larger than what is allowed by the smallest baselines. We find that stars separated 
by more than $d_{max}\approx 0.75\,\mathrm{mas}$ are not well reconstructed since the variations in the Fourier magnitude 
start to become comparable to the shortest baseline. 

\begin{figure}[h]
  \vspace{5cm}
  \begin{center}
    \rotatebox{-90}{\includegraphics[scale=0.4]{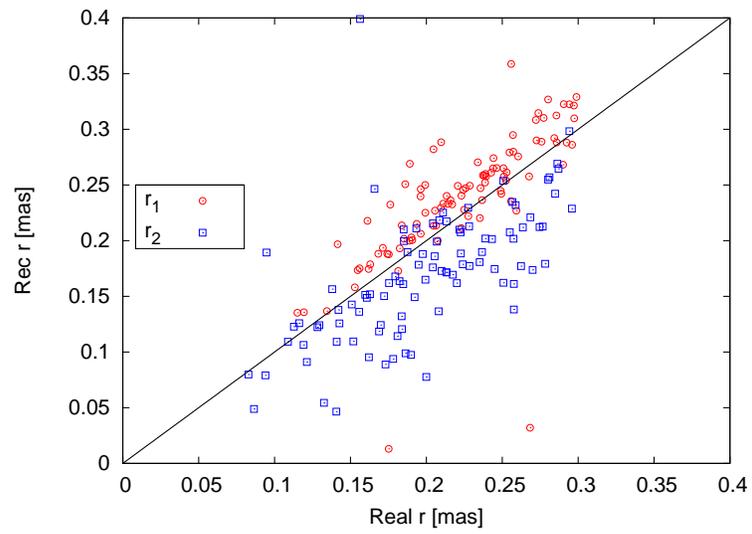}}
  \end{center}
  \vspace{0.5cm}
  \caption[ Real versus reconstructed
    radii for binary stars.]{\label{relative} Real versus reconstructed
    radii for binary stars. We only include cases where one of the members is less than 3 times brighter than the other.}
\end{figure}

In Figure \ref{relative} we show the reconstructed values of the radii as a function of their pristine 
values. We find $\sim 10\%$ uncertainties in each of the reconstructed radii. 
Aside from the angular separation, a variable that plays a role 
in successfully reconstructing pristine radii is the 
ratio of absolute brightness\footnote{Product 
of relative brightness per unit area (in arbitrary linear units between 0 and 1) and relative area of both stars} of both binary members. When 
one of the two members is more than $\sim 3$ times brighter than the other, the fainter star
is found to be smaller than its pristine value, and sometimes not found at all when 
one of the members is more than $\sim 10$ times brighter than the other. This is in part because the sinusoidal variations
in the Fourier magnitude start to become comparable to the uncertainty. For example: a binary star whose individual components cannot be resolved,
with one component 20 times brighter that the other, has relative variations of $\sim 10\%$. With all the redundant baselines,
 a few percent uncertainty in the measured degree of correlation is sufficient to accurately measure these variations. 
However, when the binary components 
are resolved, the relative variations decrease with increasing baseline and baseline redundancy is not sufficient to 
reduce the uncertainty in the measurement of the Fourier magnitude. This signal to noise issue can of course be
improved by increasing exposure time. 

There are also issues related to algorithm performance. One such problem has to do with the fit of the
data to an analytic function (see section \ref{fit}). When the scale of the fit (found by an initial Gauss fit)
is found to be too small, too many basis elements are
used to reconstruct the data, and high frequency artifacts appear in reconstructions. Small initial scales are
typically related to the binary separation as opposed to the size of individual components, and it is the latter which 
correctly sets the scale of the fit. Artifacts may be then mistaken for binary \linebreak

\pagebreak

\pagebreak

\noindent components, and incorrect reconstructed parameters may be found. Results improve significantly when either
the correct scale is set or (model-independent) image post-processing is performed (see section \ref{improved_analysis}).

\chapter{Imaging stellar features and nonuniform radiance distributions}\label{results}
%\section{Imaging stellar features and non-uniform radiance distributions}\label{results}

In this chapter, pristine images are generated with varying complexity. In section \ref{complex}, we provide 
two example reconstructions using the Cauchy-Riemann algorithm, and 
start to quantify the reconstruction capabilities. Then we introduce some postprocessing techniques and quantify the improvements in a 
more systematic way: We investigate the reconstruction capabilities for images with increasing complexity by first generating pristine images of stars 
with limb-darkened atmospheres, then we 
introduce a localized bright or dark feature, and finally increase the number of features and explore some of the parameter space, 
i.e., spot size, location, etc. 

\section{Featured images with Cauchy-Riemann\\ reconstructed images} \label{complex}

\begin{figure}
  \begin{center}
    \rotatebox{-90}{\includegraphics[scale=0.55]{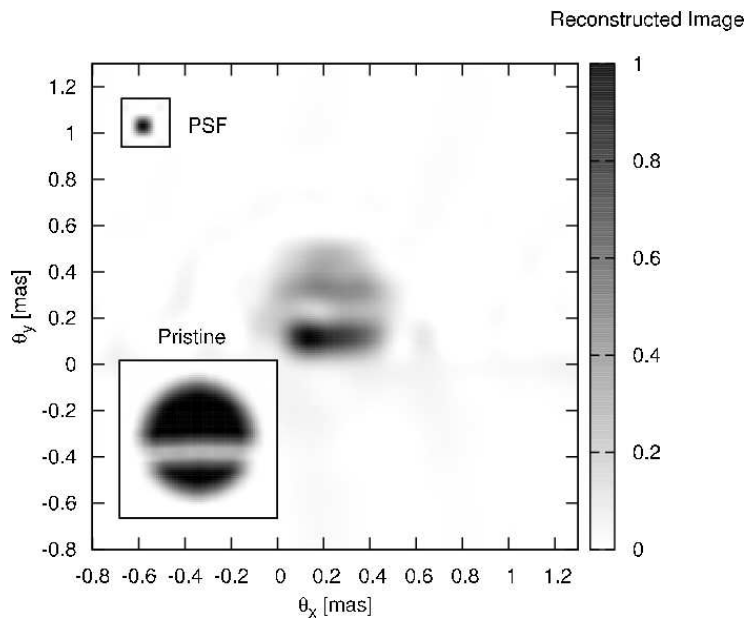}} 
  \end{center}
  \caption[Star with obscuring disk (raw reconstruction).]{\label{horizontal_disk} Star with obscuring disk (raw reconstruction). This corresponds to $4^{th}$ magnitude and 10 hrs of observation time. The correlation between the real and reconstructed image is $c=0.947\pm 0.001$. Note that an inverted gray scale is used.
}
\end{figure}

We now present two examples of more complex images, and show that the capabilities can, to a large extent, be understood from results of 
less complex images, such as uniform disks and binaries. In Figure \ref{horizontal_disk} we show the reconstruction of the image of a star with a dark 
band (an obscuring disk for example), corresponding to a $4^{th}$ magnitude star and 10 hrs of observation time. The metric used to quantify the 
agreement with the pristine image (bottom left corner of Figure \ref{horizontal_disk}) is a normalized correlation\footnote{\label{footnote_c}For two images $A_{i,j}$ and $B_{i,j}$, the normalized correlation $C_{i,j}$ is 
$C_{i,j}=Max_{k,l}\left\{N^{-2}(\sigma_A\sigma_B)^{-1} \sum_{i,j}^{N, N}(A_{i,j}-\bar{A})(B_{i+k, j+l}-\bar{B})\right\}$, where $\sigma_A$ and $\sigma_B$ 
are the standard deviations of images $A$ and $B$, and $\bar{A}$, $\bar{B}$ are the image averages.}
whose absolute value ranges between 0 (no correlation) and 1 (perfect correlation/anti-correlation). To quantify the uncertainty in the correlation, we perform the noisy simulation and reconstruction several times, and find the standard deviation of the degree of correlation. In the case of Figure \ref{horizontal_disk}, the correlation $c$ is $c=0.947\pm 0.001$. Note that the uncertainty in the correlation is small, and the image reconstruction is not perfect, which implies that the reconstruction is not only affected by the SNR level, but also by the reconstruction algorithm performance limitations. 

 In order to determine the confidence with which we can detect the feature (dark region within the disk), we quantify the difference between the reconstruction and a featureless image, i.e., a uniform disk whose radius matches the radius of the pristine image. By finding the degree of correlation of the reconstruction with a the pristine image (convolved with the PSF), and comparing it to the correlation of the reconstruction with a uniform disk, we can quantify the confidence level to which our reconstruction is not a uniform disk. We find that the correlation of the reconstructed image with a uniform disk  is $c=0.880\pm 0.001$.  This is lower than $c=0.947\pm 0.001$ by 61 standard deviations, and this difference is a measure of the confidence with which we can establish that the reconstruction does not correspond to that of a featureless star. 

\begin{figure}
  \begin{center}
    \rotatebox{-90}{\includegraphics[scale=0.55]{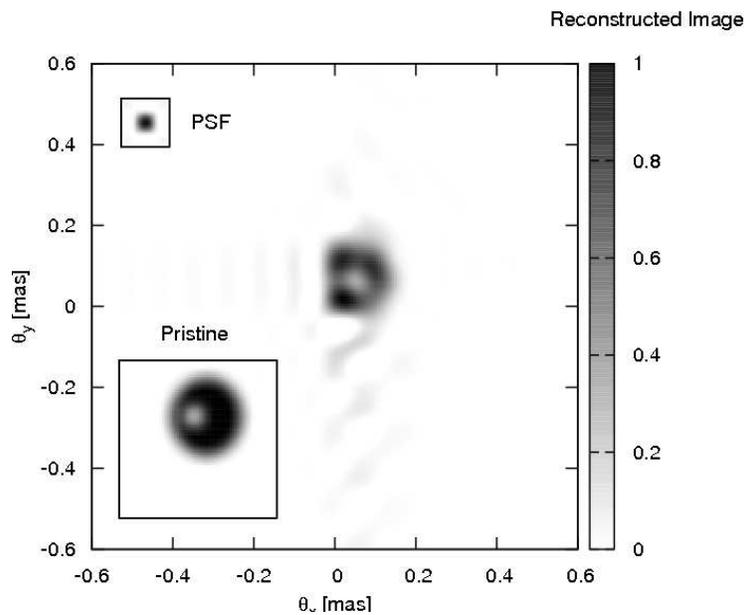}}
  \end{center}
  \caption[Star with dark spot (raw reconstruction).]{\label{dark_spot_raw} Star with dark spot (raw reconstruction). This corresponds to 
    a $4^{th}$ magnitude star, 10 hrs of observation time and a degree of correlation of $0.940\pm0.001$.}  
\end{figure}

Another example of a complex image reconstruction can be seen in Figure \ref{dark_spot_raw}. Here we show the reconstruction of a star with a dark spot, whose correlation with the pristine image is $c=0.940\pm0.001$. We can compare this correlation with the correlation of the reconstructed image and a uniform disk, which is $c=0.904\pm0.001$ (lower by 30 standard deviations). 

For both examples, we also calculate the correlation with the pristine image as a 
function of the angular scale (in $\mathrm{mas}$) of the pristine image. We then find the
degree of correlation of each reconstruction and its pristine image, and also the
degree of correlation of the reconstruction with a uniform disk.  By comparing 
these two correlation values, we can estimate the 
smallest feature (spot) that can be reconstructed\footnote{A more careful analysis would require only changing the spot size 
as opposed to scaling the whole pristine image.}. Below some point we
no longer distinguish between the reconstruction of the featured image and that 
of a uniform disk. We find that the smallest feature that 
can be reconstructed is close to $0.05\,\mathrm{mas}$. This can already be seen 
from the order of magnitude estimate made in section \ref{sensitivity} and a comparable 
value of $0.03\,\mathrm{mas}$ is found in section \ref{disks}. When pristine images 
have angular sizes greater than $\sim 0.8\, \mathrm{mas}$, the degree of 
correlation drops significantly due to lack of short baselines.  %Note that this value is higher than what is found in section \ref{disks}. This is due to higher signal to noise ratio in the simulated data corresponding to Figures \ref{horizontal_disk} and \ref{dark_spot_raw}. The resolution limits discussed in turn correspond to $\sim 16\times 16$ effective resolution elements (pixels).  

%\section{Towards constraining astrophysical models}

\section{Improved analysis and postprocessing routines}\label{improved_analysis}

The resulting reconstructed image with this estimated phase obtained from the Cauchy-Riemann algorithm is sometimes not ideal, and so is taken as a first guess
for iterative algorithms. We have explored the use of the Gerchberg-Saxton (error-reduction) algorithm described in section \ref{gerchberg-saxton}. Recall that constraints must be applied in both the Fourier domain and the image domain for this algorithm to converge. The Fourier constraint consists in replacing the Fourier magnitude of the image by that given by the data. The constraints in image space can be very general. The image constraint that we impose comprises applying a 
\emph{mask}, so that only pixels within a certain region are allowed to have positive nonzero values. For the images presented below, the mask is a circular region whose radius is typically found by measuring the radius of the first guess obtained from the Cauchy-Riemann approach. In all reconstructions where the Gerchberg-Saxton is used, we perform 50 iterations, and found that more iterations typically do not produce significant changes in the reconstruction \citep{mnras2}. 

Another postprocessing application that has been utilized is MiRA (Multi-aperture image reconstruction algorithm) \citep{Thiebaut}. MiRA
has become a standard tool for image reconstruction in amplitude (Michelson) interferometry. MiRA is an iterative procedure which slightly modifies image
pixel values so as to maximize the agreement with the data. In the image reconstruction process, additional constraints such as smoothness or compactness can be applied simultaneously, but this is something that we have not yet experimented with, i.e., the regularization parameter is set to zero for all reconstructions presented here. In the results presented below, the number of iterations is set by the default stopping criterion of the optimizer. The MiRA software only uses existing data in the $(u,v)$ plane, as opposed to using the fit of an analytic function to the data as is done in the Cauchy-Riemann and Gerchberg-Saxton routines. This results in removing artifacts in the reconstruction that can be caused by the fit of an analytic function to the data.  

%An example of image postprocessing with MiRA only, is performed on the oblate star of Figure \ref{oblate_example_c} is shown in Figure \ref{improved_oblate}. We have also performed MiRA postprocessing  on the images in Figures \ref{horizontal_disk} and \ref{dark_spot}, and we obtain the ones shown in Figures \ref{mira_image_a} and \ref{mira_image_b}. These preliminary results show the overall reduction in noise and improvement in the sharpness of the reconstructed images. 

\begin{figure}[h]
  \begin{center}
    \scalebox{1.5}{
    \begin{tikzpicture}
      %\tikzstyle{every node}=[circle, minimum size=1.5cm];
      \node(title_pristine){\textcolor{blue}{Pristine}};
      \node(pristine)[below of= title_pristine, node distance=1cm]{\includegraphics[scale=0.35]{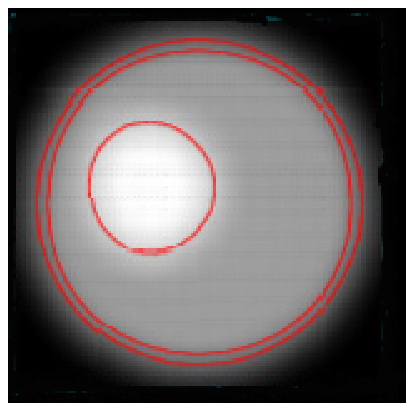}};
      \node(simul_box)[right of=pristine, node distance=3cm]{\includegraphics[scale=0.1, angle=-90]{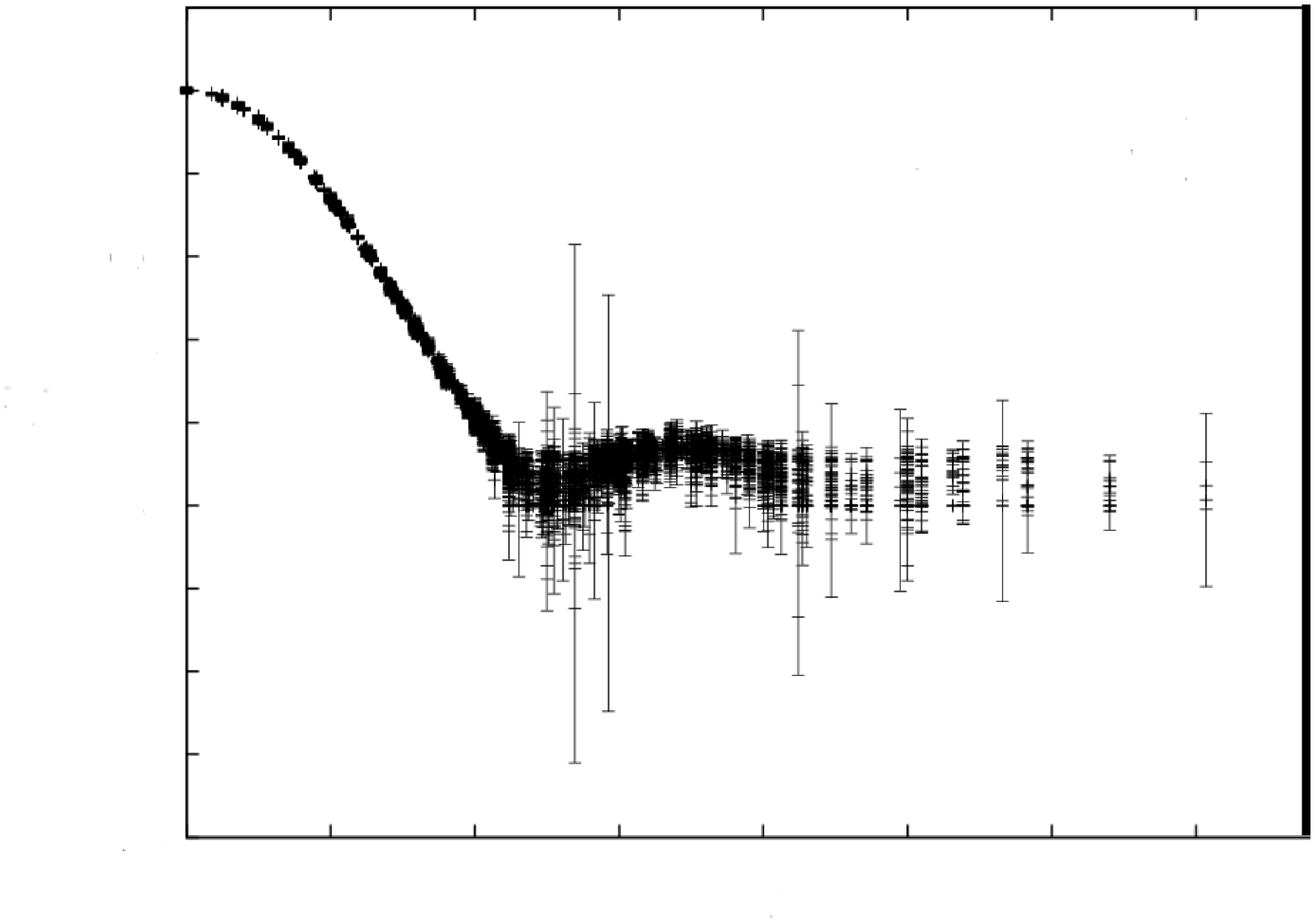}};
      %\node(gs_diagram)[below right of=pristine, node distance=5cm]{\includegraphics[scale=0.25]{/home/paul/documents/public_outreach/talk_figures/gs.pdf}};
      \node at (3.3,-6.2){\includegraphics[scale=0.25]{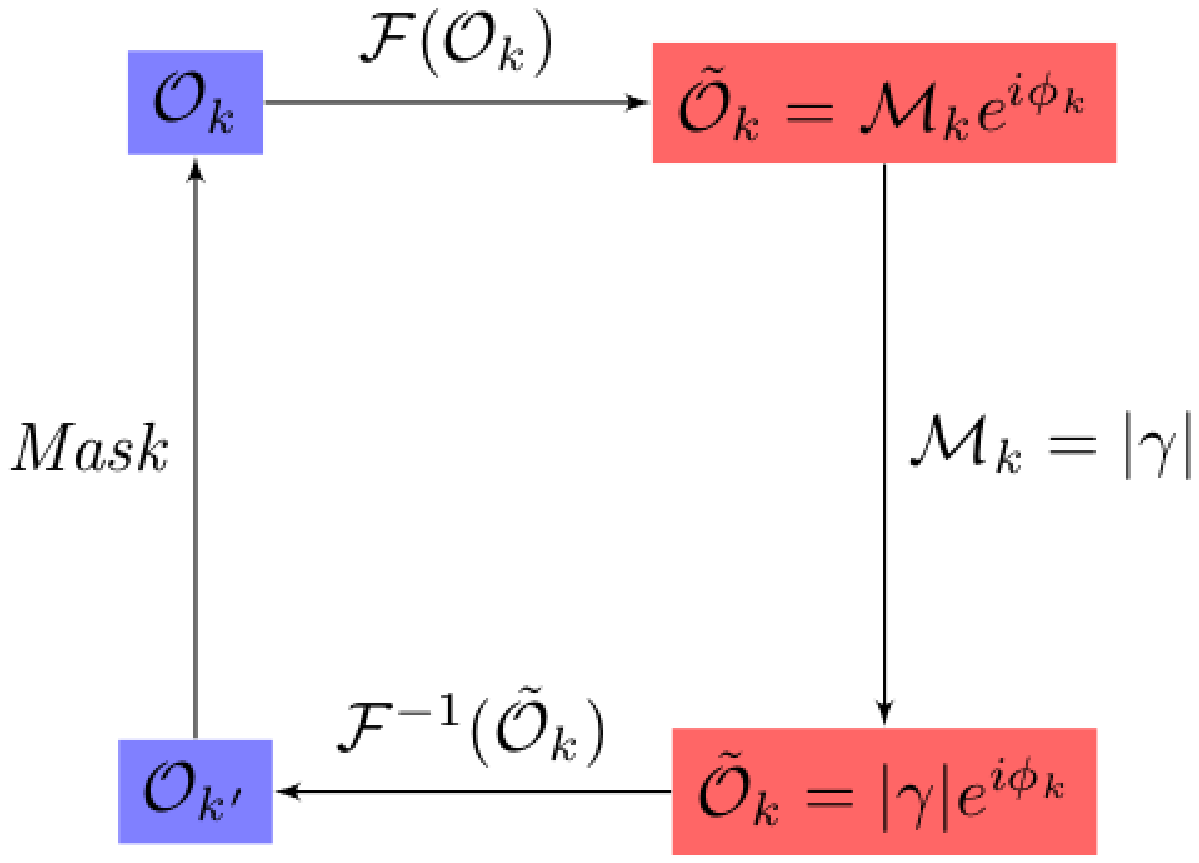}};
      \node(title_data)[above of=simul_box, node distance=1cm]{\textcolor{blue}{Data}};
      \node(Hermite_fit)[right of=simul_box, node distance=3cm]{\includegraphics[scale=0.55]{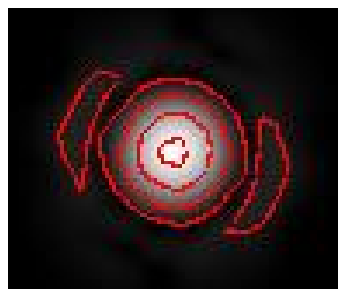}};
      \node(title_fit)[above of=Hermite_fit, node distance=1cm]{\textcolor{blue}{Hermite fit}};
      \node(cr)[below of=Hermite_fit, node distance=2.5cm]{\includegraphics[scale=0.40]{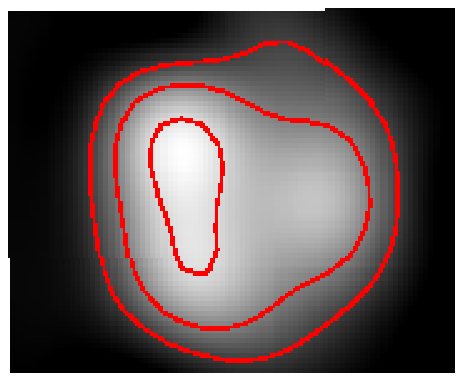}};
      \node(title_cr)[below of=cr, node distance=1cm]{\textcolor{blue}{Cauchy-Riemann}};
      \node(gs)[left of=cr, node distance=3cm]{\includegraphics[scale=0.40]{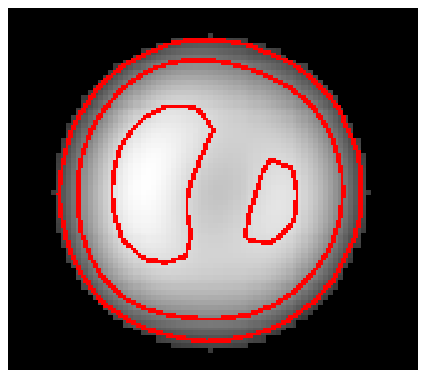}};
      \node(title_gs)[below of=gs, node distance=1cm]{\textcolor{blue}{Gerchberg-Saxton}};
      \node(mira)[left of=gs, node distance=3cm]{\includegraphics[scale=0.40]{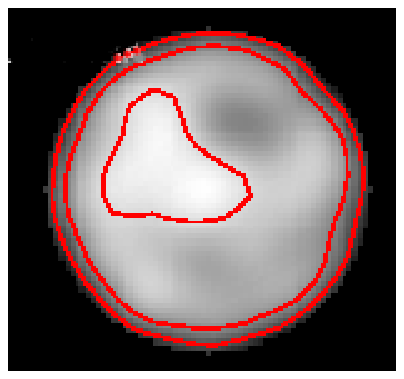}};    
      \node(title_mira)[below of=mira, node distance=1cm]{\textcolor{blue}{MiRA}};
      \node(cr_eq1)[below of=cr, node distance=2.3cm]{ $\frac{\partial\Phi}{\partial\psi}=\frac{\partial\, lnR}{\partial\xi}$ };
      \node(cr_eq2)[below of=cr_eq1, node distance=0.6cm]{ $\,\,\,\,\,\frac{\partial\Phi}{\partial\xi}=-\frac{\partial\, lnR}{\partial\psi} $ };
      \node(mira_eq)[below of=mira, node distance=2.3cm]{$arg\,min\left\{\chi^2\right\}$};
      %\node(rec)[left of=mira, node distance=2cm, fill=blue!40]{Rec. Image};
      \draw[->, red](pristine)--(simul_box);
      \draw[->, red](simul_box)--(Hermite_fit);
      \draw[->, red](Hermite_fit)--(cr);
      \draw[->, red](cr)--(gs);
      \draw[->, red](gs)--(mira);
      %\draw[->](mira)--(rec);
    \end{tikzpicture}
    }
  \caption{\label{analysis_overview}Analysis overview for data simulation and image reconstruction.}
  \end{center}
\end{figure}

A systematic study of the improvements with image postprocessing is presented in  sections \ref{limb_darkening}, \ref{single_spots} and \ref{multiple_spots}. We investigate the performance of each algorithm individually as well as the performance of linking algorithms together, particularly the Cauchy-Riemann algorithm, followed by the Gerchberg-Saxton and MiRA (Figure \ref{analysis_overview}). The order of postprocessing routines is also investigated and results are presented in section \ref{single_spots}.

\subsection{Limb-darkening}\label{limb_darkening}

\begin{figure}[h]
  \begin{center}
  \includegraphics[scale=0.55, angle=-90]{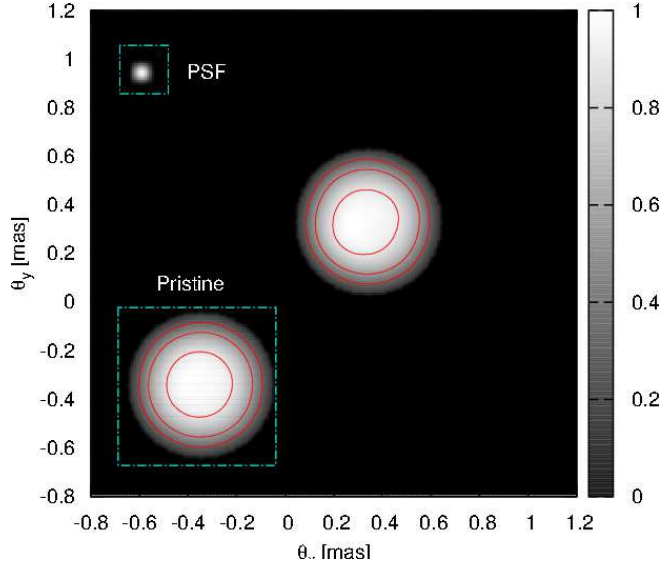}
  \end{center}
  \caption[Image reconstruction of a star with limb darkening parameter $\alpha=5$, apparent visual magnitude $m_v=3$ 
and $10\,\mathrm{hrs}$ of observation time.]{\label{limb_example} Image reconstruction of a star with limb darkening parameter $\alpha=5$, apparent visual magnitude $m_v=3$ 
and $10\,\mathrm{hrs}$ of observation time. The pristine starting image from which intensity interferometric data were simulated 
is shown in the bottom left corner with the same contour lines. The Cauchy-Riemann phase reconstruction was performed to produce a raw image, and then the Gerchberg-Saxton
routine was implemented to produce the postprocessed image shown.}
\end{figure}

\begin{figure}
  \begin{center}
  \includegraphics[scale=0.4, angle=-90]{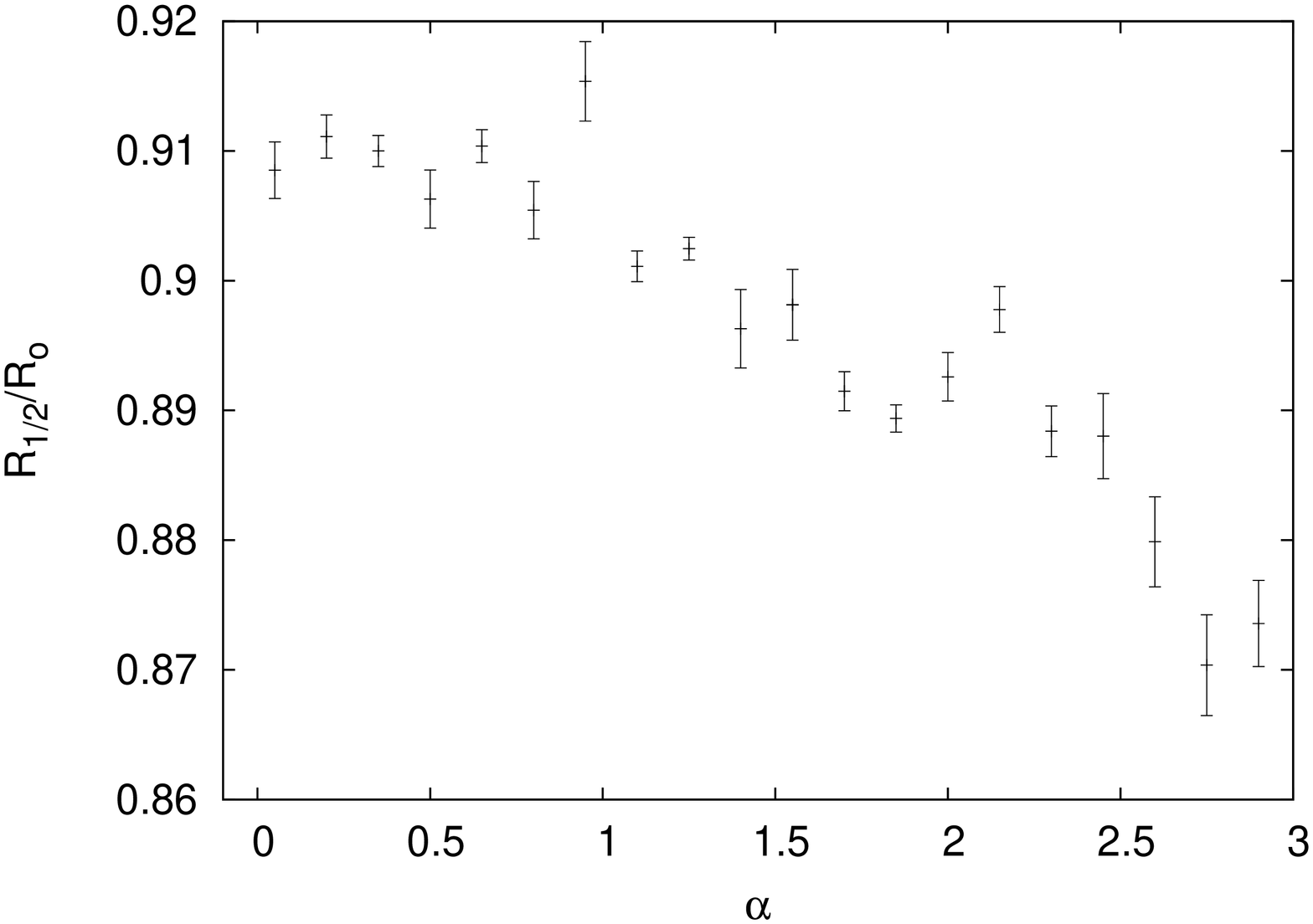}
  \end{center}
  \vspace{0.5cm}
  \caption[For each value of $\alpha$, SII data were 
simulated corresponding to stars with apparent visual magnitude $m_v=3$, $10\,\mathrm{hrs}$ of observation time and $\lambda=400\,\mathrm{nm}$.]{\label{r_vs_alpha} For each value of $\alpha$, SII data were 
simulated corresponding to stars with apparent visual magnitude $m_v=3$, $10\,\mathrm{hrs}$ of observation time and $\lambda=400\,\mathrm{nm}$. For 
each image reconstruction, a 1-dimensional profile is found by calculating the average intensity at each radial position. The 
 half height $R_{1/2}$ (angular radius at 
half maximum) is found. For each value of $\alpha$, the process was repeated several (10) times, and error bars are the standard deviation of the distribution 
of $R_{1/2}(\alpha)$. The ratio $R_{1/2}/R_o<1$ since the image reconstruction is at best a convolution with the PSF of the array. }
\end{figure}

Image reconstruction is actually not necessary for the study of limb-darkening, which can be studied directly from the knowledge of the squared degree of coherence. A direct analysis of the data, with no image reconstruction, is likely to yield better results than the ones presented below. However, it is instructive to first see this effect in reconstructed images before adding stellar features to the simulated pristine images. Limb darkening can be approximately modeled with a single parameter $\alpha$ as $I(\phi)/I_0=(\cos{\phi})^\alpha$ \citep{limb}, where $\phi$ is the angle between the line of sight and the perpendicular to the stellar surface isopotential. The values of $\alpha$ depend on the wavelength and  can be found by assuming hydrostatic equilibrium. At a wavelength of $400\,\mathrm{nm}$, $\alpha\approx 0.7$ \citep{limb_wavelength} for sun-type stars, and deviations from this value may be indicative of stellar mass loss. An example of the reconstruction of a limb darkened star with $\alpha=5$ is shown in Figure \ref{limb_example}; such a large value is chosen so that the effect is clearly visible in a two dimensional image with linear scale. More realistic values are considered below. To obtain Figure \ref{limb_example}, a first estimate of the phase was obtained from the Cauchy-Riemann algorithm and used to generate a raw image. Then the Gerchberg-Saxton postprocessing loop (Figure \ref{gs_routine}) was performed several (50) times.  In  Figure \ref{r_vs_alpha} the ratio of the average radius at half maximum $R_{1/2}$ and the nominal radius of $R_o$, is shown as a function of the limb darkening parameter $\alpha$. Here, data were simulated corresponding to stars with apparent visual magnitude $m_v=3$, temperature $T=6000^\circ\,\mathrm{K}$, radii of $R_o=0.3\,\mathrm{mas}$, $10\,\mathrm{hrs}$ of observation time and $\lambda=400\,\mathrm{nm}$. The ratio $R_{1/2}/R_o$ is less than 1, even in the absense of noise since the reconstruction is at best a convolution\footnote{A ``perfect'' reconstruction gives $R_{1/2}/R_o=0.96$ for $\alpha=0$. The extra discrepancy is due to a small hot-spot in the reconstruction. $R_{1/2}<R_o$ since the reconstruction is normalized to the highest pixel value. } with the array point-spread-function (PSF).  

From Figure \ref{r_vs_alpha} we can see  that we have some sensitivity  to changes in the limb-darkening parameter $\alpha$. Stars experiencing high mass loss rates are likely to have high values of $\alpha$, and  if we fit a uniform disk function to a limb-darkened reconstruction, the fit yields a smaller radius. For example, in the case of $\alpha=2.0$, a uniform disk fit yields an angular radius that is smaller by 7\% (still larger than the subpercent uncertainties found in radius measurements \citep{mnras}). A real example is the case of the star \emph{Deneb}, where the difference between the extracted uniform disk diameter ($\theta_{UD}=2.40\pm0.06\,\mathrm{mas}$) and the limb-darkened diameter is $0.1\,\mathrm{mas}$ \citep{deneb}, and its measured mass loss rate is $10^{-7}M_{\odot}\mathrm{yr}^{-1}$.

\subsection{Stars with single features }\label{single_spots}

\begin{figure}[!]
  \begin{center}
  \includegraphics[scale=0.6, angle=-90]{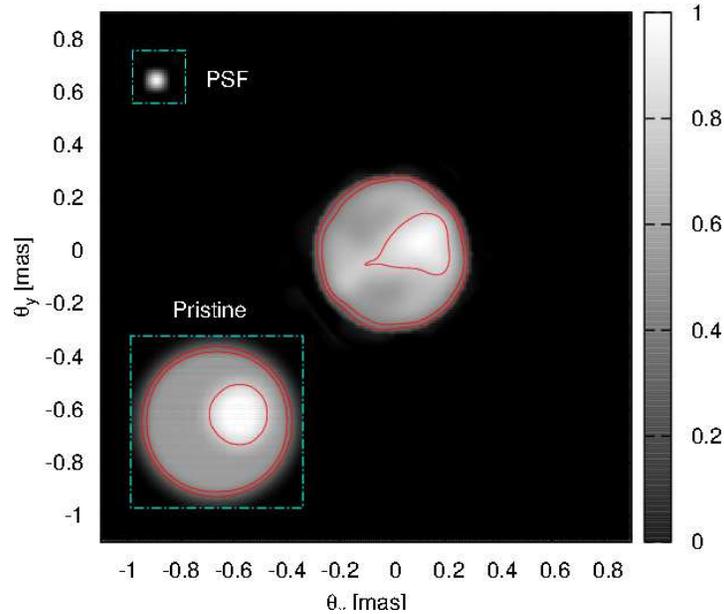}
  \end{center}
  \caption[Reconstructed bright spot.]{\label{bright_spot}Reconstructed bright spot. This simulated reconstruction corresponds to a star of $m_v=3$, $10\,\mathrm{hrs}$ 
of observation time, $T=6000\,\mathrm{K}$, and spot temperature of $T_{spot}=6500^\circ\,\mathrm{K}$.}
\end{figure}

\begin{figure}[h]
  \begin{center}
  \includegraphics[scale=0.6, angle=-90]{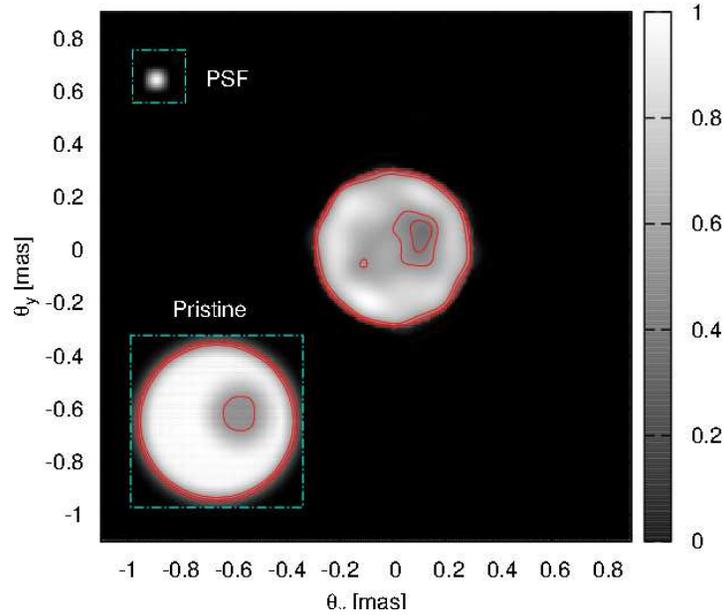}
  \end{center}
  \caption[ Reconstructed dark spot.]{\label{dark_spot} Reconstructed dark spot. This simulated reconstruction corresponds to a star of $m_v=3$, $10\,\mathrm{hrs}$ 
of observation time, $T=6000\,\mathrm{K}$, and spot temperature of $T_{spot}=5500^\circ\,\mathrm{K}$.}
\end{figure}

Stars were simulated as black bodies with a localized feature of 
a higher or lower temperature as described in section \ref{simulation}. In the simulated images, the effect 
of limb darkening is included as described in the previous section. Here the full  reconstruction analysis was used, which consists 
in first recovering a raw image from the Cauchy-Riemann algorithm. The raw image is 
then used as a starting point
for several iterations of the Gerchberg-Saxton loop (see Figure \ref{gs_routine}), and finally the output of the Gerchberg-Saxton algorithm 
is the starting image for the MiRA algorithm. Examples can be seen in 
Figures \ref{bright_spot} and \ref{dark_spot}, corresponding to the postprocessed reconstructions of bright stars
of $m_v=3$, $10\,\mathrm{hrs}$ of observation time and a temperature $T=6000^\circ\,\mathrm{K}$. In Figure \ref{bright_spot} the 
temperature of the spot is $T_{spot}=6500^{\circ}\,\mathrm{K}$, and in Figure \ref{dark_spot} the temperature of the spot is 
$T_{spot}=5500^{\circ}\,\mathrm{K}$.  % Since the last step (MiRA) only utilizes the data (and not a fitted function) for optimizing the 
%reconstruction, it has the effect of removing most artifacts that might originate from the fit of an analytic function to the data. 

\begin{figure}[h]
  \begin{center}
  \includegraphics[scale=0.45, angle=-90]{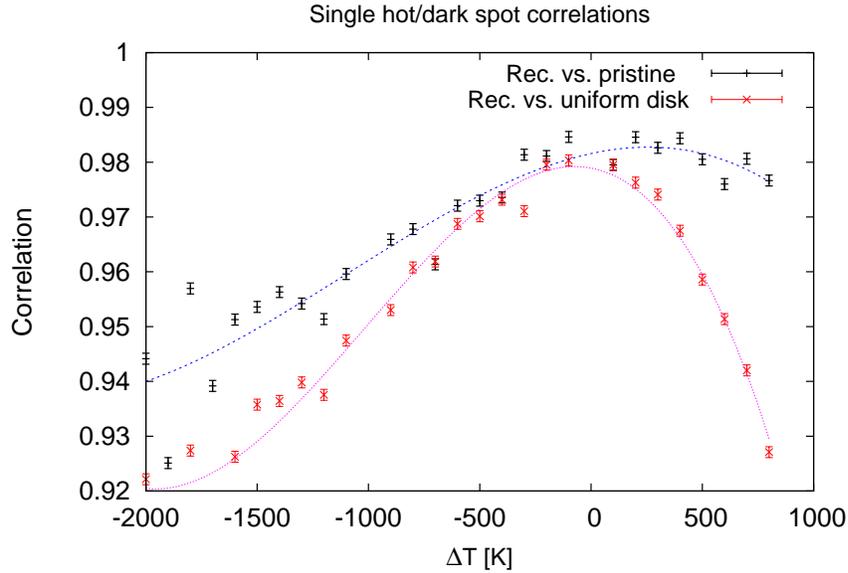}
  \end{center}
  \vspace{0.5cm}
  \caption[Quantifying reconstruction of hot/cool localized regions.]{\label{spot_correlations} The top curve and data points correspond to the correlation between reconstructed images and their corresponding
   pristine images containing spots of temperature $T+\Delta T$, where $T=6000^{\circ}\,\mathrm{K}$ ($m_v=3$, $10\,\mathrm{hrs}$ of observation time). The 
   lower curve and data points 
   are the correlation between the reconstructed image and a uniform disk of the same size as the pristine image. To estimate the uncertainties, we 
  performed several reconstructions and found the statistical standard deviation of the correlation.}
\end{figure}

%It is interesting to find what is the hottest and coolest spots that can be imaged on a simulated reconstruction such as the one shown in 
%Figure \ref{bright_spot}. The absolute number depends on several variables such as the size, location, and shape of the spot. However, we 
%can estimate it by taking pristine images like the one used for the reconstruction in Figures \ref{bright_spot} and \ref{dark_spot} and vary 
%the temperature. 
We can estimate the smallest temperature contrast that can be detected by varying the parameters in the model producing the pristine image. The performance 
in terms of temperature contrast obviously depends on several variables such as the size, location and shape of the spot. To quantify 
 the smallest detectable spot temperature contrast, we calculate the 
normalized correlation (see footnote \ref{footnote_c}) between the reconstructed image and the pristine image convolved with the array PSF. This
correlation is difficult to interpret by itself, so that it is compared  with the correlation between the reconstruction and a simulated star with 
no spots. By comparing these
two values we can have an idea of the confidence level for reconstructing spots with different temperatures. This comparison is shown in 
Figure \ref{spot_correlations}, where the top curve corresponds to the correlation as a function of spot temperature $6000^\circ\,\mathrm{K}+\Delta T$  
between the reconstructed images and the pristine images, and
the lower curve is the correlation between the reconstructed images and a spotless disk of the same size as the pristine image. A total of 26 stars 
were simulated, and the uncertainty in the correlation was estimated by performing several (10) 
reconstructions for one particular case ($\Delta T=500^\circ\,\mathrm{K}$). From the figure
it can be seen that spots are accurately imaged  when $\Delta T<- 700^\circ \, \mathrm{K}$ or 
$\Delta T> 200^\circ \, \mathrm{K}$ approximately.  For a black body of spectral density $B(T)$, a temperature difference  
$\Delta T<- 700^\circ \, \mathrm{K}$ corresponds to a flux ratio
$B(T+\Delta T)/B(T)<0.45$, and a temperature difference  $\Delta T> 200^\circ \, \mathrm{K}$ corresponds to flux ratios 
$B(T+\Delta T)/B(T)>1.2$. This asymmetry can be partly understood in terms of the brightness ratio between black bodies 
$B(T+\Delta T)/B(T)$, whose rate of change is higher when $\Delta T>0$ than when $\Delta T<0$. This however does not fully account for the asymmetry
between cool and hot spots. Most of the \linebreak

\noindent asymmetry is due to the fact that all the simulated stars in Figure \ref{spot_correlations} have the same
integrated brightness, and the radiance per solid angle is larger for a star containing a bright spot than for an annular region in a star containing
a dark spot. The same analysis can be performed by simulating stars with different integrated brightness, but the estimate becomes unnecessarily 
cumbersome and implies knowledge that we would not have access to prior to performing an image reconstruction.\footnote{For example, we would need to have information
on  the radiance per solid angle in an annular region in a star containing a dark spot} We should also not forget that this is an estimate, and in a more precise calculation we would need to consider additional variables such as spot size, position, etc.  \parskip 0pt

\begin{figure}[h]
  \begin{center}
  %\vspace{3.5cm}
  \includegraphics[scale=0.55]{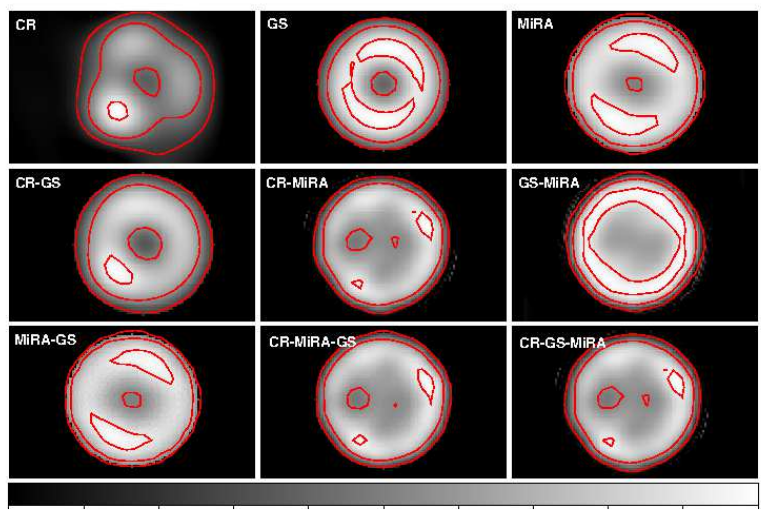}
  \end{center}
  \caption{\label{sup1}Image reconstructions for different algorithm combinations. The pristine corresponds to that of Figure \ref{bright_spot}. }
\end{figure}

\begin{figure}[h]
  \begin{center}
  \vspace{3.2cm}
    \includegraphics[scale=0.55]{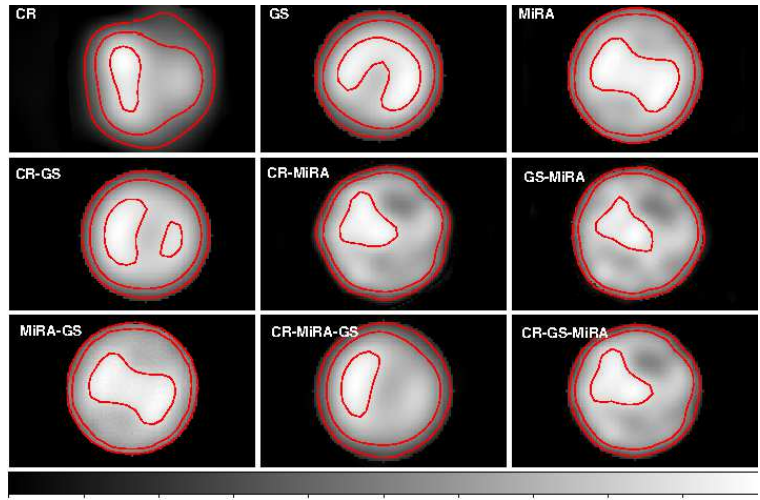}
    \end{center}
    \caption{\label{sup2}Image reconstructions for different algorithm combinations. The pristine corresponds to that of Figure \ref{dark_spot}. }
\end{figure}

To test whether the full chain of algorithms is needed to produce Figures \ref{bright_spot} and \ref{dark_spot},  and to investigate algorithm performance, we reconstruct Figures \ref{bright_spot} and \ref{dark_spot} with different algorithms and combinations of algorithms. Then we calculate the correlation of the reconstructions with the pristine image (convolved with the PSF). The results are shown in Table \ref{table}, and the reconstruction for each algorithm combination is shown in Figures \ref{sup1} and \ref{sup2}. The correlation\linebreak

\begin{table}[h]
\caption[Table comparing correlations of different algorithms.]{\label{table} Table comparing correlations of different algorithms. The algorithms are CR (Cauchy-Riemann), GS (Gerchberg-Saxton) and MiRA. The first two colums of numbers correspond to correlations of  Figure \ref{bright_spot}, and the second pair of columns corresponds to correlations of Figure \ref{dark_spot}. The bold columns correspond to correlations between the reconstructed image and the pristine image. The columns not in bold correspond to the correlation with a uniform disk from these it can be seen that bright spots are more easily detected.  The uncertainty in the correlation is $\Delta c=0.001$.}
\vspace{0.5cm}
\begin{center}
\begin{tabular}{l |l l|l l} \hline\hline
Algorithm & C (F \ref{bright_spot})& & C (F \ref{dark_spot})\\
          & Pristine & UD & Pristine & UD \\\hline
CR & \bf 0.954  &0.928 & \bf 0.942  &0.944 \\
GS & \bf 0.955  &0.928 & \bf 0.943  &0.944  \\
MiRA & \bf 0.978  &0.972 & \bf 0.974  &0.980\\
CR $\rightarrow$ GS & \bf 0.973  &0.954 &\bf 0.966  &0.968\\
CR $\rightarrow$ MiRA & \bf 0.979  &0.963 & \bf 0.973 & 0.971\\
GS $\rightarrow$ MiRA & \bf 0.976  &0.965 & \bf 0.973 & 0.978\\
MiRA $\rightarrow$ GS & \bf 0.970  &0.961 & \bf 0.965 & 0.972\\
CR $\rightarrow$ MiRA $\rightarrow$ GS & \bf 0.968 & 0.952 & \bf 0.963 & 0.961\\  
CR $\rightarrow$ GS $\rightarrow$ MiRA & \textbf{0.980}  &0.961 & \textbf{0.977}&  0.973\\
\hline\hline
\end{tabular}
\end{center}
\end{table}

\noindent is found for reconstructions using combinations of Cauchy-Riemann\footnote{It only makes sense to use the Cauchy-Riemann algorithm first, since this is not an iterative algorithm relying on a first guess.} (CR), Gerchberg-Saxton (GS), and MiRA. The single most effective algorithm for these reconstructions is MiRA, and the highest correlation is obtained by using the full analysis: Cauchy-Riemann, followed by Gerchberg-Saxton and MiRA.  When MiRA or the Gerchberg-Saxton algorithms are used directly, the reconstructed image is usually symmetric. The Cauchy-Riemann stage is usually  the one responsible for roughly reconstructing asymmetries such as the bright or dark spot displayed in Figures \ref{bright_spot} and \ref{dark_spot}. The role of Gerchberg-Saxton is more to improve the phase reconstruction. MiRA plays the important role of removing artifacts, caused for example by the data fitting in the Cauchy-Riemann phase, and  improving overall definition. Even though non-symmetric images can be reconstructed, the final product still is somewhat more symmetric than the pristine image for reasons that are still under investigation. When the correlation with the pristine image is compared to the correlation with a uniform disk, we can again see that the bright spot (Figure \ref{bright_spot}) is more easily detected than the dark spot (Figure \ref{dark_spot}).

\newpage
A better estimate of the smallest temperature difference that can be imaged requires an exhaustive exploration of parameter space, but temperature 
differences of less than $\Delta T\approx 200^\circ\,\mathrm{K}$ do not seem to be possible to image when the same brightness, temperature, exposure time,  
angular diameter, spot size and spot position as above are used. Results are likely to improve for hotter stars than those simulated above since signal-to-noise is higher and also the brightness contrast is higher for the same relative temperature differences ($\Delta T/T$). Another question related to imaging single spots is that of finding the smallest spot that can be reconstructed. In previous work (Chapter \ref{sii_with_iact}, \citep{mnras}), we show that the smallest possible spot that can be reconstructed is given by the PSF of the IACT array used in the simulations, namely $0.06\,\mathrm{mas}$.

%Why is the bright spot better reconstructed?

%does deltaT=500K correspond to something possible. What are typical values in stellar atmospheres. 
%What do we expect in mass loss and von Zeipel

\subsection{Multiple features}\label{multiple_spots}
\begin{figure}
  \begin{center}
  \includegraphics[scale=0.55, angle=-90]{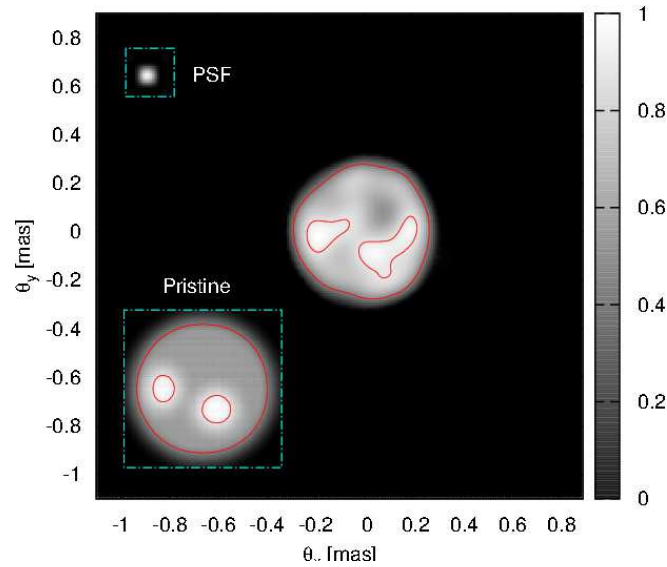}
  \end{center}
  \caption[ Reconstructed star with two hot spots.]{\label{double_spot} Reconstructed star with two hot spots. The pristine image has a temperature of $6000^\circ\,\mathrm{K}$, and each hot spot has a temperature of $6500^\circ\,\mathrm{K}$. The simulated data corresponds to $m_v=3$ and $T=10\,\mathrm{hrs}$. }
\end{figure}

\begin{figure}
  \begin{center}
  \includegraphics[scale=0.55, angle=-90]{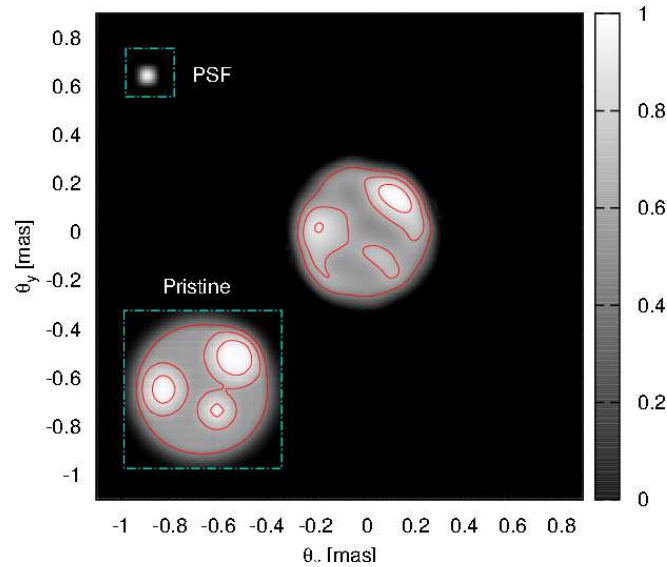}
  \end{center}
  \caption[Reconstructed star with three hot spots.]{\label{triple_spot} Reconstructed star with three hot spots. The pristine image has a temperature of $6000^\circ\,\mathrm{K}$, and the spots have temperatures of $6500^\circ\,\mathrm{K}$ (top right spot and left spot) and $6800^\circ\,\mathrm{K}$ (lower spot), The simulated data corresponds to $m_v=3$ and $T=10\,\mathrm{hrs}$. }
\end{figure}

As a natural extension to the simulations presented above, data were simulated corresponding to stars with two or more recognizable features. In Figures \ref{double_spot} and \ref{triple_spot},  reconstructions of stars containing several hot spots are shown. The brightness and exposure time are the same as those used to simulate single-spot stars ($m_v=3, 10\,\mathrm{hrs}$). A detailed investigation of reconstruction of two-spot stars was not performed, but the general behavior is similar to that presented in the section \ref{single_spots}. The reconstructions improve significantly when the pristine image has a higher degree of symmetry, e.g., when both spots lie along a line that splits the star in two. This is expected since phase reconstruction is not really necessary for centro-symmetric images. For this reason, we tested reconstructions with nonsymmetric pristine images. Even though the shape of the spots is usually not well reconstructed, the approximate position and size are reasonably accurate.

The reconstruction and identification of features degrades as the number of features in the pristine image is increased. A common characteristic of reconstructing stars with several features, is that the larger features in the pristine image are better reconstructed. This is more so in the case of stars containing darker regions. Nevertheless, information of positions, sizes and relative brightness of star spots can still be extracted. In Figure \ref{triple_spot} a reconstruction of a star containing three hot spots of different sizes and relative brightness is shown. This simulated reconstruction corresponds to the same brightness and exposure parameters as those of Figure \ref{double_spot}.

%\section{Improvement in simulations} \label{janvidas_work}

\chapter{Experimental Efforts} \label{experiment}

There are currently several ongoing efforts that aim to measure correlations of intensity fluctuations. There are two main experiments in operation at the University of Utah: A laboratory experiment and the \emph{StarBase} observatory. A short part at the end of this chapter describes the StarBase observatory, but most of it describes our laboratory efforts.% Much of the description of the electronics can be found in \citep{stephan.spie}.

\section{Laboratory efforts}

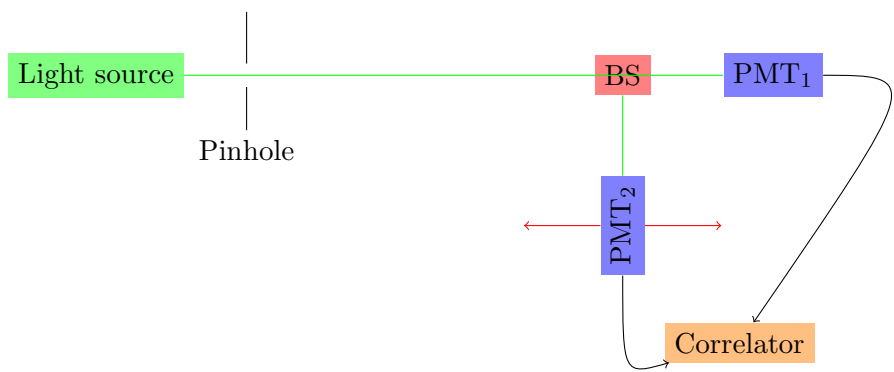
\begin{figure}[!]
        \begin{center}
  \begin{tikzpicture}
    \node(light_source)[fill=green!50]{Light source};
    \node(pinhole)[right of=light_source, node distance=2cm]{\textcolor{white}{.}};
    \node(pinhole_title)[below of=pinhole, node distance=1cm]{Pinhole};
    \node(above_pinhole)[above of=pinhole]{\textcolor{white}{.}};
    \node(bs)[right of=pinhole, node distance=5cm, fill=red!50]{BS};
    \node(pmt1)[right of=bs, node distance=2cm, fill=blue!50]{$\mathrm{PMT_1}$};
    \node(pmt2)[below of=bs, node distance=2cm, fill=blue!50, rotate=90]{$\mathrm{PMT_2}$};
    \node(r_pmt2)[right of=pmt2, node distance=1.5cm]{\textcolor{white}{.}};
    \node(l_pmt2)[left of=pmt2, node distance=1.5cm]{\textcolor{white}{.}};
    \node(correlator)[below right of=pmt2, node distance=2.2cm, fill=orange!50]{Correlator};
    %\node(image)[below  of=pmt2, node distance=2cm]{\includegraphics[scale=0.3]{/home/paul/documents/grad_seminar/lab1.pdf}};
    %\node(data)[below of=pinhole, node distance=3.5cm]{\includegraphics[scale=0.2]{/home/paul/documents/public_outreach/talk_figures/laser_preliminary_plot.pdf}};
    \draw(pinhole)--(pinhole_title);
    \draw(pinhole)--(above_pinhole);
    \draw[green](light_source)--(pmt1);
    \draw[green](bs)--(pmt2);
    \draw[->, red](pmt2)--(r_pmt2);
    \draw[->, red](pmt2)--(l_pmt2);
    \draw[->](pmt1)..controls (11,0)..(correlator);
    \draw[->](pmt2)..controls (7,-4)..(correlator);
  \end{tikzpicture}
  \end{center}
  \caption[Illustration of intensity interferometry laboratory experiment.]{\label{lab_outline}Illustration of intensity interferometry laboratory experiment. The baseline can be adjusted by moving $\mathrm{PMT_2}$. The correlation can be found via an analog system (section \ref{analog}) or digital system (section \ref{digital_system}). }
\end{figure}

The laboratory experiment consists in measuring the angular size of an artificial star using two miniature telescopes. The artificial star is a small pinhole ($<1\,\mathrm{mm}$) illuminated by an incoherent light source (see sections \ref{light_sources}). The miniature telescopes consist of two photo-multiplier tubes (PMTs) are placed $3\,\mathrm{m}$ away from the pinhole as shown in Figure \ref{lab_outline}. A beam-splitter is used to allow us to effectively place the two detectors at zero baseline. One of the PMTs is movable so that the baseline can be changed. The light collecting area of the PMTs is restricted to a couple of millimeters in diameter so that the individual PMTs do not resolve\footnote{Note that when using $400\,\mathrm{nm}$ light, the transverse coherence length is  $< 10\,\mathrm{mm}$} the pinhole. Light from the pinhole travels through a ``pipe'' and the light detectors are placed inside a metal box\footnote{The metal box should also help shield against external signals.} so that they only receive light from the pinhole. 

\section{Slow control and front end electronics}

The electronics in the laboratory are set up essentially the same way as in the camera used for on-sky observations at the StarBase observatory. The camera electronics consist of two parts. The slow control electronics provide power\footnote{All power sources are external batteries to isolate electronics.} to the front end, digitize the anode current to monitor the DC light intensity $\langle I\rangle$, provide high voltage to the photo-detector and can be used to program parameters of the front end electronics. The Slow control was developed by Jeremy Smith, Derrick Kress and Janvida Rou in the University of Utah. The front end electronics convert high frequency intensity fluctuations $\Delta I$ down to the single photon level to analog light pulses which can be transported by optical fiber with minimal bandwidth loss and attenuation over great distances. The optical fiber signals are then converted back to electrical signals which may be correlated (at the central control). A schematic of the electronics is shown in Figure \ref{electronics}. The front end electronics were developed for IACTs by \citet{rose}, and are described in detail by \citet{white}. More details of the electronics for SII are given by \citep{stephan.spie}.

\begin{figure}
  \begin{center}
  \tikzstyle{blueblock} = [rectangle, draw, fill=blue!20, 
    text width=4.5em, text centered, rounded corners, minimum height=1.5em, node distance=2.5cm]
  \tikzstyle{greenblock} = [rectangle, draw, fill=green!50, 
    text width=5em, text centered, rounded corners, minimum height=1.5em, node distance=2.5cm]
  \tikzstyle{redblock} = [rectangle, draw, fill=red!40, 
    text width=4.5em, text centered, rounded corners, minimum height=1.5em, node distance=2.5cm]
  \tikzstyle{Blueblock} = [rectangle, draw, fill=blue!40, 
    text width=4.5em, text centered, rounded corners, minimum height=1.5em, node distance=2.5cm]
  \tikzstyle{line} = [draw, -latex']
  \tikzstyle{arrow} = [draw, -latex']

  \begin{tikzpicture}[auto, scale=0.95]
    \draw [fill=orange!40] (-1.5, -5.5) rectangle (4,1);
    \node at (0,0.55) {\textcolor{red}{CAMERA}};
    \draw [fill=green!40] (5,-3) rectangle (9, -6);    
    \draw [fill=red!40] (4.5, -3) -- (7, -2) -- (9.5, -3) -- (4.5, -3);
    \draw [fill=green!30] (-1.25,-5) rectangle (3.7, -3);    
    \node [greenblock] (PMTBase){PMT Base};
    \node at (0, -4) [blueblock] (HV) {HV};
    \node [redblock, right of=PMTBase](Frontend){Front End};
    \node [blueblock, right of=HV] (ADC){ADC};
    \node at (7.7,-3.5) [redblock] (Reciever) {Receiver};
    %\node at (6.5,-5) [Blueblock](CPU){CPU};
    \node [Blueblock, below left of=Reciever, node distance=1.6cm](CPU){CPU};
    \node at (0,-4.7) {Slow Control};
    \node [ellipse, text width=3.8em, draw, fill=orange!60, below of=Reciever, node distance=1.8cm] (Correlator) {Correlator};
    
    %\draw [->, shorten >=1pt] (HV) -- (PMTBase);
    \draw[snake=coil, line after snake=0pt, segment aspect=0, blue, semithick] (-2,0) -- (PMTBase);
    \draw [arrow] (HV) -- (PMTBase);
    \draw [arrow] (PMTBase) -- (Frontend);
    %\draw [->] (ADC) to node {} node [swap] {Anode Current} (Frontend) ;
    \draw [arrow, semithick] (Frontend) -- node[above, sloped]{\tiny{Anode Current}}(ADC) ;
    \draw [arrow, blue, semithick] (Frontend) .. controls +(right:5cm) and +(up:3cm) .. node[above, sloped] {\small{Optical Fibers}}  (Reciever);
    \draw [blue, semithick](2.9, -3) .. controls (3, 0) and (7, 1) .. (CPU);
    \draw (1, -3) -- node[sloped] {\small{I2C}} (Frontend);
    \draw [arrow] (Reciever) -- (Correlator);
    \draw [line, blue, semithick] (9,1) -- (Reciever);
    \draw [arrow] (Reciever) -- (Correlator);
  \end{tikzpicture}

  \caption[Outline of electronics.]{\label{electronics}Outline of electronics. The camera, shown on the left, consists of the PMT, the slow control, and the front end electronics. The front ten electronics provide power to the PMT and read the anode current from the front end electronics. The front end electronics convert the electronic signal to an optical signal, sent to the control station (right), and converted back to an electronic signal, where it can be correlated with the signal from another telescope. }
  \end{center}
\end{figure}
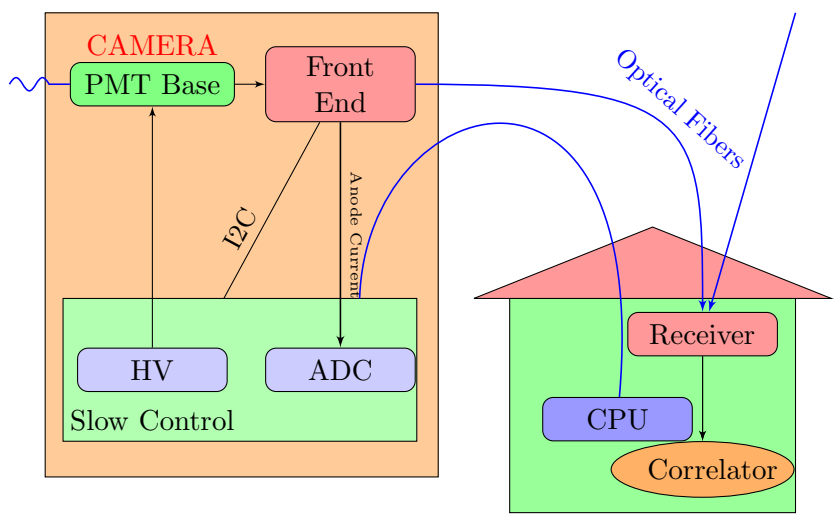

\begin{figure}
  \begin{center}
    \tikzstyle{blueblock} = [rectangle, draw, fill=blue!20, 
      text width=5em, text centered, rounded corners, minimum height=1.5em, node distance=2.5cm]
    \tikzstyle{greenblock} = [rectangle, draw, fill=green!50, 
      text width=5em, text centered, rounded corners, minimum height=1.5em, node distance=2.5cm]
    \tikzstyle{redblock} = [rectangle, draw, fill=red!40, 
      text width=5em, text centered, rounded corners, minimum height=1.5em, node distance=2.5cm]
    \tikzstyle{Blueblock} = [rectangle, draw, fill=blue!40, 
      text width=5em, text centered, rounded corners, minimum height=1.5em, node distance=2.5cm]
    \tikzstyle{line} = [draw, -latex']
    \tikzstyle{arrow} = [draw, -latex']
    
    \begin{tikzpicture}
      \node(signal_generator)[greenblock]{Signal gen. $13\,\mathrm{Hz}\,\,\pm 5\mathrm{V}\,$};
      \node(Lock_in)[below left of=signal_generator, blueblock, node distance=3cm]{Lock-in amp};
      \node(mixer)[circle, right of=Lock_in, node distance=3cm, fill=red!50, draw]{$\times$};
      \node(flipper)[right of= mixer, circle, fill=green!50, draw, node distance=3cm]{$\times \pm 1$};
      \node(s1)[right of=flipper, node distance=3cm]{$I_1$};
      \node(s2)[below of=s1, node distance=2.5cm]{$I_2$};
      \node(result)[below of=Lock_in, node distance=2cm]{$\left\langle \Delta I_1 \Delta I_2\right\rangle$};
      \draw[arrow](s1)--(flipper);
      \draw[arrow](flipper)--(mixer);
      \draw[arrow](mixer)--(Lock_in);
      \draw[arrow](s2)--(mixer);
      \draw[arrow](signal_generator)--(flipper);
      \draw[arrow](Lock_in)--(result);
      \draw[arrow](signal_generator)--node[sloped, above]{Ref.}(Lock_in);
    \end{tikzpicture}
  \end{center}
  \caption[Schematic of analog system for measuring the correlation between input signals $I_1$ and $I_2$.]{\label{analog_system} Schematic of analog system for measuring the correlation between input signals $I_1$ and $I_2$. The polarity of one of the signals is periodically flipped, so that the output of the linear mixer periodically changes sign, and can be detected with a lock-in amplifier.}
\end{figure}
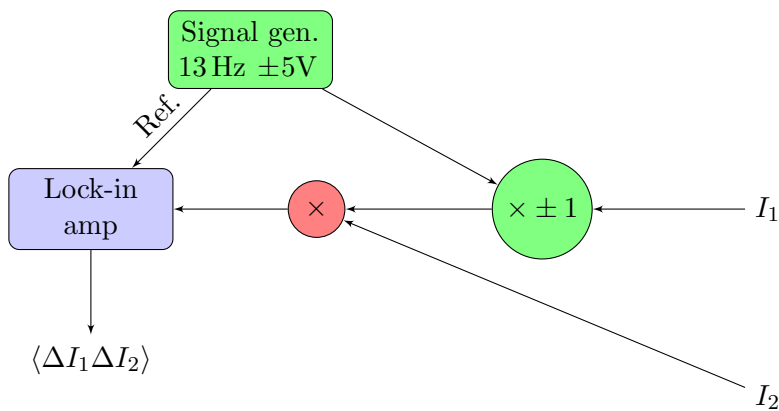

\begin{figure}
  \begin{center}
    \includegraphics[scale=0.5]{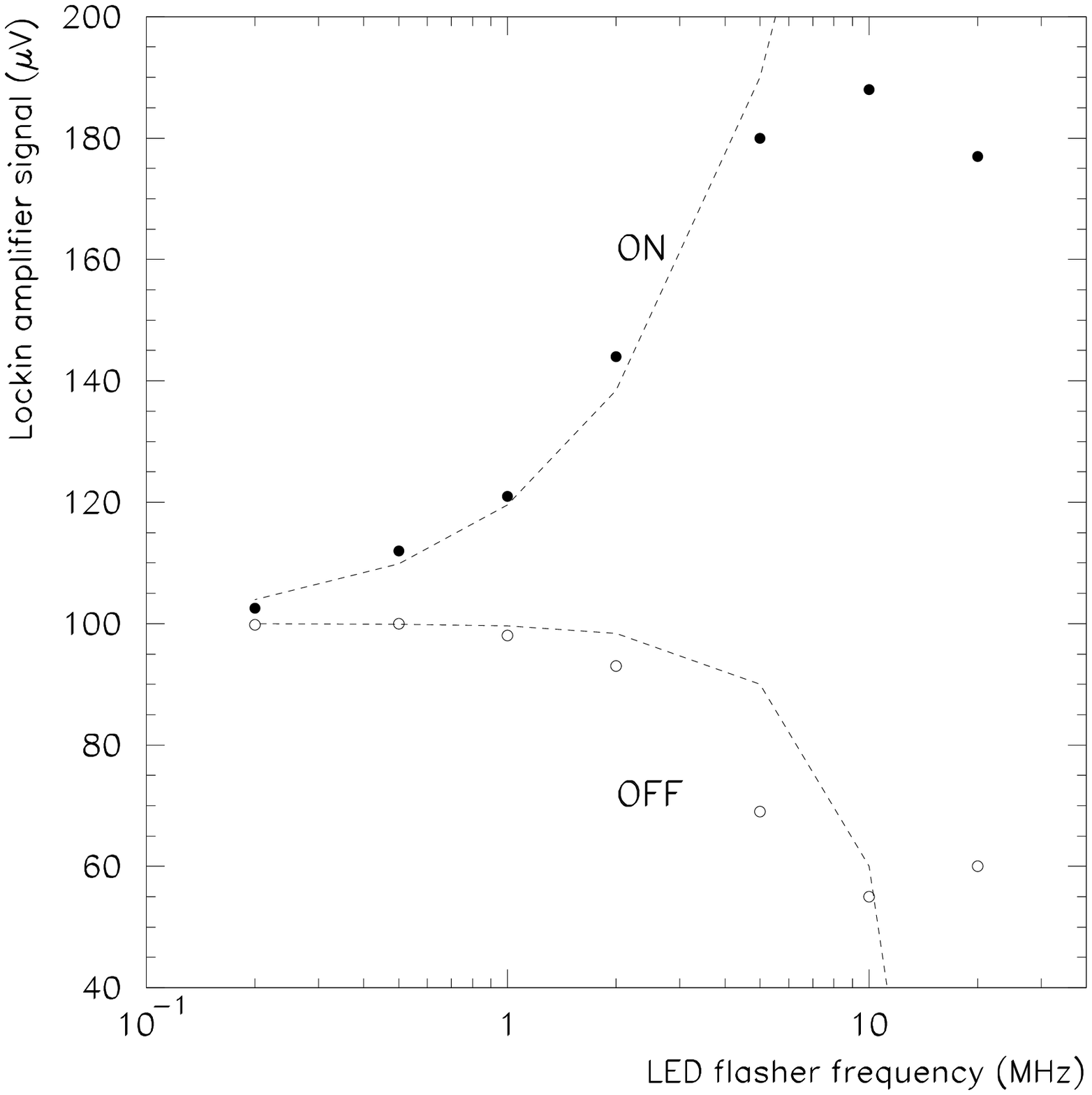}
  \end{center}
  \caption[Measured correlation for pulses with a duration of $20\,\mathrm{ns}$. ]{\label{led} Measured correlation for pulses with a duration of $20\,\mathrm{ns}$. The integration time of the lock-in amplifier was set to $10\,\mathrm{s}$. The top curve corresponds to the pulses being in phase, and the bottom curve corresponds to the pulses being out of phase. Both of these curves are compared to the theoretical prediction.}
\end{figure}

\section{Correlators}

We have experimented with a couple of approaches to measure the intensity correlations: one is with an analog system, similar to that used by Hanbury-Brown, and another is with a signal digitizer so the correlation is obtained off-line.

\subsection{Analog system}\label{analog}

In the analog system, one of the output signals undergoes a periodic polarity flip (phase switch), and then both signals are fed into a analog mixer (multiplier). The output of the mixer is fed into a phase sensitive (lock-in) amplifier, whose reference frequency is provided by the signal that controls the phase switching. If there is a correlation between the two signals, then the output of the linear mixer displays a periodic change from correlation to anti-correlation at the phase switching frequency amplified by the lock-in amplifier. A schematic of this system can be seen in Figure \ref{analog_system}. The time constant of the lock-in amplifier determines the integration time of the correlation, and the measurement of $|\gamma|^2$ can be normalized by finding the DC signal provided by the slow control.

The functionality of the analog system was demonstrated by using a pair of fast blinking light emitting diodes (LED), which provided short ($20\,\mathrm{ns}$) and faint ($1\,\mathrm{mV}$) light pulses correlated between the two channels. The output of the lock-in amplifier as a function of pulse frequency is shown in Figure \ref{led}. The top curve corresponds to the pulses being correlated, and the bottom curve corresponds to the signals being anti-correlated (since signals are AC coupled). Measurements can be compared to the theoretical prediction, which can be calculated as follows. Assume two periodic pulses, with period $\mathcal{T}$, and pulse duration $\Delta t$, defined as 

\begin{equation}
  s(t)=
  \begin{cases}
    s_0\,\,& \text{when}\,\,\,0\leq t \leq \mathcal{T}-\Delta t\\
    s_1\,\,& \text{when}\,\,\,\mathcal{T}-\Delta t< t \leq \mathcal{T}.\\
  \end{cases}
\end{equation}

Now, noting that the average signal is 

\begin{equation}
  \langle s \rangle = s_0+\frac{\Delta t}{\mathcal{T}}(s_1-s_0),
\end{equation}
then the correlation is

\begin{eqnarray}
c=\left\langle \delta s^2\right\rangle &=& \frac{1}{\mathcal{T}}\int_0^\mathcal{T}(s-\langle s \rangle)^2 dt\\
 &=& (s_1-s_0)^2\frac{\Delta t(\mathcal{T}-\Delta t)}{\mathcal{T}^2}\\
 &=& \delta s^2\frac{\Delta t(\mathcal{T}-\Delta t)}{\mathcal{T}^2}.
\end{eqnarray}

When a time delay  $\Delta \mathcal{T}$, equal to the length of the pulse, is introduced between the two signals, it is straightforward to show that the (time-delayed) correlation $c_{\Delta t}$ is 

\begin{equation}
  c_{\Delta  \mathcal{T}}=-\delta s^2\frac{ \mathcal{T}}{\mathcal{T}}.
\end{equation}

In order to compare the expected behavior of $c$ and $c_{\Delta  \mathcal{T}}$ to measurements, they need to be scaled accordingly, i.e., the units of the expected values are expressed in square volts, whereas the read out of the lock-in amplifier is in volts. There is also an offset (of unknown origin) of $100\,\mu\mathrm{V}$ in the correlation that needs to be added to the expected curves in order to compare them with measurements. The expected curves are shown in Figure \ref{led}, and deviations from the data can be seen when the LED frequency approaches the reciprocal of the pulse width ($(2\Delta t)^{-1}=25\,\mathrm{MHz}$) as expected.

\subsection{Digital system}\label{digital_system}

Recent advancements in high speed data acquisition now allow us to continuously digitize data at high frequencies while recording the traces in disk. The correlation between intensity signals is then obtained off-line by data analysis. An advantage of this very flexible approach is that noise from narrow-band sources, such as cell-phones, motors, etc., can be removed through signal processing techniques. Since measuring correlations for thermal light sources requires long integration times ($\gg1\,\mathrm{s}$), the disadvantage is that the vast amounts of data  generated ($\sim 4\,\mathrm{Gbs^{-1}}$) can cause computational difficulties. Much of the work related to digitizing the data efficiently has been done by David Kieda, and is described by \citep{stephan.spie}. The two electronic signals can be sampled at $250\,\mathrm{MHz}$ with 12 bit resolution using a National Instruments PXIe-5122 high speed digitizer. This digitizing system has proven to be capable of streaming data for several hours. 

The functionality of the digitizing system was demonstrated by flashing LED pulses at $1\,\mathrm{MHz}$. The signals fed to the LEDs consisted of a floor of light of $820\,\mathrm{mV}$ and periodic pulses of $840\,\mathrm{mV}$ that were $8\,\mathrm{ns}$ wide. By averaging many traces recorded with the oscilloscope, we estimated the number of additional photons associated with the pulse to be 1/5 per pulse.  This means that\pagebreak

\begin{figure}[h]
  \begin{center}
    \includegraphics[scale=0.5, angle=-90]{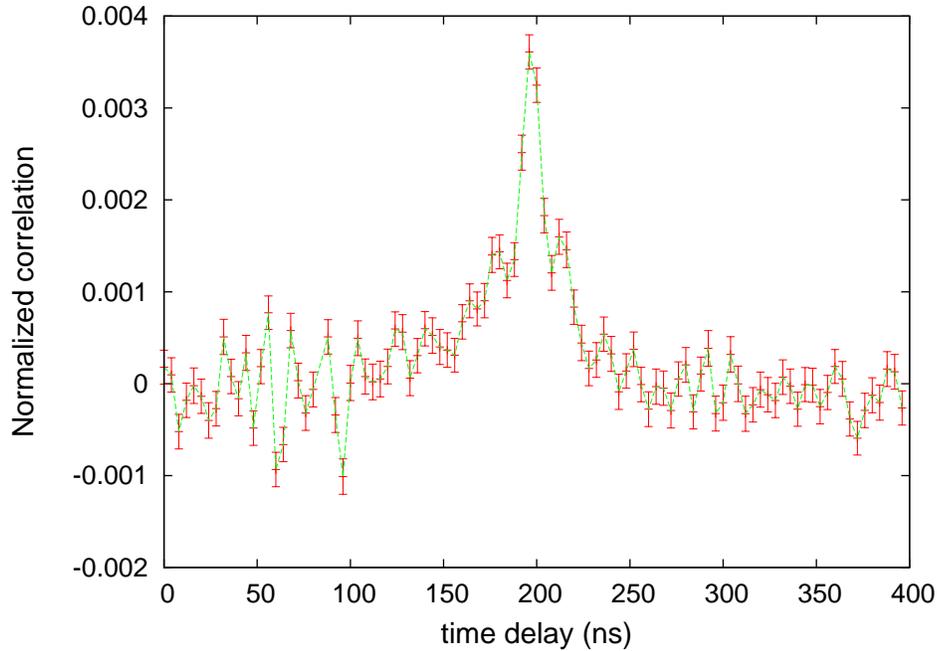}
  \end{center}
  \caption[ Degree of correlation as a function of the time lag between two digitized LED signals.]{\label{led_delay} Degree of correlation as a function of the time lag between two digitized LED signals. LEDs were flashed at $1\,\mathrm{MHz}$ and consist of a floor of light of $820\,\mathrm{mV}$ and $8\,\mathrm{ns}$ pulses of $840\,\mathrm{mV}$ height. The data sample is $1\,\mathrm{s}$ long, and the uncertainty in the correlation was found by evaluating the correlation every $4\,\mathrm{ms}$ and then calculating the statistical standard deviation. }    
\end{figure}

\noindent every 25 pulses on average we have two photons in coincidence between the two channels. A time lag of $200\,\mathrm{ns}$ was introduced between the two LEDs so the correlation as a function of the time lag should be maximal when signals are brought back in time in the data analysis. This maximal correlation at $200\,\mathrm{ns}$ can clearly be seen in Figure \ref{led_delay}, which corresponds to $1\,\mathrm{s}$ of data. The erratic behavior at short time delays ($<150\,\mathrm{ns}$) may be indicative of PMT cross talk or external noise.

\newpage
%\section{Thermal and pseudo-thermal light sources} \label{light_sources}

\subsection{Thermal light source} \label{light_sources}

The light source must provide incoherent light with a high spectral density and a manageable\footnote{A very low count rate would require very long integration times to detect a correlation (see SNR estimate below).} photon count rate.  A mercury arc lamp of wavelength $435\,\mathrm{nm}$ produces incoherent light similar to what can be detected from a hot star with an interferometric filter ($\Delta \lambda\sim 10\,\mathrm{nm}$). With such a source the coherence time is clearly much shorter than the electronic resolution time, and in this regime, the SNR given by equation \ref{snr2}, i.e.,

\begin{equation}
  SNR=NA\alpha|\gamma|^2\sqrt{T_0\Delta f/2}. \nonumber
\end{equation}

To estimate the $SNR$, we can first find the photon rate $NA\alpha\delta\nu$ by reading the typical anode current\footnote{The slow control actually provides a DC voltage which can be converted to a current by finding an equivalent resistance of the electronics. Using a signal generator to provide a calibration signal, we have found the equivalent resistance to be $12.2\,\mathrm{k}\Omega$.} from the PMT photo-cathode, which is typically of the order order of $I\approx 10\,\mu \mathrm{A}$. The gain of the PMTs is of the order of $G\approx 10^5$, therefore, the number of photons per unit time is

\begin{equation}
  NA\alpha\Delta\nu\approx  \frac{I}{eG}\approx 6\times  10^8\, \mathrm{s}^{-1}.
\end{equation}

Since the optical bandwidth is $\Delta\nu=c\Delta\lambda/\lambda^2\approx 10^{13}\mathrm{Hz}$ and the electronic bandwidth is (pessimistically) $\Delta f\approx 100\,\mathrm{MHz}$, then a $SNR\approx 3$ requires an integration time of $\sim 3\,\mathrm{min}$.

Several attempts have been made to measure correlations with a thermal source. In these attempts, the analog system was used as described in section \ref{analog}. Measuring intensity correlations has proven to be difficult with a thermal source. It is advantageous to perform an intensity interferometry experiment where long integration times are not needed. The electronic time resolution or the spectral density cannot be significantly improved. However, in the next section we describe an experiment in which an artificial light source with an extremely long coherence time is used for intensity interferometry experiments.

\subsection{Pseudo-thermal source}

\begin{figure}[h]
  \begin{center}
    \includegraphics[scale=0.1]{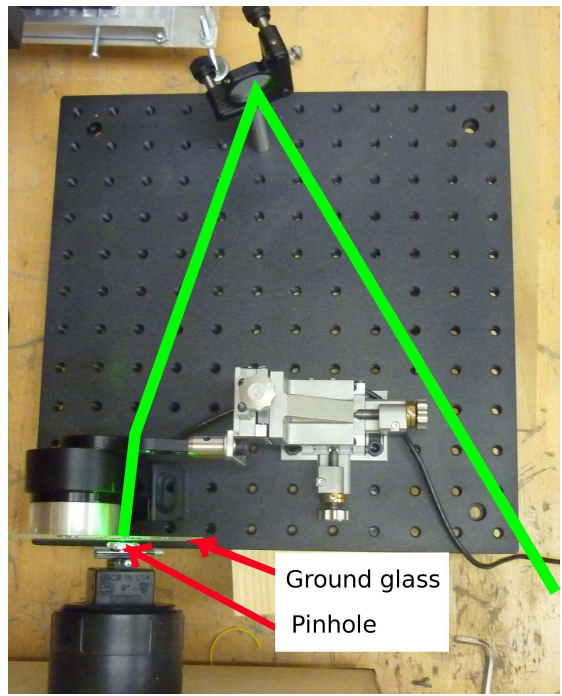}
  \end{center}
  \caption{\label{pseudo_thermal} Laser scattered light through rotating ground glass produces a pseudo-thermal light source.}
\end{figure}

An incoherent light source with a coherence time  which is much longer than the electronic resolution has been constructed as proposed by \citet{martienssen}. A pseudo-thermal light source can be easily created by scattering coherent light (e.g., laser light) off a medium that continuously changes with time, therefore producing a time-varying speckle pattern. This can be accomplished by shinning laser light ($\lambda=534\,\mathrm{nm}$) through a rotating sheet of ground glass as shown in Figure \ref{pseudo_thermal}. The rotating ground glass is put as close as possible to the pinhole to create the artificial star, and a rapidly changing speckle pattern can be seen where the light is detected by the PMTs. 

The typical time over which the speckle pattern changes sets the coherence time of the source, and can be adjusted by changing the speed of the ground glass, or by using a different sized glass grain.  The photon detection statistics generated by such a source are very similar to those of a thermal light source \citep{morgan-mandel}, except that the detected light fluctuation are now dominated by the wave noise instead of the shot noise (see section \ref{variance}), that is, the fluctuations are proportional to the number of detected photons per resolution time. The coherence area is essentially the size of a speckle, and when photo-detectors are separated by less than the typical size of a speckle, i.e., within the coherence volume, then their signals are correlated. A pseudo-thermal source then allows us to do an intensity interferometry experiment in ``slow motion.''

\section{Results with pseudo-thermal light source}

Correlations were measured between two PMTs receiving light from an artificial star (pinhole) that is illuminated by pseudo-thermal light.

\subsection{Individual signals from each channel}

\begin{figure}[h]
  \begin{center}
    \includegraphics[scale=0.5, angle=-90]{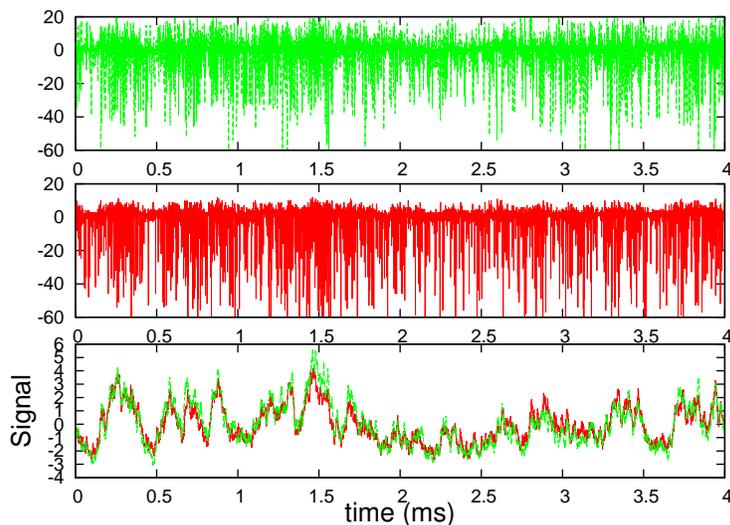}
    \vspace{2cm}
  \end{center}
  \caption[Sample PMT traces.]{\label{small_trace} The top two traces correspond to the raw data obtained with the digitizing system in two separate channels. The signal is plotted every 1000 points so that the plot is not saturated. In order to maximize the degree of correlation, the envelope of each curve is found before the correlation is calculated. To find the envelope, the absolute value of the signal is found before a lo pass filter ($\tau<0.01\,\mathrm{ms}$) is applied. The bottom figure shows the two filtered signals, which are clearly correlated. }
\end{figure}

A sample trace is shown in Figure \ref{small_trace}. The top curve in Figure \ref{small_trace} is the raw data obtained by the digitizer. Since speckles have a typical duration of a fraction of a millisecond (described below in section \ref{temporal_coherence}), the measurement of the normalized degree of correlation is maximal when the envelopes of the traces are used as opposed to the raw traces. Since the measured signals are AC coupled, the absolute value is taken (middle curve in Figure \ref{small_trace}) before a low pass filter is applied to find the envelope. A filtered trace envelope is displayed in the bottom of Figure \ref{small_trace}. The filtering time constant must be large enough so as to maximize the correlation, but still much smaller than the coherence time. Therefore, it is useful to measure the coherence time before finding the normalized degree of correlation.

\subsection{Temporal coherence \label{temporal_coherence}}

\begin{figure}[h]
  \begin{center}
    \includegraphics[angle=-90, scale=0.5]{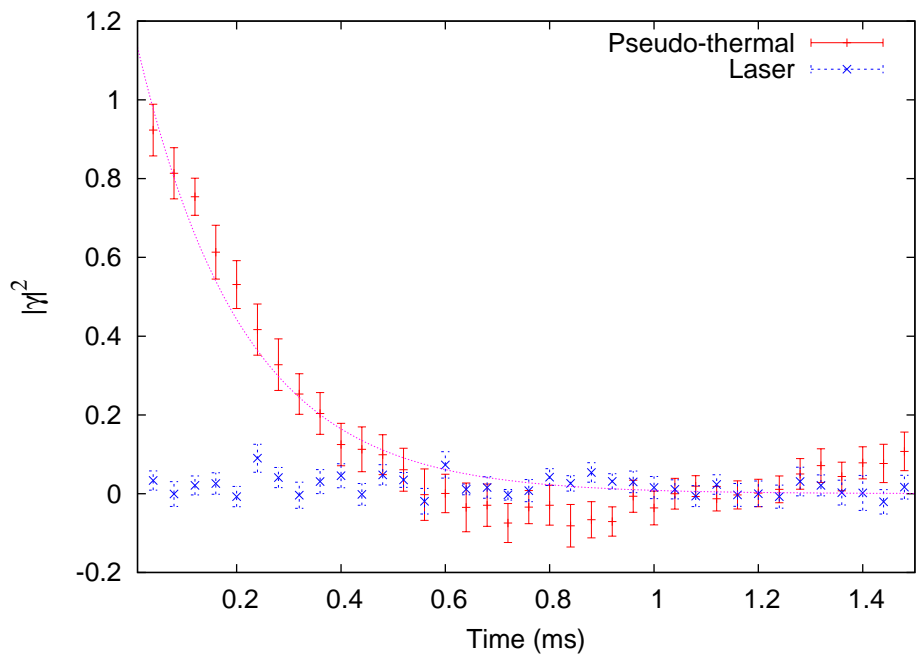}
    \vspace{0.2cm}
  \end{center}
    \caption[Temporal correlation for pseudo-thermal light compared to laser light.]{\label{auto_correlation} Temporal correlation for pseudo-thermal light compared to laser light. A time delay between the two (nonfiltered) PMT signals was introduced off-line. The ``photon-bunching'' region can be seen below $\sim 0.6\,\mathrm{ms}$. The estimation of the error-bars is described in section \ref{uncertainty_estimation}.}
\end{figure}

In order to find the coherence time of the pseudo-thermal light source, the correlation is found as a function of a time displacement introduced between the two channels (at zero baseline). Since data were taken with the digital system described in section \ref{digital_system}, the time delay is introduced off-line. Correlations corresponding to a small sample of $2\,\mathrm{ms}$ are shown in Figure \ref{auto_correlation}, where the temporal coherence is measured for both the pseudo-thermal light source, and the laser, i.e., rotating ground glass compared to nonrotating ground glass. \pagebreak

\pagebreak

An ideal laser should not display temporal correlations since it has negligible intensity fluctuations and emits Poisson distributed light (section  \ref{photon_statistics}). The laser used in this experiment actually contains several modes, with a mode spacing of $303\,\mathrm{MHz}$, and should experience intensity modulations with this frequency, but these are under-sampled at $250\,\mathrm{MHz}$ with the digitizing system used in this experiment.  Even though an ideal laser emits coherent light, intensity correlations (or their absence) do not have a relation to the angular radiance distribution of the source.\footnote{The relation between intensity correlations and the degree of coherence (eq. \ref{boxed_result}) only exists in light whose electric field tends towards being a Gaussian random variate.}

In Figure \ref{auto_correlation}, the ``photon-bunching'' region is clearly seen for delay times smaller than $\sim 0.4\,\mathrm{ms}$, and the coherence time $\tau$ can be estimated by fitting a Lorentzian, which yields $\tau=0.23\,\mathrm{ms}$. Therefore, a filtering time constant of $0.01\,\mathrm{ms}$ is appropriate in order to maximize the degree of correlation (see section \ref{uncertainty_estimation} for more details). On a thermal source, the type of plot shown in Figure \ref{auto_correlation} contains information on the spectrum, and the inverse of the coherence time would essentially be the optical bandwidth of the source. For the case of the pseudo-thermal source, this contains information on the time-scale in which the speckles pass through the detector. If the speckles only originate from the pinhole, then they should all be of similar size, i.e., of the size of the coherence area, and therefore, the curve shown in Figure \ref{auto_correlation} should have a slope of zero near the origin. The fact that the curve does not level out implies that we may also be observing smaller speckles. Smaller speckles would need to originate from sources that have a large angular size, and therefore it is possible that they originate from internal reflections in our experimental set up (e.g., the metal box where the PMTs are placed). However, we have not established this rigorously. 

\subsection{Uncertainty in the correlation \label{uncertainty_estimation}}

\begin{figure}[h]
  \begin{center}
    \includegraphics[scale=0.5, angle=-90]{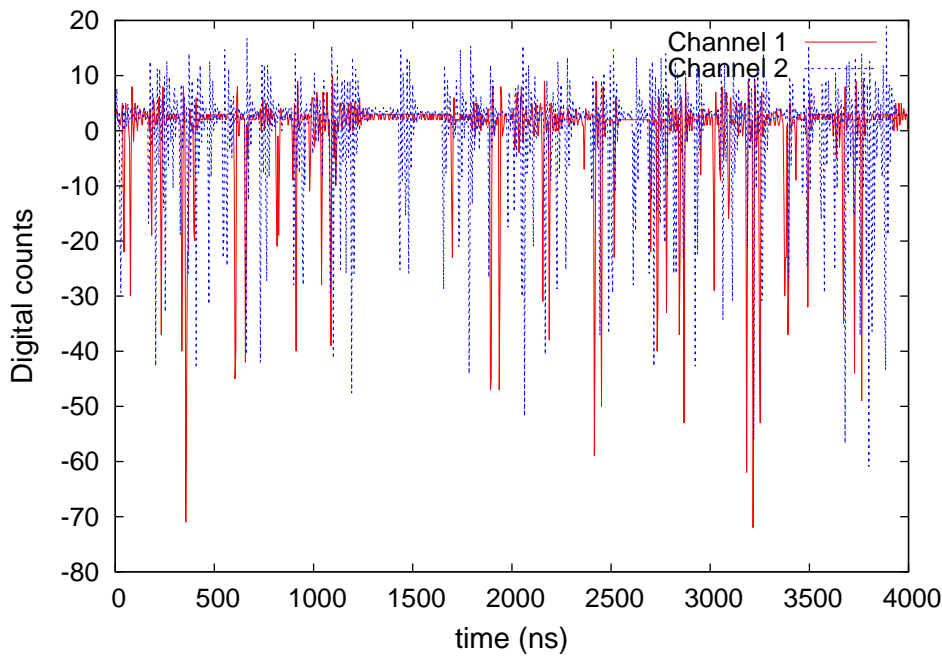}
  \end{center}
  \vspace{0.3cm}
  \caption[Digital counts for each channel.]{\label{tiny_trace} Digital counts for each channel. According to this trace, the number of photons per unit time is $\sim 8\times 10^7\,\mathrm{s}^{-1}$. Since the electronic resolution time is $4\,\mathrm{ns}$, there are $ 8\times 10^7\,\mathrm{s}^{-1}\times 4\,\mathrm{ns}=0.3$ photons per resolution time. No obvious correlation can be seen a these time-scales.}
\end{figure}

A $4000\,\mathrm{ns}$ sample trace is shown in Figure \ref{tiny_trace}, where individual photons can be seen. At these short time-scales no correlation is expected, but by counting the number of photons per unit time ($\left\langle n \right\rangle\approx 8\times10^7\,\mathrm{s}^{-1}$), the expected SNR in a correlation measurement (with the pseudo-thermal source) can be estimated as follows. Recall from section \ref{variance} that the variance of the number of photo-electrons $n$ contains a Poisson term (shot noise) and a term that is proportional to the variance of the time integrated light intensity $\mu$ (eq. \ref{variance_general}), i.e.,

\begin{equation}
\left\langle \Delta n^2 \right\rangle = \left\langle n \right\rangle + \left\langle \Delta\mu^2 \right\rangle,  \nonumber
\end{equation}
where 
\pagebreak

\pagebreak
 
\begin{equation}
  \mu\equiv \eta\int_t^{t+T}I(r, t')dt'. \nonumber
\end{equation}

In this expression, $I$ is the light intensity integrated over the optical bandwidth and $\eta$ is a constant that characterizes the detector. When the electronic time resolution $T$ is much smaller than the coherence time, the integrated light intensity is approximately $\mu\approx\eta I(t)T$. Since there is essentially no light between adjacent speckles, the variance of the integrated light intensity is approximately 

\begin{equation}
\left\langle \Delta\mu^2 \right\rangle\approx (\left\langle I(t)\right\rangle T)^2.
\end{equation}

Therefore, the SNR achieved in one resolution time, is equal to the number of photons per resolution time, i.e., $\eta\left\langle I(t)\right\rangle T=NA\alpha\Delta\nu T\approx 0.3$ from Figure \ref{tiny_trace}. For an observation time $T_0$, the SNR increases as the square root of the observation time, i.e.,

\begin{equation}
  SNR = NA\alpha\Delta\nu \Delta T|\gamma|^2 \sqrt{T_0/T}. 
\end{equation}

Therefore, when we integrate for $\sim 1 \,\mathrm{s}$, we should expect a $SNR\approx 4\times 10^3$ . This value should be compared to the measured SNR, described below. 

\begin{figure}[h]
  \vspace{5cm}
  \begin{center}
    \includegraphics[scale=0.5, angle=-90]{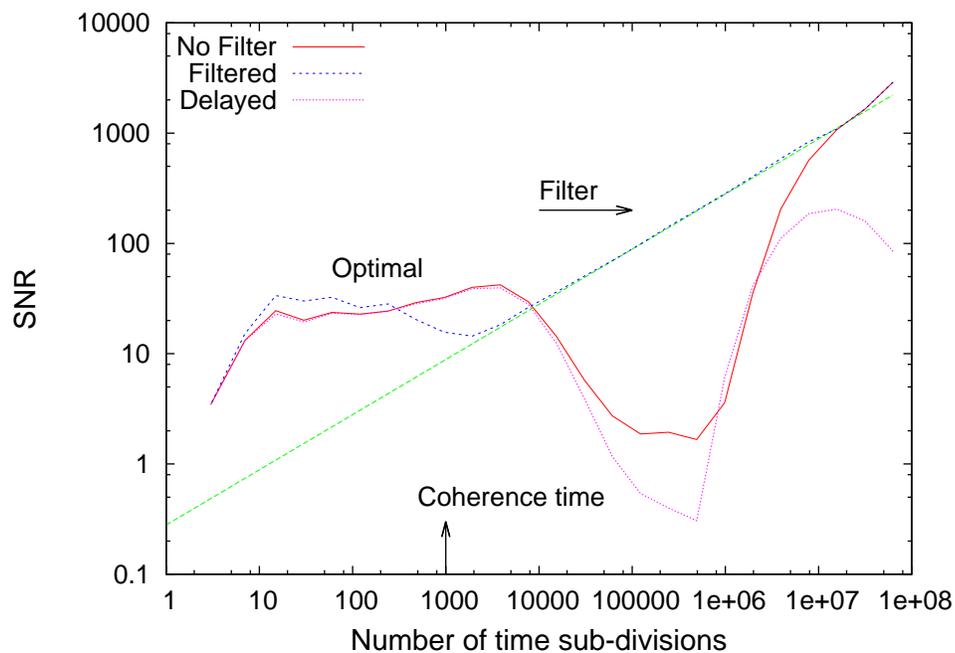}    
    \vspace{0.5cm}
  \end{center}
  \caption[Ratio of the correlation its statistical standard deviation as a function of the number of time subdivisions ($\propto 1/\mathrm{Time\,\,window}$) in a $0.2\,\mathrm{s}$ run.]{\label{time_window}Ratio of the correlation its statistical standard deviation as a function of the number of time subdivisions ($\propto 1/\mathrm{Time\,\,window}$) in a $0.2\,\mathrm{s}$ run. The red curve corresponds to the the SNR with no filtering applied to the data. The blue curve has been obtained by applying a low-pass filter with $\tau=0.01\,\mathrm{ms}$. The purple curve corresponds to no filtering, but a time delay of $8\,\mu\mathrm{s}$ was applied to test for possible cross talk between PMTs.}
\end{figure}

\begin{figure}[h]
  \begin{center}
    \includegraphics[scale=0.5, angle=-90]{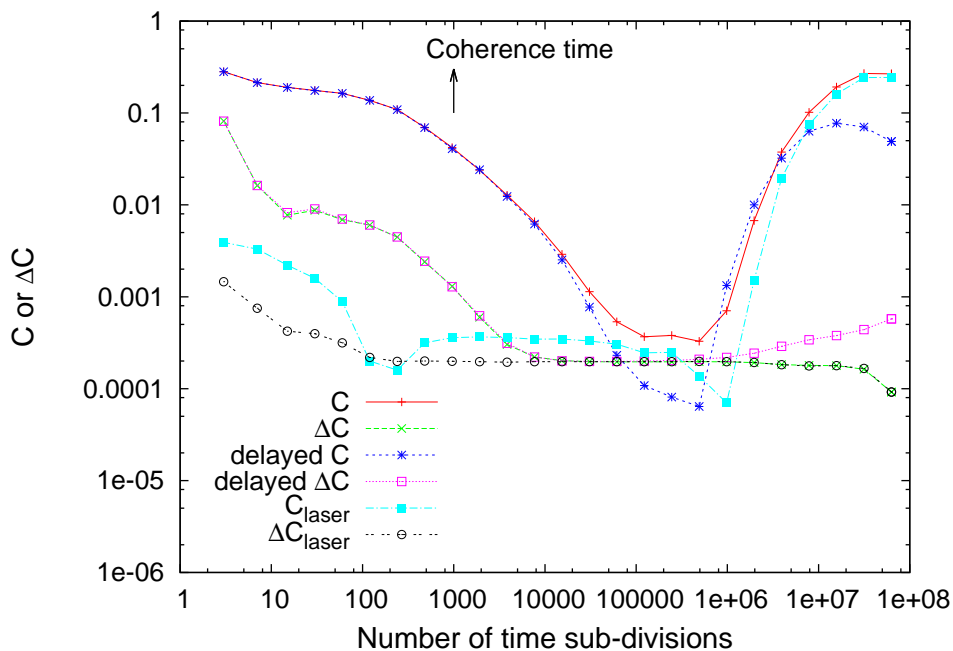}    
    \vspace{0.1cm}
  \end{center}
  \caption[Correlation and its statistical standard deviation as a function of the number of time subdivisions ($\propto 1/\mathrm{Time\,\,window}$) in a $0.2\,\mathrm{s}$ run.]{\label{signal_and_noise}Correlation and its statistical standard deviation as a function of the number of time sub-divisions ($\propto 1/\mathrm{Time\,\,window}$) in a $0.2\,\mathrm{s}$ run. The red curve corresponds to the the SNR with no filtering applied to the data.  To obtain the purple curve,  a time delay of $80\,\mathrm{ns}$ was applied to test for possible cross talk between PMTs. Also shown is the correlation for pure laser light (nonrotating ground glass), which displays the same behavior at small time windows as the pseudo-thermal light, and shows no significant correlations when the averaging time window is much longer than the electronic resolution time.}
\end{figure}

In order to evaluate the uncertainty in the correlation, the data run is subdivided in many small time windows. In each time window, the correlation is found, and then the statistical standard deviation of the correlation is found. The duration of each time subdivision must be much longer than the coherence time so that it includes the passing of many speckles. To find the optimal duration of each time subdivision, we perform a study of the ratio of the degree of coherence and its standard deviation ($SNR=C/\Delta C$) as a function of the number of sub-divisions (Figures \ref{time_window} and \ref{signal_and_noise}) . In Figure \ref{time_window}, we can see that the SNR is of the order of $\sim 30$ when the time averaging time window is much longer than the coherence time and when more than a few time samples are used (for sufficient statistics). Since the coherence time is $0.2\,\mathrm{ms}$, a time window of $2\,\mathrm{ms}$ is appropriate, and yields a SNR of $\sim 30$. The fact that the measured SNR is two orders of magnitude smaller than expected indicates that this experiment is limited by electronic noise rather than photon noise. 

When no filtering is applied to the data (red curve in Figure \ref{time_window}), the SNR decreases as soon as the time window is smaller than the coherence time. It is surprising to see that when no filtering is applied (red curve in Figure \ref{time_window}), the signal, as well as the SNR, increase when the length of the time window approaches the electronic resolution (Figure \ref{signal_and_noise}). The electronic resolution time is actually close to the laser mode spacing of $303\,\mathrm{MHz}$, which is under-sampled at $250\,\mathrm{MHz}$ with the digitizing system. These high frequency correlations are possibly due to\linebreak
\pagebreak

\pagebreak

\noindent beating between modes. The correlation of pure laser light (nonrotating ground glass) displays the same behavior at very small time windows (light blue curve in Figure \ref{signal_and_noise}). However, we have not rigorously proven this hypothesis and we cannot rule out high frequency noise pick-up in the electronics.

\subsection{Simulation of slow electronics}

We also study the behavior of the SNR ($C/\Delta C$) as a function of the filtering time constant. Recall that when the coherence time is much shorter than the electronic resolution time, the SNR is diluted by the ratio of the coherence time and the electronic resolution time. When the electronic time resolution is much smaller than the coherence time (current regime with the pseudo-thermal light source), the SNR is independent of the electronic time resolution. By increasing the filtering time constant, the electronics can be made artificially slower. 

\begin{figure}[h]
  \vspace{-0.8cm}
  \begin{center}
    \includegraphics[scale=0.5, angle=-90]{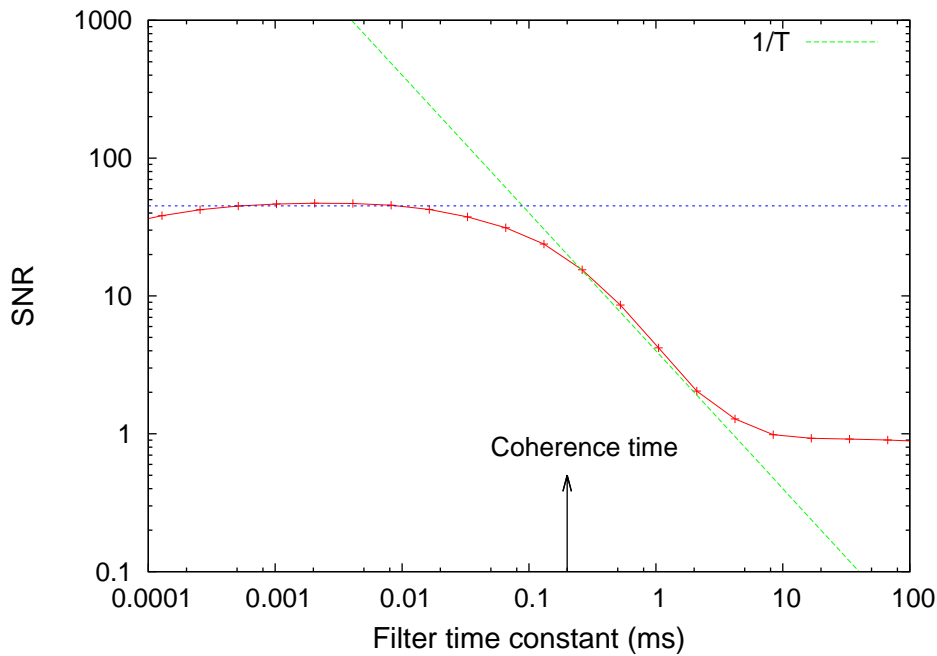}    
    \vspace{0.2cm}
  \end{center}
  \caption[Ratio of the degree of correlation and its standard deviation as a function of the filtering time constant.]{\label{snr_f} Ratio of the degree of correlation and its standard deviation as a function of the filtering time constant. The averaging time window is $8\,\mathrm{ms}$.}
\end{figure}

Figure \ref{snr_f} shows the behavior of the SNR as a function of the filtering time constant. When the filtering time constant is much smaller than the coherence time, the SNR is essentially constant. When the filtering time constant is larger than the coherence time, the SNR is inversely proportional to the Filtering time constant as expected. This is experimental evidence of what is described in section \ref{snr_section}, which essentially says that when electronics are ``slow'', or light intensity fluctuations are very fast (such as with a thermal source), then the ratio of the wave-noise (signal) and the shot noise is proportional to the ratio of the coherence time and the electronic resolution time.

\subsection{Spatial coherence and angular diameter measurements}

The most relevant result for SII is the measurement of spatial coherence. The diameters of different pinholes emitting pseudo-thermal light were found by measuring the normalized intensity correlation as a function of the PMT separation. Since the coherence time is much longer than the resolution time, we in fact find the correlation between the envelopes of the signals. To find the envelope, take the absolute value and then filter out frequencies larger than $25\,\mathrm{kHz}$ from the individual signals. 

\begin{figure}[h]
  \vspace{-1cm}
  \begin{center}
    \includegraphics[angle=-90, scale=0.5]{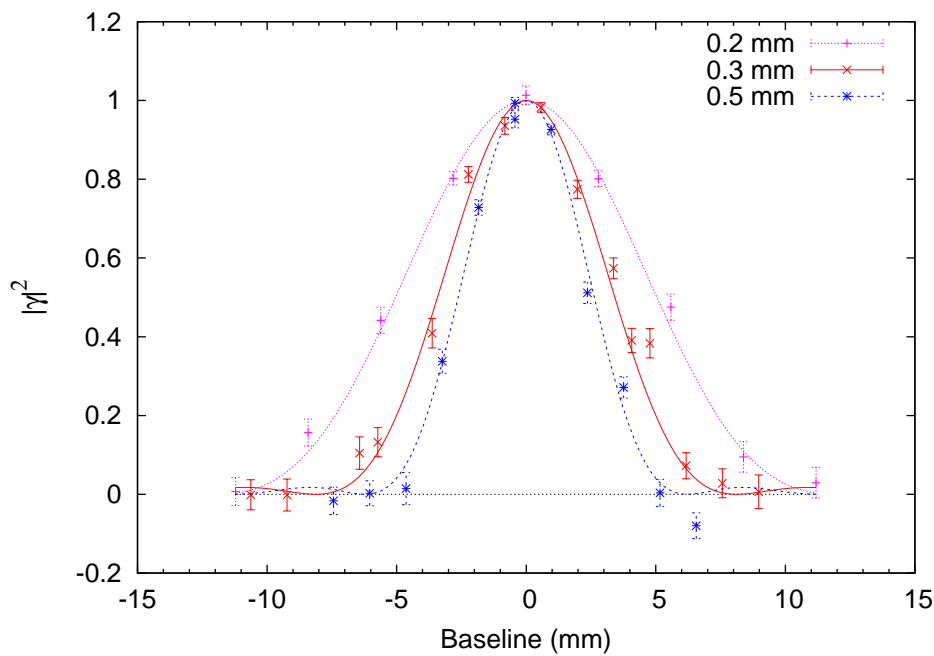}
    \vspace{1cm}
    \caption[Normalized degree of correlation as a function of PMT separation for three different pinhole sizes.]{\label{pinholes} Normalized degree of correlation as a function of PMT separation for three different pinhole sizes. The curves corresponding are the best fit curves.}
  \end{center}
\end{figure}

It is reasonable to assume that the pinholes are uniformly illuminated. The diameter $\Delta \theta$ of a uniformly illuminated pinhole can be extracted by fitting the correlations to an Airy function, and by recalling that $\Delta\theta=1.22 \lambda/D$, where $D$ is the first zero of the Airy function. The pinhole diameters have  nominal values of $0.2\,\mathrm{mm}$, $0.3\,\mathrm{mm}$  and $0.5\,\mathrm{mm}$, although the $0.5\,\mathrm{mm}$ pinhole was made by piercing a sheet of aluminum foil with a metal wire and is therefore only approximately this size. Data are presented in Figure \ref{pinholes} along with the best fit curves, and the measured pinhole diameters are $0.17\pm 0.02\,\mathrm{mm}$, $0.24\pm 0.03\,\mathrm{mm}$ and $0.32\pm 0.05\,\mathrm{mm}$.  
%$0.19\pm 0.01\,\mathrm{mm}$, $0.27\pm 0.03\,\mathrm{mm}$ and $0.38\pm 0.05\,\mathrm{mm}$. %ACCORDING TO FIT THAT IS MISSING A FACTOR
\pagebreak

\pagebreak

%It is worth noting that the uncertainties in the results shown in Figures \ref{auto_correlation} and \ref{pinholes} are considerably smaller than those estimated in the end of section \ref{light_sources} (by roughly 2 orders of magnitude!). 

\section{Image reconstruction from autocorrelation data}

In the experiment with the pseudo-thermal light source, time varying speckle patterns can be seen at the detectors, and correlations can be measured when both detectors lie on average within the same speckle. In view of the experiment with the pseudo-thermal light source, we can understand a thermal light source as a completely rough diffracting screen at each instant in time. 

In the case of the pseudo-thermal light source experiment, the time integrated correlation (as a function of PMT separation) is essentially an autocorrelation of the speckle pattern. Therefore, one can in principle take an image of a speckle pattern, compute its autocorrelation, which is equivalent to the squared modulus of the Fourier transform of the artificial star, and apply the methods developed in Chapter \ref{phase_recovery} for reconstructing the image of the artificial star. 

\begin{figure}[h]
\vspace{-1.2cm}
\begin{center}
\begin{tikzpicture}  
    \node(laser)[fill=green!50]{Laser};
    \node(be)[right of=laser, fill=red!50, node distance=3cm]{Beam expander};
    \node(mask)[right of=be, node distance=3cm]{\textcolor{white}{.}};
    \node(above_mask)[above of=mask]{Mask};
    \node(below_mask)[below of=mask]{\textcolor{white}{.}};
    \node(paper)[right of=mask, node distance=3cm]{\textcolor{white}{.}};
    \node(above_paper)[above of=paper]{Rough surface};
    \node(below_paper)[below of=paper]{\textcolor{white}{.}};
    \node(ccd)[below of=mask, fill=blue!50, rotate=20, node distance=2cm]{CCD};
    \draw[green](laser)--(be);
    \draw[line width=10pt, green!50](be)--(mask); 
    \draw[line width=5pt, dashed, green!50](mask)--(paper); 
    %\draw[line width=5pt, densely dotted, green!50](paper)--(ccd); 
    \node(above_left_paper)[above left of=paper, node distance=1.4cm]{\textcolor{white}{.}};
    \node(below_left_paper)[below left of=paper, node distance=1.4cm]{\textcolor{white}{.}};
    \node(left_paper)[left of=paper, node distance=1.4cm]{\textcolor{white}{.}};
    \node(cpu)[left of=ccd, fill=red!80, node distance=2cm]{CPU};
    \draw(above_mask)--(below_mask);
    \draw(above_paper)--(below_paper);
    \fill[green](paper) circle (0.1cm);
    \draw[->, line width=5pt, dashed, green!50](paper)--(above_left_paper);
    \draw[->, line width=5pt, dashed, green!50](paper)--(below_left_paper);
    \draw[->, line width=5pt, dashed, green!50](paper)--(left_paper);
    \draw(ccd)--(cpu);
\end{tikzpicture}
\end{center}
\vspace{-0.5cm}
\caption[Schematic of speckle experiment.]{\label{speckle_set_up} The laser beam passes through a beam expander and then through a mask of any desired shape (e.g., single or multiple circular holes). The rough paper screen reflects a speckle pattern which is captured with a CCD camera. The autocorrelation of the speckle pattern is the squared magnitude of the Fourier transform of the image on the rough screen.}
\end{figure}
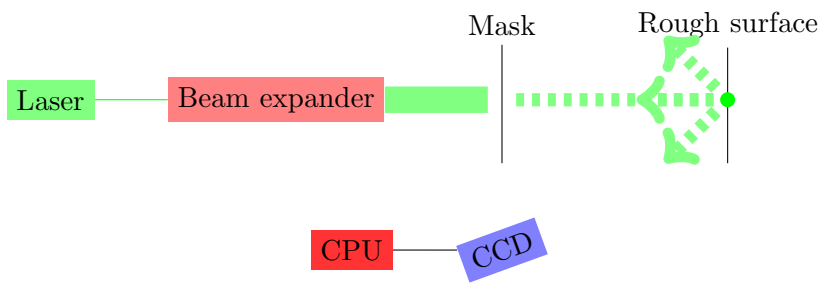

This type of experiment is currently in progress as shown schematically in Figure \ref{speckle_set_up}. An artificial star is created by shinning laser light through a mask of any desired shape. Then the spatial coherence is broken  by scattering the light off a rough (paper) screen. The resulting speckle pattern does not change with time, and can then be recorded with a CCD camera.

\begin{figure}[h]
\vspace{-0.3cm}
\begin{center}
    \begin{tikzpicture}
    \node(speckle_image){\includegraphics[scale=0.15]{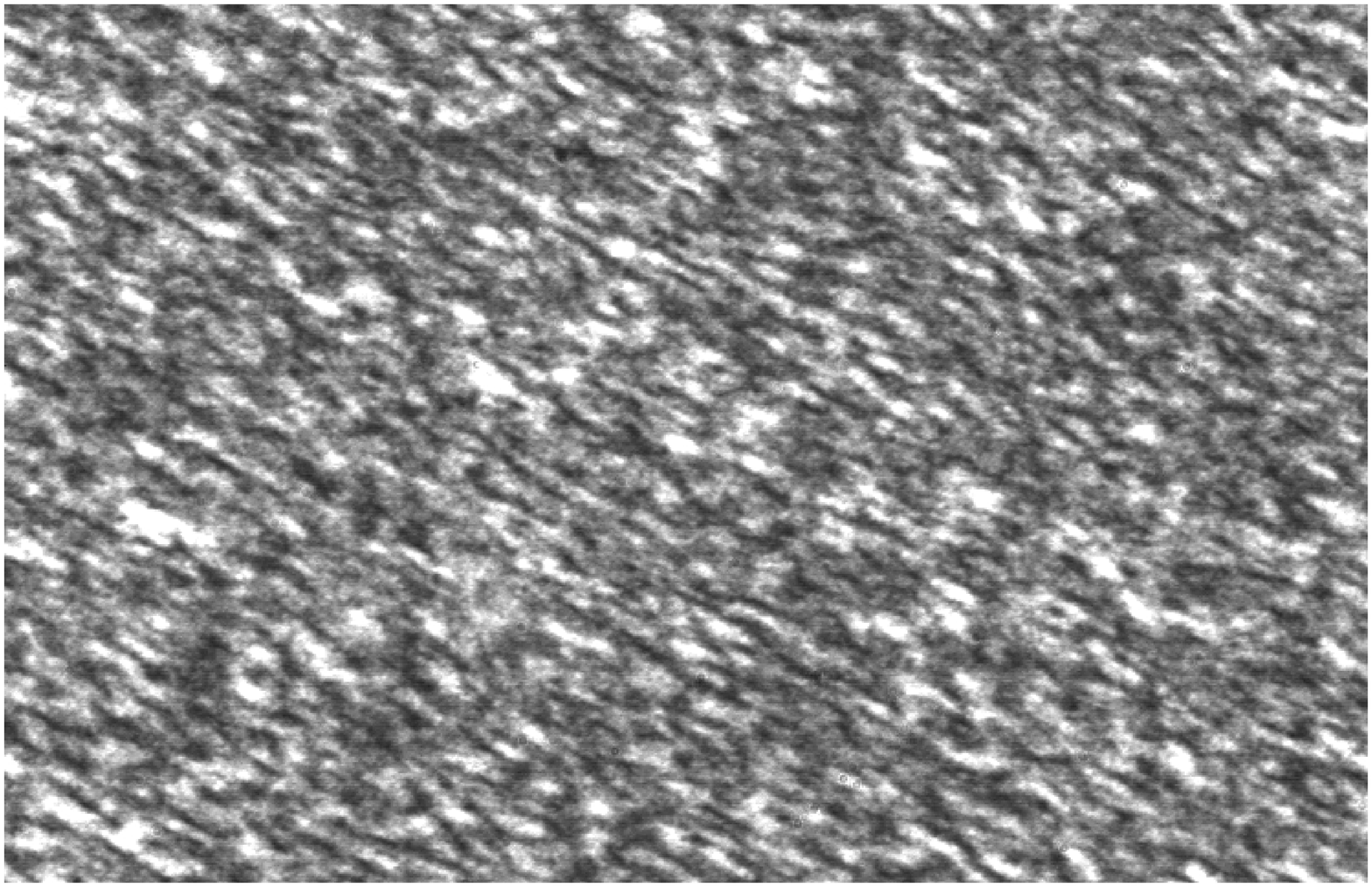}};
    \node[above of=speckle_image, node distance=1.5cm]{\textcolor{red}{Speckle image}};
    \node(triangle)[right of=speckle_image, node distance=4.0cm]{\includegraphics[scale=0.35]{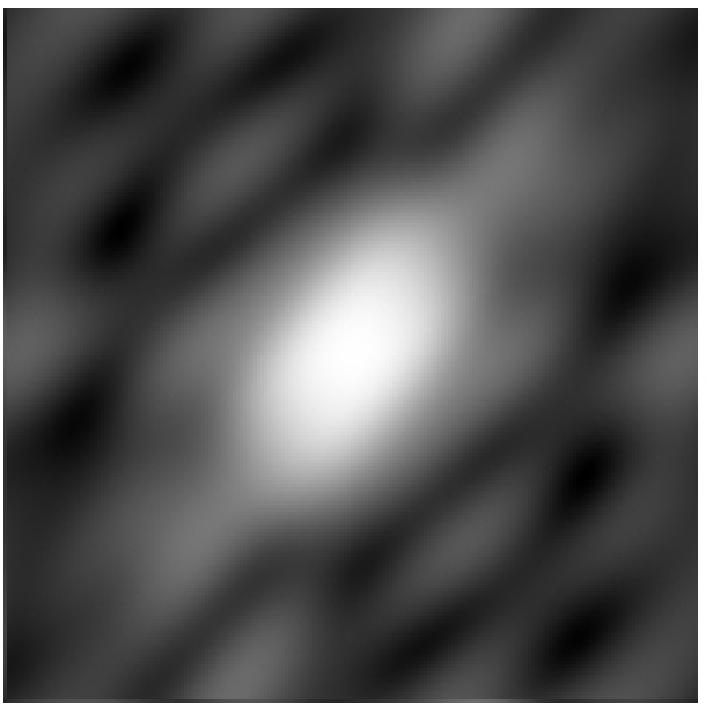}};
    \node[above of=triangle, node distance=1.5cm]{\textcolor{red}{Autocorrelation}};
    \node[above of=triangle, node distance=1.5cm]{\textcolor{red}{Autocorrelation}};
    \node(rec)[right of=triangle, node distance=4.0cm]{\includegraphics[scale=0.45]{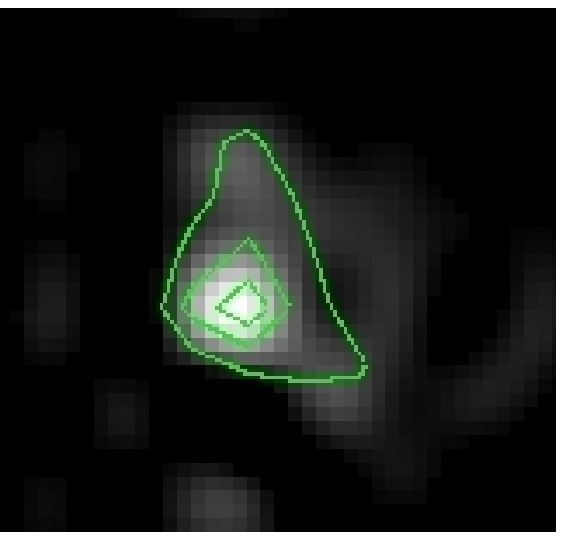}};
    \node[above of=rec, node distance=1.5cm]{\textcolor{red}{Reconstruction}};
    \draw[->, red](speckle_image)--(triangle);
    \draw[->, red](triangle)--node[above, sloped]{\small{Analysis}}(rec);
    \end{tikzpicture}
\end{center}
    \caption[Example image reconstruction from speckle data.]{\label{prelim_speckle} The autocorrelation is found for a speckle image obtained with a triangular-shaped mask. Image reconstruction techniques allow us to obtain the righ-most image.}
\end{figure}

Some preliminary results can be seen in Figure \ref{prelim_speckle}. In this example, a triangular-shaped mask is used. The figure shows the speckle pattern, its autocorrelation and corresponding image reconstruction. To obtain this image, the Cauchy-Riemann algorithm was applied, followed by 100 iterations of the Gerchberg-Saxton algorithm\footnote{The constraint applied in image space is that in each iteration, all pixels below 0.02/1.0 are set to zero.}. The experiment, which has been done in collaboration with Ryan Price and Erik Johnson, will be performed more carefully so that actual angular scales of the image can be obtained. 

\section{Starbase}

\begin{figure}[h]
  \begin{center}
    \includegraphics[scale=0.6]{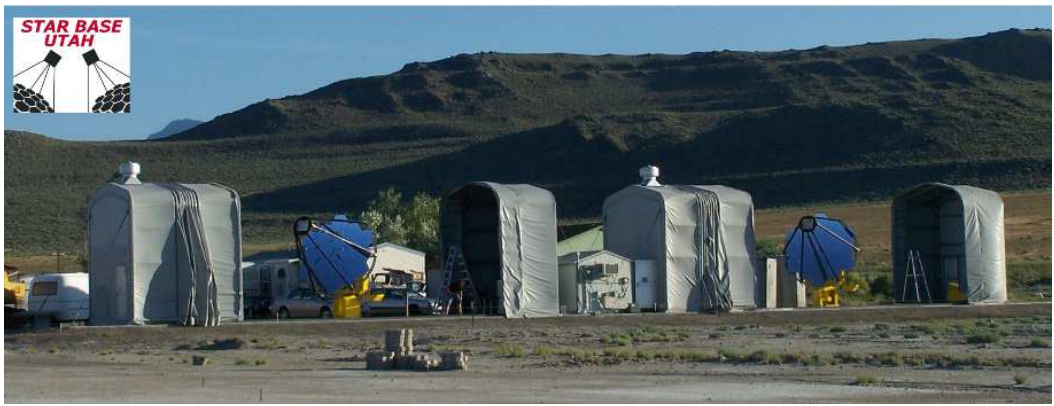}
  \end{center}
  \caption[StarBase telescopes.]{\label{starbase} The StarBase $3\,\mathrm{m}$ telescopes are protected by buildings which can be rolled open for observation. The control room is located in a smaller building located between the two telescopes.
  }
\end{figure}

As a first test toward implementing SII with IACT arrays, pairs of $12\,\mathrm{m}$ telescopes in the VERITAS array at the Fred Lawrence Whipple Observatory in Arizona were interconnected through digital correlators \citep{dravins_2008}. These tests were made during  nights shared with VHE observations with a very temporary setup and established the need for a dedicated test bench on which various options of secondary optics and electronics could be evaluated in a realistic environment. In order to satisfy this requirement, the two StarBase telescopes were deployed on the site of the Bonneville Seabase diving resort in Grantsville, Utah, 40 miles west from Salt Lake City. The two telescopes (Figure \ref{starbase}) are on a $23\,\mathrm{m}$ East-West base line. The telescopes had earlier been used in the Telescope Array experiment1 operated until 1998 on the Dugway proving range. Each telescope is a $3\,\mathrm{m}$, f/1 Davies-Cotton light collector \linebreak

\pagebreak

\noindent composed of 19 hexagonal mirror facets $\sim 60\,\mathrm{cm}$ across. This design is typically used for IACT and secondary optics tested on the StarBase telescopes could be used directly on the VERITAS telescopes for larger scale tests.

%\subsection{Science goals}

\subsection{Sensitivity and outlook}

Using Equation \ref{snr2} with conservative parameters for the StarBase telescopes ($A = 6\,\mathrm{m}^2$ , $\alpha = 0.2$ and $\Delta f = 100\,\mathrm{MHz}$), the 5 standard deviation measurement of a degree of coherence $|\gamma(d)|^2=0.5$ will require an observation time $T\approx 10\,\mathrm{min}\times 2.5^{2m_v}$ where $m_v$ represents the visual magnitude and where we have made the crude approximation $n = 5 \times 10^{−5} \times 2.5^{-m_v} \mathrm{m}^{−2} \mathrm{s}^{−1}  \mathrm{Hz}^{−1} $ . This corresponds to one hour for $m_v=1$ and 6.5 hours for $m_v=2$ and when considering the measurement of $|\gamma(d)|^2\approx1$, these observation times should be divided by four. 

 The first objective will be the detection of the degree of coherence for an unresolved object ($|\gamma(d)|^2\approx1$). The distance between the two telescopes being $23\,\mathrm{m}$ (smaller baselines can be obtained during observations to the east and to the west due to the projection effect), at $\lambda=400\,\mathrm{nm}$ the stars have to be smaller than typically $\sim 3\,\mathrm{mas}$ in diameter. An essentially unresolved star suitable for calibration should be less than $\sim 1\,\mathrm{mas}$ in diameter. Good candidates for this are in increasing order of magnitude $\alpha$-Leo, $\gamma$-Ori, $\beta$-Tau or even $\eta$-UMa which, should be observable as a unresolved object for calibration within 50 minutes \cite{stephan.spie}. Alternatively, it will be possible to measure any star as an unresolved object by correlating the signals from two channels mounted on the same telescope by means of the camera beam splitter. These observations should allow us to establish methods for adjusting the signal time delays optimally and also to identify the most effective correlator. A next phase will be dedicated to the measurement of a few bright stars in order to further demonstrate the technique. This second phase will possibly include the observation of coherence modulation resulting from orbital motion in  the binary star Spica with a $1.5\,\mathrm{mas}$ semi major axis and $m_v=1.0$, or even, possibly Algol ($2.18\,\mathrm{mas}$ semi major axis, $m_v=2.1$). 

\subsection{First data sample}

\begin{figure}[h]
        \begin{center}
        \includegraphics[scale=0.5, angle=-90]{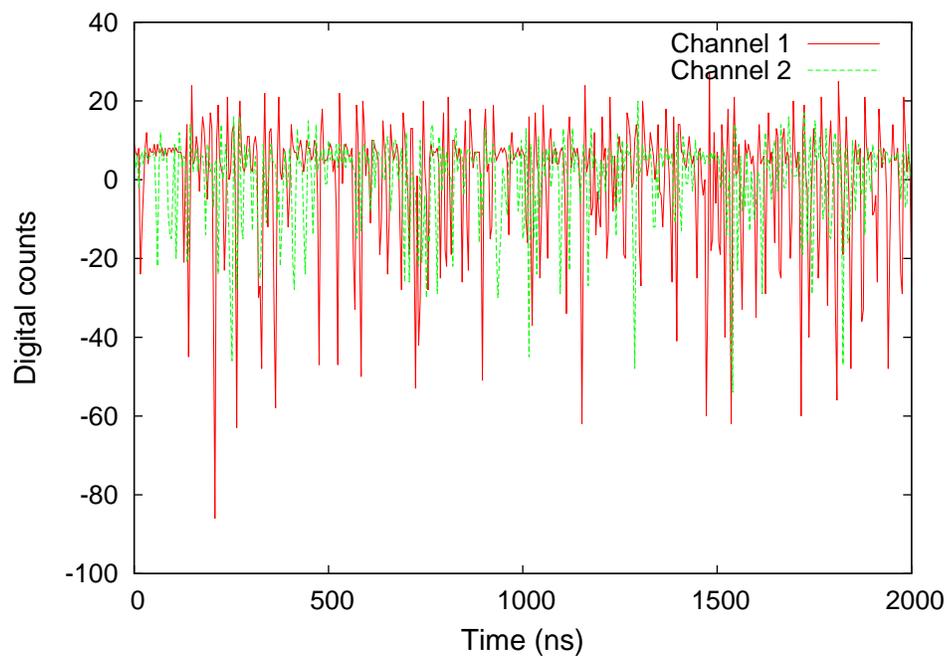}
        \end{center}
        \vspace{0.5cm}
        \caption{\label{spica_traces} The first individual signals obtained from the binary star \emph{Spica}.}
\end{figure}

\begin{figure}[h]
        \vspace{5cm}
        \begin{center}
        \includegraphics[scale=0.5, angle=-90]{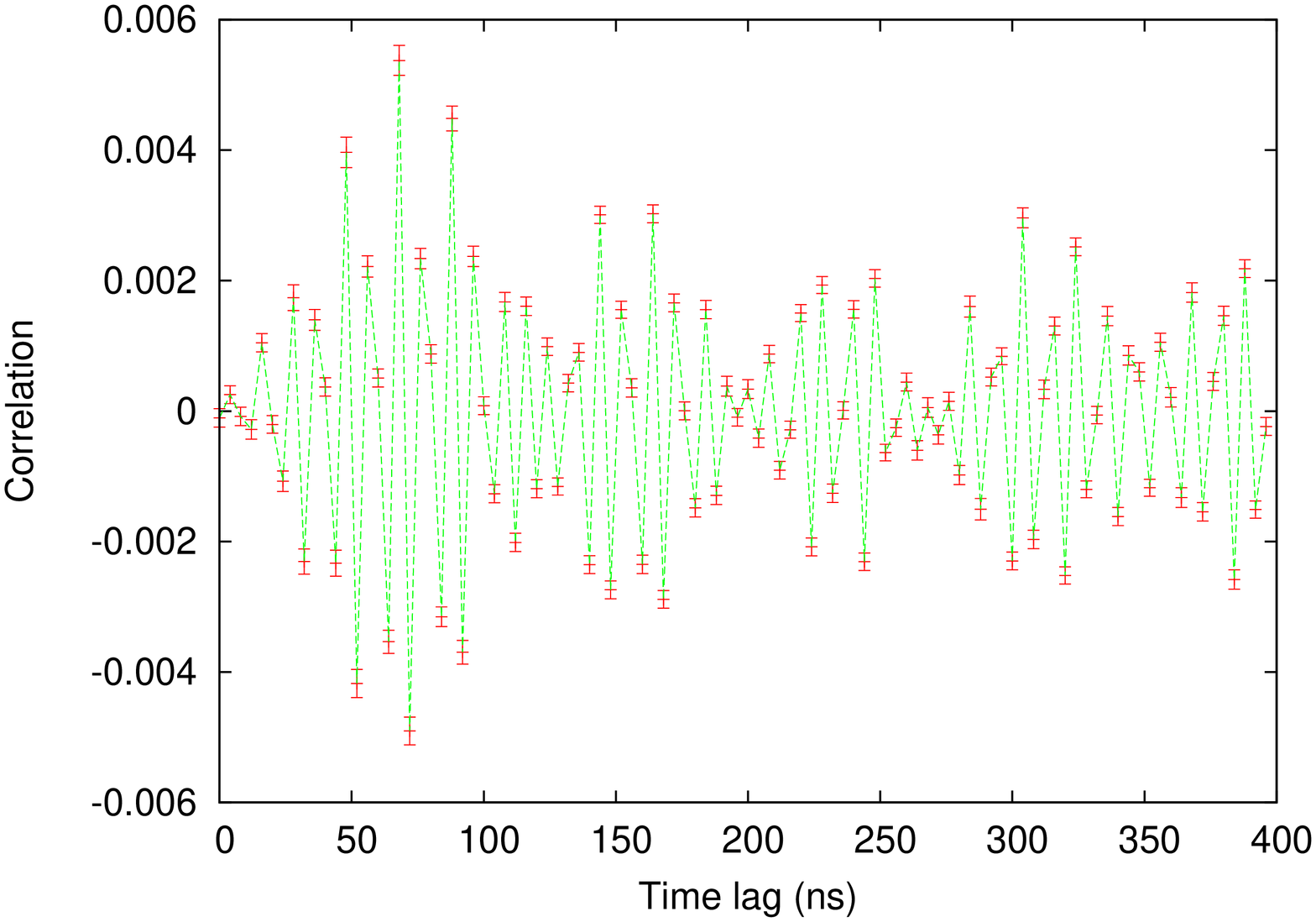}
        \end{center}
        \vspace{0.5cm}
        \caption[Degree of correlation for $1\,\mathrm{s}$ of data on \emph{Spica}]{\label{spica_corr} Degree of correlation as a function of the time lag between the two channels. This corresponds to $1\,\mathrm{s}$ of data.}
\end{figure}

The StarBase observatory recently began taking data of the binary star \emph{Spica}. Figure \ref{spica_traces} shows the digitized signal of each telescope. We have so far only analyzed $1\,\mathrm{s}$ of data and calculated the normalized correlation as a function of a time lag between the two signals as shown in Figure \ref{spica_corr}. At this very preliminary stage we do not expect to see a signal since an integration time of a few hours is necessary to detect a correlation (eq. \ref{snr2}). From Figure \ref{spica_corr}, we note that the fluctuations in the degree of correlation are comparable to what was obtained in the laboratory with the LED experiment when the LEDs were out of phase (section \ref{digital_system}). \linebreak

\pagebreak

\pagebreak

\noindent A next test at StarBase could be to perform the same LED experiment as was done in the laboratory.

\chapter{Conclusions}

Gamma-ray astronomy is typically not used for stellar astrophysics, but can still further our understanding of some isolated stellar systems. This is the case of $LSI+61^{\circ}303$, consisting of a main sequence Be star and a compact object, which has been detected in the TeV range with VERITAS. This object showed a clear intensity modulation as a function of the orbital phase. We describe a gamma-ray attenuation model and apply it to this system. With this model, we are able to constrain fundamental parameters of the system such as the mass of the compact object and the density of circumstellar matter around the Be star. However, important details of this source, such as the circumstellar matter distribution, may only be obtained from model-independent imaging at high angular resolution. Interferometry observations have already allowed us to reconstruct high angular resolution images of a few stars at optical wavelengths, but most bright and nearby stars cannot be resolved with current facilities. 

The recent success of gamma-ray astronomy, as well as the advancements in instrumentation and computing technology since the days of the Narrabri intensity interferometer, have prompted a revival of optical SII. Kilometric scale arrays of many large light collectors will allow an improvement in angular resolution by nearly an order of magnitude when compared with current optical amplitude interferometers. Several thousand stars will be resolved with such large arrays used as intensity interferometers,  and due to the very dense sampling of the $(u,v)$ plane, new mathematical (phase retrieval) algorithms will allow for high angular resolution images to be reconstructed from SII data. 

We performed a simulation study of the imaging capabilities at $400\,\mathrm{nm}$ of an IACT array consisting of 97 telescopes separated  up to  $1.4\,\mathrm{km}$. This is a preliminary design for the CTA project, expected to be operational in 2018. Our method uses a model-independent algorithm to recover the phase from intensity interferometric data. We tested the method on images of increasing degrees of complexity, parameterizing the pristine image, and comparing the reconstructed parameter with the pristine parameter. We now summarize our results and briefly discuss how fundamental stellar parameters can be constrained with the methods described in this thesis. 

We found that for bright disk-like stars ($m_v=6$, $T>6000^{\circ}\,\mathrm{K}$), radii are well reconstructed from $0.03\,\mathrm{mas}$ to $0.6\,\mathrm{mas}$.  Even though using a phase retrieval approach to recover images might not be the most efficient way to measure stellar radii, such a study  starts to quantify the abilities of measuring other scale parameters in more complicated images (e.g. oblateness, distance between binary components, star spots, etc.). The range of angular radii that can be measured with a CTA-like array ($0.03\,-\,0.6\,\mathrm{mas}$) will complement existing measurements ($2\,-50\,\mathrm{mas}$) \citep{Haniff}. With the aid of photometry, the effective temperature\footnote{Defined as $T_{eff}=\left(\frac{L}{4\pi R^2\sigma}\right)^{1/4}$, where $R$ is the radius, $L$ is the luminosity, and $\sigma$ is the Stephan-Boltzmann constant.} scale of stars within $0.03\,-\,0.6\,\mathrm{mas}$ can be extended.  %(later) Multi-wavelength angular diameter measurements will also reveal the wavelength dependence of limb-darkening \citep{Aufdenberg}, and such measurements can be used to constrain energy transport models as is done in amplitude interferometry \citep{procyon}. 

Binary stars are well reconstructed when one of the members is not much brighter (three times as bright) than the other, and when they are not too far apart ($\leq 0.75\,\mathrm{mas}$). As with amplitude interferometry, SII, along with spectroscopy, will allow us to determine the masses and orbital parameters in binary stars. If measured with enough precision ($\leq 2\%$) \citep{Andersen}, the determination of the mass can be used to test main sequence stellar models. An advantage of using an array such as the one used in this study, is that individual radii can be resolved. An interesting phenomena to be studied with interacting binary stars is mass transfer (e.g., \citet{beta_lyrae}), and capabilities for imaging this phenomena can be further investigated. 

For oblate stars, results similar to those obtained for disk-like stars are found. Due to the relative ease of SII to observe at short ($\sim 400\,\mathrm{nm}$) wavelengths, measuring fundamental parameters of hot B type stars is possible. B stars are particularly interesting since rapid rotation, oblateness, and mass loss are a common feature. We show that oblateness can be accurately measured, and the next step is to quantify the capabilities of imaging realistic surface brightness distributions in hot stars. By imaging brightness distributions, we will be able to study effects such  as limb darkening and mass loss in hot massive stars \citep{ridgway}, as well as gravity darkening \citep{von_zeipel}. 

To image nonuniform brightness distributions we recur to postprocessing routines that significantly increase the quality of the reconstructed images. Since the Cauchy-Riemann algorithm provides a reasonable first guess of a reconstruction, postprocessing routines consist of convergent iterative algorithms such as the Gerchberg-Saxton algorithm and the MiRA (Multi-aperture image Reconstruction Algorithm). The postprocessing significantly improves the image reconstruction, but the postprocessing routines by themselves are usually not sufficient for performing reconstructions, especially when the pristine image is not centro-symmetric. 

A study of the imageability of limb-darkening, which is related to the mass loss rate in hot stars ($T\sim 10,000^{\circ}\,\mathrm{K}$), indicates that realistic mass loss rates of the order of $10^{-7}M_{\odot}\mathrm{yr}^{-1}$ can be imaged. By simulating data corresponding to stars containing bright or cool spots,  we find that features with sizes of $\sim 0.05\,\mathrm{mas}$ and temperature differences of a few hundred $^\circ\mathrm{K}$ can be reconstructed. Photometric variability (at the $\sim10\%$ level) that coincides with the rotational period would provide indirect evidence of the presence of such stellar hot spots, and in fact, this type of photometric variability has been reported in the past \citep{be_variability}. The origin of such variability is still highly debated, and actual images would provide another mean to study this phenomenon. 

Several experiments are currently performing intensity correlation measurements as a preparatory stage for the use of IACT arrays as intensity interferometers. These efforts include the StarBase observatory and a laboratory experiment in the University of Utah. The Starbase telescopes have recently begun taking data and testing electronics. An intensity interferometry experiment that measures correlations from simulated stars emitting pseudo-thermal light (essentially incoherent light with long coherence times compared with thermal light)  has been performed. This experiment allowed us to measure angular diameters as well as to see the limitations of the electronics. However, the value of this experiment lies in that it enables an intuitive understanding of intensity interferometry in terms of speckles, i.e., when observing stars at a narrow optical bandwidth, detectors that are closer to each other than the typical size of a speckle, display a higher degree of correlation since they are on average detecting light corresponding to the same speckle. 

There are opportunities to further investigate SII simulations and experiments. For example, the simulated data used in this study to reconstruct images assumes point-like telescopes and does not include the effects of electronic noise or night-sky background. Some of these effects are currently being investigated \citep{janvida}. It is also possible to use more realistic pristine images that may be provided from Monte-Carlo radiative transfer models, such as those used for modeling the atmospheres of hot \emph{Be} stars \citep{carciofi}. Reconstructions from these pristine images may be used to investigate how radiative transfer models can be constrained. On the experimental side, we would like to make further attempts to measure correlations with an actual thermal source, using both the analog and digital systems. The StarBase telescopes have recently begun operating, and correlation measurements from small stars and binary systems will soon be performed. The lessons learned from these experiences will allow us to achieve the ultimate goal: To view stars as not mere unresolved point sources, but as the fascinating objects they truly are.

\newpage
\thispagestyle{empty}
\textcolor{white}{.}
\vspace{0.4in}
\begin{center}
\textbf{REFERENCES}
\end{center}
\bibliographystyle{apj}
\bibliography{thesis}

\end{document}